\documentclass[12pt,openany,oneside]{book}

\usepackage{amsxtra,amssymb,amsfonts,amsthm,amsmath,latexsym,verbatim,epsfig,stmaryrd}
\usepackage{diagrams}
\renewcommand{\baselinestretch}{1}

\theoremstyle{plain}
\newtheorem{definition}{Definition}%[section]
\newtheorem{theorem}{Theorem}%[section]
\newtheorem*{theorem*}{Theorem}
\newtheorem{conjecture}{Conjecture}%[section]
\newtheorem{lemma}{Lemma}%[section]
\newtheorem{corollary}{Corollary}%[section]
\newtheorem{remark}{Remark}%[section]
\newtheorem{example}{Example}%[section]
\newcommand{\refT}[1]{Theorem~\ref{T:#1}}

\newcommand{\refL}[1]{Lemma~\ref{L:#1}}
\newcommand{\refC}[1]{Corollary~\ref{C:#1}}
\newcommand{\refCj}[1]{Conjecture~\ref{Cj:#1}}

\newcommand{\refD}[1]{Definition~\ref{D:#1}}
\newcommand{\refE}[1]{Example~\ref{E:#1}}

\newcommand{\refS}[1]{Section~\ref{S:#1}}
\newcommand{\refCh}[1]{Chapter~\ref{S:#1}}

\def\ve{{\varepsilon}}

\def\lra{\longrightarrow}
\def\hra{\hookrightarrow}
\def\xra{\xrightarrow}

\def\lla{\longleftarrow}
\def\ovs{\overset}
\def\lrhu{\rightharpoonup}

\def\pr{\mathop{\rm pr}\nolimits}
\def\R{{\mathbb R}}
\def\T{{\mathbb T}}

\def\Z{{\mathbb Z}}
\def\C{{\mathbb C}}

\def\H{{\mathbb H}}
\def\E{{\mathcal E}}
\def\O{{\mathcal O}}

\def\g{{\mathfrak g}}
\def\h{{\mathfrak h}}

\def\su{\mathfrak{su}}
\def\uu{{\mathfrak u}}
\def\nn{{\mathfrak n}}

\def\m{{\mathfrak m}}
\def\g{{\mathfrak g}}
\def\h{{\mathfrak h}}

\def\Sum{\mathop\Sigma}
\def\i{\imath}
\def\pfi{\varphi}

\def\oH{\buildrel\circ\over H}
\def\oH1{\buildrel\circ\over H\kern-.02in{}^1}
\def\hookuparrow{{\cup\kern-.04in{}^\uparrow}}

\def\vert{\Vert}

\def\b{\mathbf b}

\def\Im{\mathop{\rm Im}\nolimits}

\def\Ker{\mathop{\rm Ker}\nolimits}

\def\Stab{\mathop{\rm Stab}\nolimits}
\def\End{\mathop{\rm End}\nolimits}
\def\Ext{\mathop{\rm Ext}\nolimits}
\def\Hom{\mathop{\rm Hom}\nolimits}
\def\pr{\mathop{\rm pr}\nolimits}
\def\Ad{\mathop{\rm Ad}\nolimits}
\def\ad{\mathop{\rm ad}\nolimits}
\def\Aut{\mathop{\rm Aut}\nolimits}

\def\id{\mathop{\rm id}\nolimits}
\def\pt{\mathop{\rm pt}\nolimits}
\def\CP{\C{\bf P}}

\def\tr{\mathop{\rm tr}\nolimits}
\renewcommand{\Re}{\mathop{\rm Re}\nolimits}

\def\bee{\begin{equation}}
\def\eee{\end{equation}}
\def\be{\begin{equation*}}
\def\ee{\end{equation*}}
\def\bal{\begin{aligned}}
\def\eal{\end{aligned}}

\begin{document}

\pagestyle{empty}

{\bf

\vspace{-.3 in}

\begin{center} {\Large  HOMOGENEOUS SPACES\\
and\\
                    FADDEEV-SKYRME MODELS} \end{center}

%\vspace{.15 in}

\begin{center} by \end{center}

\begin{center}  SERGIY KOSHKIN\end{center}

\begin{center} B.S., National Technical University of Ukraine, 1996\end{center}

\begin{center} M.S., National Academy of Ukraine Institute of Mathematics, 2000\end{center}

%\vspace{.1 in}

\begin{center} \underline{$\hspace{3.8 in}$} \end{center}

%\vspace{.1 in}

\begin{center} A DISSERTATION \end{center}

%\vspace{.1 in}

\begin{center} submitted in partial fulfillment of the \end{center}

\vspace{-.2 in}

\begin{center} requirements for the degree \end{center}

\vspace{.1 in}

\begin{center} DOCTOR OF PHILOSOPHY \end{center}

%\vspace{.1 in}

\begin{center} Department of Mathematics \end{center}

\vspace{-.2 in}

\begin{center} College of Arts and Sciences \end{center}

\vspace{.1 in}

\begin{center} KANSAS STATE UNIVERSITY \end{center}

\vspace{-.2 in}

\begin{center} Manhattan, Kansas \end{center}

%\vspace{.1 in}

\begin{center} 2006 \end{center}

\vspace{.1 in}

$\hspace{3.7 in}$ Major Professor

%\vspace{-.1 in}

$\hspace{3.7 in}$ David Auckly

}

\par\vfill\pagebreak

%%%%%%%%%%%%%%%%%%%%%%%%%%%%%%%%%%%%%%%%%%%%%

%ABSTRACT

%%%%%%%%%%%%%%%%%%%%%%%%%%%%%%%%%%%%%%%%%%%%%

\thispagestyle{empty}

\begin{center} {\bf \large ABSTRACT}\end{center}

\vspace{.5in}

We study geometric variational problems for a class of models in quantum field theory known as Faddeev-Skyrme models. Mathematically one considers minimizing an energy functional on homotopy classes of maps from closed 3-manifolds into homogeneous spaces of compact Lie groups. The energy minimizers known as Hopfions describe stable configurations of subatomic particles such as protons and their strong interactions. The Hopfions exhibit distinct localized knot-like structure and received a lot of attention lately in both mathematical and physical literature.

High non-linearity of the energy functional presents both analytical and algebraic difficulties for studying it. In particular we introduce novel Sobolev spaces suitable for our variational problem and develop the notion of homotopy type for maps in such spaces that generalizes homotopy for smooth and continuous maps. As the spaces in question are neither linear nor even convex we take advantage of the algebraic structure on homogeneous spaces to represent maps by gauge potentials that form a linear space and reformulate the problem in terms of these potentials. However this representation of maps introduces some gauge ambiguity into the picture and we work out 'gauge calculus' for the principal bundles involved to apply the gauge-fixing techniques that eliminate the ambiguity. These bundles arise as pullbacks of the structure bundles $H\hra G\to G/H$ of homogeneous spaces and we study their topology and geometry that are of independent interest.

Our main results include proving existence of Hopfions as finite energy Sobolev maps in each (generalized) homotopy class when the target space is a symmetric space. For more general spaces we obtain a weaker result on existence of minimizers only in each 2-homotopy class.

\par\vfill\pagebreak
%\thispagestyle{empty}

%%%%%%%%%%%%%%%%%%%%%%%%%%%%%%%%%%%%%%%%%%%%%

%TABLE OF CONTENTS

%%%%%%%%%%%%%%%%%%%%%%%%%%%%%%%%%%%%%%%%%%%%%

\pagestyle{plain}

\frontmatter

%\vspace*{-1cm}
\renewcommand{\baselinestretch}{1.295}
\tableofcontents
\renewcommand{\baselinestretch}{1.5}

\par\vfill\pagebreak

%%%%%%%%%%%%%%%%%%%%%%%%%%%%%%%%%%%%%%%%%%%%%

%ACKNOWLEDGEMENTS PAGE

%%%%%%%%%%%%%%%%%%%%%%%%%%%%%%%%%%%%%%%%%%%%%

\addcontentsline{toc}{chapter}{Acknowledgements}

\begin{center} {\bf \large ACKNOWLEDGEMENTS}\end{center}

I would like to thank my advisor D.Auckly for suggesting the problems studied in this thesis and patiently teaching me the tools such as obstruction theory, gauge theory, etc. that were necessary to solve them.

\par\vfill\pagebreak

%%%%%%%%%%%%%%%%%%%%%%%%%%%%%%%%%%%%%%%%%%%%%

%DISSERTATION TEXT

%%%%%%%%%%%%%%%%%%%%%%%%%%%%%%%%%%%%%%%%%%%%%

%\pagenumbering{arabic}\setcounter{page}{1}

\mainmatter

\chapter{Introduction}

\section{Preliminaries}\label{S:01}

The subject of this thesis is a mathematical study of a class of
non-linear $\sigma$--models that arise in quantum field theory. We
call them {\it Faddeev-Skyrme models} although other names are also
used in the literature \cite{GP,Mn}. Mathematically one has a
variational problem with topological constraints for maps from a
$3$--manifold into homogeneous spaces. The solution requires some
extensive incursions into geometry and topology of such maps that
are of independent interest. This section gives some historical
perspective on the problem and its mathematical treatment.

In 1961 an English physicist T.H.R. Skyrme introduced a new model
describing strong interactions of quantum fields corresponding to
mesons. The fields of the model are maps from $\R^3$ into $S^3$. The
$3$--sphere is interpreted as the group $SU_2$ of unimodular unitary
complex $2\times2$ matrices and only maps converging to the identity
matrix at infinity are considered. Skyrme's idea was to add to the
standard Dirichlet energy
$$
E_2(\psi):= \frac{1}{2} \int_{\R^3} |d\psi |^2 dx
$$
an additional stabilizing term $$ E_4(\psi):= \frac{1}{4}
\int_{\R^3} |d\psi\wedge d\psi|^2 dx
$$
that would prevent stationary fields from being singular as it
happens for harmonic maps. Here the derivative $d\psi$ takes values
in the corresponding matrix Lie algebra $\su_2$ and the wedge
product $d\psi\wedge d\psi:=\Sum_{i<j} \frac{\partial\psi}{\partial
x_i}\frac{\partial\psi}{\partial x_j} dx^i\wedge dx^j$ is defined
using the matrix multiplication. Because of the condition at
infinity the maps $\psi$ can be identified via the stereographic
projection with maps from $S^3$ to $S^3$ and one can talk about
their topological degree. This degree serves as a constraint when
minimizing the Skyrme functional
\begin{equation}\label{e0.1}
E(\psi)=\int_{\R^3} \frac12 |d\psi|^2\, +\,\frac14 |d\psi\wedge
d\psi|^2\;dx\,,
\end{equation}
without a constraint constant maps are obviously the only absolute
minimizers.

If the $\R^3$ above is replaced by $\R^2$ and only maps with a
certain symmetry are considered the Euler-Lagrange equations for the
Skyrme functional are related to the sin-Gordon equation that admits
solitons as solutions \cite{DFN}. It was expected that solitonic
behavior is preserved in the $3$--dimensional case as well. Skyrme
conjectured that the solitons should be interpreted as combinations
of baryonic particles (protons, neutrons, etc.) and the degree of a
map gives the number of such particles, the baryonic number.
Solitonic behavior is then explained by topological reasons --
evolution (i.e. a homotopy) does not change the degree of a map.
Solitons of this kind are now called topological \cite{MS}. After
the appearence of the Standard Model of quantum interactions and
some experimental evidence the Skyrme model became accepted as an
effective description of meson-baryon interaction.

The Skyrme model was later generalized to consider maps from $\R^3$
into $G$, where $G$ is a compact semisimple Lie group \cite{DFN}.
$G$ is represented by unitary or orthogonal matrices and the
functional has the same form (\ref{e0.1}). If the metric on $G$ is
bi-invariant $|d \psi|=|d\psi\,\psi^{-1}|$ and $|d\psi\wedge
d\psi|=|d\psi\,\psi^{-1}\wedge d\psi\,\psi^{-1}|$ so the functional
can be written more intrinsically as
\begin{equation}\label{e0.2}
E(\psi)=\int_{\R^3} \frac12 |d\psi\,\psi^{-1}|^2\, +\,\frac1{16}|[d
\psi\,\psi^{-1},d \psi\,\psi^{-1}]|^2\;dx\,.
\end{equation}
where  $[a,a]$ is the Lie bracket of $\g$--valued forms ($\g$ is the
Lie algebra of $G$ ). In this form it is explicitly independent of a
matrix representation of $\g$.

More Skyrme-type models emerge if one considers maps $\R^3\overset
{\psi}\lra G/H $ into the coset space of $G$ by a closed subgroup
$H$. The first model of this kind was introduced by L.D.Faddeev in
1975 \cite{Fd1,Fd2}. In his case $G/H=SU_2/U_1\simeq S^2$ and one
can define energy by simply restricting (\ref{e0.1}) to the
$S^2$--valued maps via the equatorial embedding $S^2\hra S^3$. 
As in the case of maps $S^3\to S^3$ whose homotopy type is
characterized by a single number the homotopy type of maps $S^3\to
S^2$ is given by the Hopf invariant. It was expected that this model
will also exhibit solitonic behavior for the same topological
reasons. Moreover, unlike in the case of the original Skyrme model
the center of a soliton would be not a single point but a closed
loop, possibly knotted (recall that the Hopf invariant of a map is
given by the linking number of the preimages of two generic points
in $S^3$ \cite{Ha}). This remained a conjecture until 1997 when Faddeev 
and A.Niemi used computer modelling to show that energy minimizers 
of the Faddeev functional do have knot-like structure \cite{FN1}. 
Their result was later confirmed by more extensive computations in \cite{BS1}.

In 1980-s physicists began to consider models for maps taking values
in more general homogeneous spaces (see historical remarks in
\cite{BMSS}). They were motivated by attempts to construct
'effective' theories that describe the behavior of the Standard
Model fields in asymptotic situations. For instance, the hypothesis
of Abelian Dominance suggested by G.'tHooft \cite{'tH} leads to
effective theories for maps taking values in a coset space $G/\T$
with $\T$ a maximal torus of $G$. E.Witten and his collaborators
\cite{ANW,Wt1,Wt2} studied models with $G/H$ being symmetric spaces.
Based on some earlier work of Y.M. Cho \cite{Cho1,Cho2} Faddeev and
Niemi conjectured in 1997 that the low-energy limit of $SU_N$
Yang-Mills theory is described by an $SU_N/\T$ Skyrme-type model
\cite{FN1,FN2}. Since then the Faddeev-Niemi conjecture has received
considerable attention in the physics literature
\cite{Fd3,CLP,Sh1,Sh2}.

Mathematical treatment of the Skyrme model and its generalizations
has not been very extensive. Skyrme suggested to look for minimizers
that have some special symmetry, the so-called hedgehog ansatz (see
\cite{GP}). In 1983 L.Kapitansky and O.Ladyzhenskaya proved the
existence of minimizers among maps with such symmetry for the Skyrme
model on $\R^3$. In two papers \cite{Es1,Es2} M.Esteban apllied the
concentration-compactness method of P.-L.Lions \cite{Ln} to prove
existence of minimizers among maps of the degree $\pm 1$. There was
a gap in her proofs that was fixed later \cite{Es3,LY2}. As for the
energy minimizers (Skyrmions) with higher topological degrees their
existence remains elusive to this day (see the discussion in
\cite{LY2}). On the other hand, if one replaces $\R^3$ in
(\ref{e0.1}),(\ref{e0.2}) by a closed $3$-manifold $M$ the problem
becomes more tractable. Existence of minimizers in all homotopy
classes has been established in \cite{Kp} for maps $M\to S^3$ and
more generally for maps $M\to G$ in \cite{AK1}.

In the case of the Faddeev model the story is even shorter. Back in
1979 L.Kapitansky and A.Vakulenko proved a low energy bound for
Skyrme energy of maps in terms of their Hopf invariant which was
later improved by several authors \cite{MRS,Wr}. An existence theory
for this model has been developed in \cite{LY1} on $\R^2$ and
\cite{LY2} on $\R^3$. The authors use the concentration-compactness
method and the following two-sided inequality
\begin{equation*}
 C^{-1}|Q_\psi|^{3/4}\leq E(\psi)\leq C|Q_\psi|^{3/4}
\end{equation*}
that complements previously known lower bounds by an upper bound
($Q_\psi$ is the Hopf invariant of $\psi$). Sublinear growth of
energy along with existence of minimizers for $Q=\pm 1$ ensures that
there are minimizers with arbitrarily large Hopf numbers (although
for every concrete value, say $Q=2$ one can not tell if a minimizer
exists). For the original Skyrme model the energy growth in terms of
the degree is linear \cite{GP} and one can not apply the same
argument. As before the situation improves when $\R^3$ is replaced
by a closed $3$-manifold $M$. Existence of minimizers in every
homotopy class of maps $M\to S^2$ is proved by D.Auckly and
L.Kapitansky in \cite{AK2}.

For more general homogeneous target spaces $X=G/H$ it is not
immediately obvious how to generalize the functionals (\ref{e0.1}),
(\ref{e0.2}). N.Manton suggested to interpret $d\psi\wedge d\psi$
simply as an element of $\psi^*TX\otimes\psi^*TX$ in which case
(\ref{e0.1}) makes sense for an arbitrary Riemannian manifold $X$ as
a target \cite{Mn}. However, this functional does not coincide with
the usual Skyrme functional (\ref{e0.2}) for Lie groups except
in the case of $SU_2$. Faddeev and Niemi suggested a version of the
functional for the flag manifold $SU_N/\T$ in \cite{FN2} but their
way of introducing it only works for this particular case. To the
best of our knowledge the existence of minimizers for such models
was not considered in the literature. In fact, the only  result in
this direction is a generalization of the low energy bound to
$SU_N/\T$ model by S.Shabanov \cite{Sh2}.

There is however a natural generalization of
(\ref{e0.1}),(\ref{e0.2}) that works for arbitrary homogeneous
spaces and reduces to the previously considered functionals in the
cases of Lie groups and flag manifolds. If $dg\,g^{-1}$ is the
Maurer-Cartan form on $G$ then
$d\psi\,\psi^{-1}=\psi^*(dg\,g^{-1})$. Let $\h^\perp$ be the orthogonal
complement to the Lie algebra of $H$ with respect to some invariant
metric on $\g$ (e.g. the Cartan-Killing metric). One can see that
the form $g\pr_{\h^\perp}(g^{-1}\,d g)g^{-1}$ is horizontal and
invariant under the left action of $H$ on $G$ and therefore descends
to a $\g$--valued form $\omega^\perp$ on $G/H$. More precisely, if
$G\overset{\pi}{\lra}G/H$ is the quotient map we define
\begin{equation}\label{e0.3}
\pi^*\omega^\perp:=g\pr_{\h^\perp}(g^{-1}\,d
g)g^{-1}=\Ad_*(g)\pr_{\h^\perp}(g^{-1}\,d g)
\end{equation}
and call $\omega^\perp$ {\it the coisotropy form} of $G/H$.
Obviously when $H$ is trivial $\omega^\perp$ reduces to
$dg\,g^{-1}$. Hence for a map $M\overset{\psi}{\lra}G/H$ the
Faddeev-Skyrme energy can be defined as
\begin{equation}\label{e0.6}
E(\psi)=\int_M\frac12|\psi^*\omega^\perp|^2\,
+\,\frac14|\psi^*\omega^\perp\wedge\psi^*\omega^\perp|^2\;dm\,.
\end{equation}
and it turns into (\ref{e0.2}) for Lie groups. In this work we refer
to minimization problems for the functional (\ref{e0.6}) on
homotopy classes of maps $M\to G/H$ as {\it Faddeev-Skyrme models}.

The kinds of difficulties we encounter and the kinds of methods we
use are very different from those in the recent papers
\cite{LY1,LY2} on the Faddeev model. We do not have to deal with
effects at infinity since the domain $M$ is compact but the topology
of a general $3$--manifold is more complicated than that of $\R^3$
or $S^3$. Much work is required to describe the homotopy properties
of maps $M\to G/H$ in a way that relates them to the functional
(\ref{e0.6}). In this endeavor we follow the ideas of \cite{AK1,AK2}
on the Skyrme and Faddeev models. In particular, we represent maps
by connections and use formalism of the gauge theory to analyze
them.

\section{Main results}\label{S:02}

We consider Faddeev-Skyrme models for $M$ being a closed
$3$-manifold and $X=G/H$ being a simply connected homogeneous space
of a compact Lie group $G$. Mathematically we wish to minimize the
functional (\ref{e0.6}) on a homotopy class of maps.
As might be expected the space of continuous maps is insufficient to
contain  minimizers and has to be enlarged. Before we can describe
the suitable class of admissible maps we need as in \cite{AK1,AK2} a
description of the homotopy classes more 'explicit' than the one given
in the algebraic topology.

If $H^2(M,\Z)\neq0$ homotopy classes of maps $M\to X$ are no longer
indexed by a single invariant such as the degree or the Hopf number.
By the Postnikov classification theorem \cite{Bo,Ps,WJ} there is a
primary invariant (the $2$-homotopy type) defined for any map and a
secondary invariant defined only for pairs of maps that have the
same primary invariant. It turns out that if $X$ is simply connected
it admits a representation $X=G/H$, where $G,H$ are connected and
$G$ is compact and simply connected. Using such a representation we
have
\begin{theorem}\label{T:01}
Two continuous maps $M\overset{\pfi,\psi}{\lra}X$ are $2$-homotopic
if and only if there exists a continuous map $M\overset{u}{\lra}G$
such that $\psi=u\pfi$.
\end{theorem}
Now the secondary invariant can be defined explicitly in terms of
$u$. Since $G$ is simply connected and $\pi_2(G)=0$ for any Lie
group one has $\pi_3(G)\simeq H_3(G,\Z)$ by the Hurewicz theorem.
Let $\b_G\in H^3(G,\pi_3(G))$ denote  {\it the basic class}  of $G$,
i.e. the one that corresponds to every homology $3$--cycle in $G$
its image in $\pi_3(G)$ under the Hurewicz isomorphism
\cite{St,DK,MT}. Then $u^*\b_G$ is the secondary invariant for the
pair $\pfi,\psi$.

If $H^2(M,\Z)=0$ as for example in the case of $M=S^3$ then
\refT{01} says that any two maps are related by a map into $G$. In
particular we can choose $\pfi$ to be the constant map and define
the secondary invariant for a single map $\psi$ instead of a pair.
One can view it as a generalization of the Hopf invariant.

In general it is not necessary that the secondary invariant vanish
for $\pfi$ and $\psi$ to be homotopic. In fact there are maps
$M\overset{w}{\lra}G$ with $w^*\b_G\neq0$ but $w\pfi=\pfi$. For a
correct statement we have to factor out the subgroup generated by
such maps:
\begin{equation}\label{e0.8}
\O_\pfi:=\{w^*\b_G\mid w\pfi=\pfi\}<H^3(M,\pi_3(G)).
\end{equation}
In the case of the classical Hopf invariant this subgroup is
trivial.
\begin{theorem}\label{T:02}
Let $M\overset{\pfi}{\lra}X$ and $M\overset{u}{\lra}G$ be continuous
maps. Then  $\pfi$ and $\psi=u\pfi$ are homotopic if and only if
$u^*\b_G\in\O_\pfi$. The subgroup $\O_\pfi$ only depends on the
$2$-homotopy type of $\pfi$ and not on the map itself.
\end{theorem}
To get an integral representation for the secondary invariant we
need a deRham representative for the basic class $\b_G$. This has
been worked out in \cite{AK1} and we briefly recall the construction
here. If $G$ is a simple group then $H^3(M,\pi_3(G))\simeq\Z$ and
$\b_G$ is represented by an integral real-valued form $\Theta$ on
$G$. Explicitly 
$$
\Theta:=c_G\tr(g^{-1}dg\wedge g^{-1}dg \wedge g^{-1}dg ),
$$
where $c_G$ are numerical coefficients computed in
\cite{AK1} for every simple group. Thus
\begin{equation}\label{e0.9}
u^*\Theta=c_G\tr(u^{-1}du\wedge u^{-1}du \wedge u^{-1}du).
\end{equation}
In general if $G$ is compact and simply connected then
$G=G_1\times\dots\times G_N$, where $G_k$ are simple groups. Since
$\pi_3(G)=\pi_3(G_1)\oplus\dots\oplus\pi_3(G_N)\simeq\Z^N$:
$$
H^3(M,\pi_3(G))\simeq
H^3(M,\Z)\otimes\pi_3(G)\simeq\Z\otimes\Z^N\simeq\Z^N
$$
and we identify $H^3(M,\pi_3(G))$ with $\Z^N$. Therefore $\b_G$ is
represented by an integral vector-valued form, namely
$\Theta:=(\Theta_1,\dots,\Theta_N)$, where
$$
\Theta_k:=c_{G_k}\tr(\pr_{\g_k}(g^{-1}dg)\wedge\pr_{\g_k}(g^{-1}dg)\wedge\pr_{\g_k}(g^{-1}dg))
$$
and $\g_k$ are the Lie algebras of $G_k$. Accordingly $\O_\pfi$ from
(\ref{e0.8}) becomes a subgroup of $\Z^N$.

We can now handle Sobolev maps by picking a smooth reference map
$\pfi$ to fix a $2$-homotopy type and allowing $u$ to be a Sobolev
map. To fix a homotopy type we require in addition that $\int_M
u^*\Theta\in\O_\pfi$.

The next step is to relate our topological description to the
functional (\ref{e0.6}). It helps to restate the minimization
problem in terms of $u$ and $\pfi$. To this end consider the following
{\it isotropy subbundles} of $M\times G$:
\begin{equation}\label{e0.10}
\begin{aligned}
H_\pfi & :=\{(m,\gamma)\in M\times G\mid\pfi(m)=gH,\ g^{-1}\gamma g\in H\},\\
\h_\pfi & :=\{(m,\xi)\in M\times\g\mid\pfi(m)=gH,\ g^{-1}\xi
g\in\h\}.
\end{aligned}
\end{equation}
Sections of $M\times G$ are just maps from $M$ to $G$ and one can
see that sections of $H_\pfi$ are exactly the maps from the {\it
stabilizer} of $\pfi$ (cf. (\ref{e0.8})):
\begin{equation}\label{e0.11}
\Stab_\pfi:=\{{w\colon}M\to G\mid w\pfi=\pfi\}.
\end{equation}
For $\g$--valued forms $\alpha$ we get the corresponding {\it
isotropy decomposition}:
\begin{equation}\label{e0.12}
\alpha=\pr_{\h_\pfi}(\alpha)+\pr_{\h^\perp_\pfi}(\alpha)=:\alpha^\vert+\alpha^\perp.
\end{equation}
Following \cite{AK1,DFN} we introduce the {\it potential} of $u$ by
$a:=u^{-1}du$. This is indeed the gauge potential of a flat
connection on the trivial bundle $M\times G$ \cite{MM}. Define
$$
D_\pfi a:=a^\perp+\pfi^*\omega^\perp
$$ 
then the Faddeev-Skyrme functional (\ref{e0.6}) for $\psi=u\pfi$ becomes
\begin{equation}\label{e0.13}
E_\pfi(a)=\int_M\frac12|D_\pfi a|^2\, +\,\frac14|D_\pfi a\wedge
D_\pfi a|^2\;dm.
\end{equation}
Note also that $u^*\Theta$ in (\ref{e0.9}) also has a very simple
expression in terms of $a$:
\begin{equation}\label{e0.14}
u^*\Theta=c_G\tr(a\wedge a\wedge a)
\end{equation}
and this is the Chern-Simons invariant of $a$ since $da=-a\wedge a$.

Let us consider the spaces of maps and potentials suitable for
minimizing the functional (\ref{e0.13}). We use two such spaces. The
first is the space $\E(M,G)$ of {\it admissible maps} $u$ described
in terms of their potentials $a=u^{-1}du$ as follows:
\begin{equation}\label{e0.17}
\begin{aligned}
& {\rm 1)}\ a^\perp\in L^2(\Lambda^1M\otimes\g);\\
& {\rm 2)}\ a^{\perp}\wedge a^{\perp}\in L^2(\Lambda^2M\otimes\g);\\
& {\rm 3)}\ a^\vert\in W^{1,2}(\Lambda^1M\otimes\g).
\end{aligned}
\end{equation}
The second is the sequentially weak closure $\E'(M,G)$ of
$C^\infty(M,G)$ in $\E(M,G)$ with respect to the following weak
convergence:
\begin{equation}\label{e0.18}
\begin{aligned}
& {\rm 1)}\ u_n\overset{W^{1,2}}\lrhu u;\\
& {\rm 2)}\ a_n^{\perp}\wedge a_n^{\perp}\overset{L^2}\lrhu a^{\perp}\wedge a^{\perp};\\
& {\rm 3)}\ a_n^\vert\overset{W^{1,2}}\lrhu a^\vert,
\end{aligned}
\end{equation}
where of course $a_n=u_n^{-1}du_n$ and $a=u^{-1}du$.

In view of \refT{01} we say that a Sobolev map
$M\overset{\psi}{\lra}X$ is in the {\it $2$-homotopy sector} of
$\pfi$ if $\psi=u\pfi$ for $u\in\E(M,G)$ (if $\psi$ happens to be
continuous it will indeed be $2$-homotopic to $\pfi$). Maps $M\to X$
that are in a $2$-homotopy sector of some smooth map are also called
{\it admissible}.
\begin{theorem}\label{T:03}
Every $2$-homotopy sector of admissible maps $M\to X$ has a
minimizer of the Faddeev-Skyrme energy.
\end{theorem}

As far as the secondary invariant (\ref{e0.14}) is concerned note
that if $u\in\E(M,G)$ we only know that $a\in
L^2(\Lambda^1M\otimes\g)$ and  $a\wedge a\wedge a$ is not defined
even as a distribution. However $a=a^\vert+a^\perp$ and due to the
cyclic property of traces one has for smooth forms
\begin{equation*}
c_G\tr(a\wedge a\wedge a)=c_G(\tr(a^\vert)^{\wedge
3}+3\tr((a^\vert)^{\wedge 2}\wedge
a^\perp)+3\tr(a^\vert\wedge(a^\perp)^{\wedge
2})+\tr(a^\perp)^{\wedge 3}).
\end{equation*}
By (\ref{e0.17}) the righthand side is in $L^1(\Lambda^3M)$ and we
take it as the {\it definition} of $u^*\Theta$ for $u\in\E(M,G)$ and
a simple group $G$.  Applying the above decomposition to each simple
component one can define $u^*\Theta$ in the general case as well.

A Sobolev map $M\overset{\psi}{\lra}X$ is in the {\it homotopy
sector} of $\pfi$ if $\psi=u\pfi$ for $u\in\E'(M,G)$ and
$\int_Mu^*\Theta\in\O_\pfi$. By \refT{02} this does mean 'homotopic'
if $\psi$ is continuous. Maps $M\to X$ that are in a homotopy sector
of some smooth map are called {\it strongly admissible}.
\begin{theorem}\label{T:04}
Let $X$ be a symmetric space. Then every homotopy sector of strongly
admissible maps $M\to X$ has a minimizer of the Faddeev-Skyrme
energy.
\end{theorem}
Note that it is quite possible that admissible and strongly
admissible maps are the same class (that may also coincide with the
class of $W^{1,2}$ maps with finite Faddeev-Skyrme energy). This is
a question that we do not address in this work. It is related to 
difficult problems of approximating Sobolev maps into
manifolds by smooth maps \cite{Bt,HL1,HL2} and establishing integrality of 
cohomological invariants for Sobolev maps and connections 
\cite{AK3,EM,LY2,Ul2}. 

Let us say a few words about the role the gauge theory plays in
proving Theorems \ref{T:03}, \ref{T:04}. When we attempt to minimize
(\ref{e0.13}) the following problem presents itself. The choice of
$u$ in \refT{01} is not unique: without changing $\psi$ it can be
replaced by $uw$, where $w$ is an element of the stabilizer
$\Stab_\pfi$. Since the functional (\ref{e0.13}) only depends on
$\psi$ it remains invariant under this change and therefore admits a
non-compact group of symmetries as a functional of $u$ (or $a$). As
a result sets of maps with bounded energy are not weakly compact in
any reasonable sense. This sort of problem is well known in the
gauge theory, where the group of symmetries is the gauge group of a
principal bundle acting on connections. The gauge theory also gives
a way out: one has to {\it fix the gauge} \cite{FU,MM}. This is more than a mere
analogy, the entire problem of minimizing (\ref{e0.13}) can be
reduced to a gauge theory problem and solved as such. We give some
details below.

The isotropy subbundles admit the following gauge-theoretic
interpretation. Consider the quotient bundle of a homogeneous space:
$H\hra G\to G/H$. This is a smooth principal bundle, call it $P$
and so is its pullback $\pfi^*P$ under a map $M\overset{\pfi}\lra
G/H$. Then one has the bundles $\Ad(\pfi^*P)$ (gauge group bundle)
and $\Ad_*(\pfi^*P)$ (gauge algebra bundle) associated to it in the
usual way \cite{FU,MM}. In the next theorem we combine several results
from Chapter 2 ($\Gamma(Q)$ denotes sections of a bundle $Q$):

\begin{theorem}\label{T:05}

\noindent {\rm (i)} The bundles $H_\pfi$ and $\Ad(\pfi^*P)$ are
isomorphic and identify gauge transformations on $\pfi^*P$ with
maps from $\Stab_\pfi$.

\noindent {\rm (ii)} The bundles $\h_\pfi$ and $\Ad_*(\pfi^*P)$ are
isomorphic. This isomorphism induces isomorphisms on differential
forms under which gauge potentials and curvatures of connections on
$\pfi^*P$ are identified with $\h_\pfi$--valued (and hence
$\g$--valued) forms.

\noindent {\rm (iii)} Under the above identifications the gauge
action of $w\in\Stab_\pfi$ on $b\in
\Gamma(\Lambda^1M\otimes\h_{\pfi})$ is:
\begin{equation}\label{e0.15}
b^w=w^{-1}bw+w^{-1}dw+(w^{-1}(\pfi^*\omega^\perp)w-\pfi^*\omega^\perp)
\end{equation}
and the curvatures of $b$, $b^w$ are:
\begin{equation}\label{e0.16}
\begin{aligned}
F(b)&=db+b\wedge b-[b,\pfi^*\omega^\perp]-(\pfi^*\omega^\perp\wedge\pfi^*\omega^\perp)^\vert\\
F(b^w)&= w^{-1}F(b)w,
\end{aligned}
\end{equation}
where we set $[\alpha,\beta]:=\alpha\wedge\beta+\beta\wedge\alpha$ (plus!) for $1$--forms $\alpha,\beta$.
\end{theorem}
\noindent If $\pfi$ is a constant map then $\pfi^*\omega^\perp=0$
and the formulas for gauge action and curvature reduce to the
familiar ones for trivial bundles \cite{DFN,FU,MM}.

It turns out that the {\it isotropic part}
$a^\vert:=\pr_{\h_\pfi}(a)$  gives the gauge potential of a
connection on the subbundle $\pfi^*P\subset M\times G$ under the
identification of \refT{05}{\rm (ii)}. Moreover, if $u$ is replaced
by $uw$ and hence $a$ is replaced by $a^w:=(uw)^{-1}d(uw)$ then
$(a^w)^\vert=(a^\vert)^w$, where on the right we have the expression
from (\ref{e0.15}). In other words, as far as the isotropic parts
are concerned {\it the action of $\Stab_\pfi$ on maps $M\to G$ is
conjugate to the action of the gauge group $\Gamma(\Ad(\pfi^*P))$
on connections}. \refT{05}{\rm (iii)} along with the flatness of $a$
implies that
\begin{equation}\label{e0.19}
F(a^\vert)=d(\pr_{\h_\pfi})\wedge a^\perp-(a^\perp\wedge a^\perp)^\vert
-(\pfi^*\omega^\perp\wedge\pfi^*\omega^\perp)^\vert
\end{equation}
and $a^\perp$, $a^\perp\wedge a^\perp$ are bounded in $L^2$ by the
functional (\ref{e0.13}). This {\it is} the relation we needed
between the geometry/topology of the maps and the Faddeev-Skyrme
functional. Recall that {\it the Uhlenbeck compactness theorem} says
that a sequence of gauge potentials with bounded curvatures is gauge
equivalent to a weakly precompact one \cite{Ul1,We}. Therefore
$a^\vert$ can be controlled by fixing the gauge in
$\Ad_*(\pfi^*P)$. In terms of maps this means that we replace $u$
by a suitable $uw$ when representing $\psi$ in the minimization
process.

It is interesting to note that $D_\pfi a$ transforms as curvature in
(\ref{e0.16}), i.e. 
\begin{equation}\label{gaugeD}
D_\pfi(a^w)=w^{-1}(D_\pfi a)w.
\end{equation} 
This brings us to the subject of {\it coset models} (see \cite{BMSS} and references
therein). In general in a coset model one considers a pair
consisting of a principal $G$ bundle and its $H$ subbundle. In our
case $M\times G$ and $\pfi^*P$ form such a pair. As in the standard
gauge theory fields are connections on the $G$ bundle but they are
identified only up to gauge transformations on the $H$ subbundle
(the gauge symmetry is 'broken to $H$' in physics lingo). Energy
functionals have to be invariant under the gauge group of the
subbundle. For our pair it means that they can only depend on
$F(a^\vert)$ and $D_\pfi a$. Obviously, the functional (\ref{e0.13})
gives an example of such a model. That Faddeev-Skyrme models can be
recast in these terms underscores the fact that they exhibit both
'string-theoretic' traits as non-linear $\sigma$--models 
and 'gauge-theoretic' traits as coset models.

\section{Short summary}\label{S:03}

In \refCh{1} we develop a homotopy classification of maps from a $3$--dimensional manifold into a compact simply connected homogeneous space in terms suitable for analytic applications. This classification is obtained mostly by applying the classical obstruction theory to the bundle of shifts. In \refS{1.1} we review classical results on low-dimensional homotopy groups of homogeneous spaces. The bundle of shifts is introduced in \refS{1.2}. In \refS{1.3} we prove that two maps $\psi,\pfi$ are $2$--homotopic if and only if they are related as $\psi=u\pfi$ and in \refS{1.4} we give a necessary and sufficient condition on $u$ to make them homotopic.

\refCh{2} develops the ideas of \cite{AK2} on representing maps into homogeneous spaces by connections. In particular a map $2$--homotopic to $\pfi$ can be represented by the pure--gauge connection $u^{-1}du$. This representation is not unique but the ambiguity admits a nice description in terms of gauge theory on coset bundles. \refS{2.1} is a review of the theory of connections and gauge transformations on principal bundles including some useful facts and formulas for matrix--valued and Lie algebra--valued differential forms that are scattered in the literature. In \refS{2.2} we study the coisotropy form of a homogeneous space which appears in the formulas for gauge action and curvature on coset bundles and also in the Faddeev-Skyrme functional. Coset bundles are introduced in \refS{2.3} and we develop 'gauge calculus' for them that is necessary to prove our minimization results in Chapter 3.

In \refS{3.1} we define the Faddeev-Skyrme functional for maps into arbitrary homogeneous spaces and its equivalent version for connections. Then we introduce some Sobolev spaces of maps suitable for the minimization problems involving this functional and extend the notion of $2$--homotopy type to such maps. We prove the existence of minimizers of the Faddeev-Skyrme functional in each $2$--homotopy sector in \refS{3.2}, and in each homotopy sector in \refS{3.3} when the target homogeneous space is symmetric. Both proofs rely on the fundamental gauge-fixing result of K.Uhlenbeck \cite{Ul1} to eliminate the ambiguity introduced by representing maps as connections.

On the first reading one may skip  \refCh{1} entirely, look through last two sections of \refCh{2} for notational conventions and proceed directly to \refCh{3} turning to the preceeding sections for reference wherever necessary.

\chapter{Maps into homogeneous spaces}\label{S:1}

In this chapter we describe 2 and 3 homotopy types of maps $M
\overset {\psi} \lra G/H $ in terms of liftings to the group of
motions $G$. The idea comes from a well known construction  in
algebraic topology -- so called Whitehead tower. In it a topological
space $X$ (usually a CW complex) is included into a tower of
fibrations $X$ where each $X^n $ is n-connected and a map $M
\overset {\psi} \lra X^n $ is n-nullhomotopic if and only if it
admits a lift $M \overset {\widetilde{\psi}_n} \lra X^n $ to the
n-th floor of the tower. If $X=G/H$ is simly connected then $X^1  =
X$ and if $G$ is simply connected then it is in fact 2-connected
since $\pi_2 (G)=0$ for any Lie group. Therefore the quotient bundle $G
\overset {\pi} \lra G/H $ can be seen as a surrogate of the second
floor of the Whitehead tower and one may expect that $M \overset {\psi}
\lra G/H $  is 2-nullhomotopic if and only if it admits a
lift
$$
\begin{diagram}
 &    &G   \\
 & \ruTo^{\widetilde{\psi}}   &\dTo{\pi}    \\
M   & \rTo{\psi}&    G/H
\end{diagram}
$$
This is indeed the case and moreover it turns out that since $G$ is
a group not only 2-nullhomotopy but even 2-homotopy type can be
characterized similarly: two maps $M \overset{\pfi ,\psi} \lra G/H
$ are $2$--homotopic if and only if there is a 'relative' lift $M
\overset {u}\lra G $ such that $\psi=u\pfi$ (\refT{1.2}). A further result  
states that they are in fact homotopic if and only if $u^*\b_G$ takes 
values in a prescribed subgroup of $H^3(M,\pi_3(G))$ 
(here $\b_G$ is the basic class of $G$, see \refD{basic}).

\section{Topology of homogeneous spaces}\label{S:1.1}

In this section we recall basic facts about topology of homogenous
spaces. A smooth manifold is called homogenous under an action of a
Lie group $G$ if the action is transitive. If $x_0 \in X$ is a point
the subgroup $H_{x_0} < G$ that fixes it is called the {\it isotropy
subgroup} of $x_0$. Isotropy subgroups of different points are
conjugate and therefore isomorphic to each other. There is a 1-1
correspondence between points of $X$ and cosets in $G/H$. If $G$ is
a compact Lie group then $H_{x_0} < G$ is closed and by a theorem of
Chevalley \cite{Ch} $G/H_{x_0}$ is equipped with a natural structure of smooth
manifold so the above correspondence becomes a diffeomorphism. In
other words, as far as compact Lie groups are concerned
consideration of homogeneous spaces is equivalent to that of coset
spaces $G/H$, where $H<G$ is a closed subgroup.

We are mostly interested in simply connected homogeneous spaces:
$\pi_1 (G/H)=0 $. By a theorem of D.Montgomery \cite{Mg} if a Lie group
$G$ acts transitively on a simply connected space then so does its
maximal compact subgroup $K(G)$, i.e. $G/H\simeq K(G)/(K(G)\cap H)$
($\simeq$ means diffeomorphic). If $G_0$, $\widetilde{G}$ denote the
identity component and the universal cover of $G$ respectively it is
easy to see directly that $G/H\simeq G_0/(G_0 \cap H)$ and $G/H\simeq
\widetilde{G}/\check{H}$ where $\check{H}:=\pi^{-1}(H)$ under
$\widetilde{G} \xrightarrow{\pi} G$. Combining these facts we
conclude that for simply connected homogeneous spaces $X=G/H$ we may
assume without loss of generality that $G$ is compact, connected and
simply connected. Indeed, if $G$ is not compact we replace it by the
maximal compact subgroup $K(G)$. If that is not connected we replace
it by its identity component, which is still compact (and which we
still denote $G$ by abuse of notation). Hence now $G$ is compact and
connected. If $G$ is not simply connected we take $\widetilde{G}$.
It may not be connected but by the classification theorem of compact
Lie groups $\widetilde{G}=\widetilde{G}_1\times ...\times
\widetilde{G}_m\times \mathbb{R}^n,$ where $\widetilde{G}_k$ are
simple, connected and simply connected [BtD], Applying the Montgomery
theorem once again we replace $\widetilde{G}$ by
$K(\widetilde{G})=\widetilde{G}_1\times ...\times \widetilde{G}_m$
that has all the required properties.

\begin{example}\label{E:CPn}
$\mathbb{C} P^{n-1}$ can be presented as a coset space
$GL_n(\mathbb{C})/P$, where $P$ is a parabolic subgroup of
invertible $n\times n$ complex matrices of the type 
\[ 
\left(
\begin{array}{cccc}
* & * & . & *\\
0 & * & . & *\\
. & . &   & .\\
0 & * & . & *
\end{array} \right)
\]
Following the above algorithm we take
$K(GL_n(\mathbb{C}))=U_n(\mathbb{C})$ while $P$ is replaced by
$(U_1\times U_{n-1})(\mathbb{C})$. The unitary group is
already connected so we skip taking the identity component but
$\widetilde{U_n}(\mathbb{C})=SU_n(\mathbb{C})\times\mathbb{R}$ and
$K(U_n(\mathbb{C}))=SU_n(\mathbb{C})$. The subgroup in the meantime
is replaced by $(U_1\times U_{n-1})(\mathbb{C})$ matrices with
determinant 1 which is isomorphic to $U_{n-1}(\mathbb{C})$. Thus $\mathbb{C}P^{n-1}\simeq
SU_n(\mathbb{C})/U_{n-1}(\mathbb{C})$and $SU_n$ is compact,
connected and simply connected.
\end{example}

From this point on we assume that in $X=G/H$ the group $G$ is compact,
connected and simply connected. By the same theorem of Chevalley \cite{Ch} $G\ovs{\pi}\longrightarrow G/H$ is a fiber bundle 
(in fact, a principal bundle) and we can apply the exact homotopy sequence:
\begin{equation}\label{e0.20}
\dots \lla \pi_k(G/H)\overset{\pi_*}\lla \pi_k(G)\overset{\i_*}\lla
\pi_k (H)\overset{\partial} \lla \pi_{k+1} (G/H) ...
\end{equation}
where $H \xrightarrow{\i}G$ is the inclusion and $\partial$ is
the connecting homomorphism. Since $\pi_0(G)=\pi_1(G)=0$ we have
\bee \label{e0.21}
0=\pi_0(G)\lla\pi_0(H)\lla\pi_1(G/H)\lla\pi_1(G)=0 \eee and
$\pi_0(H)=\pi_1(G/H)=0$, i.e. $H<G$ is connected. Furthermore, since
$\pi_2(G)=0$ for any Lie group
\bee\label{e0.22}
0=\pi_1(G)\lla\pi_1(H)\overset{\partial}\lla\pi_2(G/H)\lla\pi_2(G)=0
\eee
and $\pi_2(G/H)\simeq\pi_1(H)$ by the connecting homomorphism.
Finally, from the next segment of the sequence: $\pi_3(G/H)\simeq
\pi_3(G)/\i_*\pi_3(H)$. Summarizing the discussion of this section
we get the following

\begin{corollary}\label{C:1.1}
Any compact simply connected homogeneous space $X$ admits a coset
presentation $X=G/H$, where $G$ is compact, connected and simply
connected and $H<G$ is closed and connected.
\end{corollary}

\begin{remark} By a result of Mostow \cite{Ms} the Klein bottle $\mathbb{K}$ is a
homogeneous space of a Lie group but not of a compact one. Its
fundamental group is $\pi_1(\mathbb{K})\simeq
\mathbb{Z}\rtimes\mathbb{Z}_2$(semi-direct product) and this shows
that simple connectedness of $G/H$ is essential in \refC{1.1}.
\end{remark}

\section{The bundle of shifts}\label{S:1.2}

We assume that $X=G/H$ is a compact simply connected homogeneous
space presented as in Corollary 1, $M$ is a $CW$ complex (e.g., a smooth manifold) and consider continuous maps $M\xra X$.
Characterization of homotopy type will follow from the homotopy
lifting property in a certain bundle that we call
the bundle of shifts. A particular case of this bundle is used in \cite{AS} for similar purposes. 
\begin{definition}[The bundle of shifts]\label{D:shifts}
The bundle of shifts of a homogeneous space
$G/H=X$ is the fiber bundle $Q$ over $X\times X$ given by:
\begin{equation}\label{shifts}
\begin{aligned}
X\times G\overset{\alpha} \lra X\times X.\\ 
(x,g)\longmapsto (x,gx)
\end{aligned}
\end{equation}
\end{definition}
To prove that this is indeed a fiber bundle we need some facts from the theory of principal and associated bundles
\cite{BC,Hus,St}.
\begin{definition}[Principal bundles]\label{D:princp}
Let $P$ be a topological space and $H$ a Lie group
that acts on $P$ on the right: $\begin{aligned}P\times H & \lra
P\\(p,h) & \longmapsto ph\end{aligned}$\ . This action is called a
principal map if it is free and proper. The set of orbits $X:=P/H$
is then equipped with a natural topology and $\begin{aligned}P &\overset{\pi}\lra X\\ p & \longmapsto pH\end{aligned}$ 
is a fiber bundle called a principal bundle with the structure group $H$.
\end{definition}
If $P$ is a manifold and the action is smooth then $X$ also obtains
a smooth structure and the projection $\pi$ is smooth. Taking $P=G$
a compact Lie group and $H<G$ a closed subgroup we get by the Chevalley
theorem a smooth principal bundle $G\overset{\pi}\lra G/H$ called the
{\it quotient bundle}, where the principal map $G\times H\lra G$ is
just the group multiplication.

Let $F$ be another topological space (respectively, smooth manifold),
where the structure group $H$ acts on the left
$\begin{aligned}H\times F & \lra F\\(h,f) & \longmapsto \mu
(h)f\end{aligned}$\ . One can form a set of equivalence classes
\begin{equation}\label{Borel}
P\times_\mu F:=\{[p,f]\in P\times F| (p,f)\sim (ph,\mu(h^{-1})f)\}
\end{equation}
that receives a natural structure of a topological space
(a smooth manifold). It turns out that $\begin{aligned}P\times_\mu F & \lra X\\
[p,f] & \longmapsto\pi(p)\end{aligned}$ is a bundle projection that turns
$P\times_\mu F$ into a fiber bundle over $X$ called {\it the Borel
construction} from $P$ and $\mu$ \cite{Hus}.

\begin{definition}
Let $E_1\overset{\pi_1} \lra X$, $ E_2\overset{\pi_2}
\lra X$ be two fiber bundles over $X$. A continuous (smooth) map
$E_1\overset{\mathcal{F}}\lra E_2$ is a bundle map if the diagram
$$
\begin{diagram}
E_1&& \rTo{\mathcal{F} }        &&E_2      \\
         & \rdTo_{\pi_1}   &                &\ldTo_{\pi_2}&                \\
   &   &        X      &
\end{diagram}
$$
commutes, and it is a bundle isomorphism if its inverse is also a
bundle map. A bundle $E\lra X$ is called associated to a principal
bundle $P\lra X$ if it is bundle isomorphic to a Borel
construction $E\overset{\mathcal{F}} \simeq P\times_{\mu} F$ for
some $ \mu $ ,$F$ and $\mathcal{F}$.
\end{definition}

Note that if $E_1\overset{\pi_1}\lra X$ is a fiber bundle and
$E_2\overset{\pi_2}\lra X$ is a map such that for some invertible
$E_1\overset{\mathcal{F}}\lra E_2$ the diagrams 
\bee\label{iso}
\begin{diagram}
E_1&& \rTo{\mathcal{F} }        &&E_2, &   E_2 && \rTo{\mathcal{F}^{-1}}   && E_1   \\
         & \rdTo_{\pi_1}   &  &\ldTo_{\pi_2}&   & & \rdTo_{\pi_1}   &                &\ldTo_{\pi_2}&\\
   &   &        X      & &   &  &                 &         X       &             &     
\end{diagram}
\eee
commute then $E_2$ is also a fiber bundle and $E_2\simeq E_1$.

Along with a quotient bundle $G\overset{\pi} \lra G/H=X$ consider its
Cartesian double $G\times G \overset{\pi\times\pi}\lra X\times X $.
This is also a quotient (and hence principal) bundle with
$\widehat{G}:=G\times G$ and $\widehat{H}:=H\times H<G\times
G=\widehat{G}$, which is its structure group.

\begin{lemma}\label{L:1.1}
Let $G$ be a compact Lie group, $H<G$ a closed
subgroup and $G\overset{\pi}\lra X=G/H$ the corresponding coset
bundle. Then the bundle of shifts $Q\overset{\alpha}\lra X\times X$ \eqref{shifts} is
a fiber bundle associated to the quotient double $G\times
G\overset{\pi\times \pi}\lra X\times X$.
\end{lemma}

\begin{proof}
We will construct an explicit isomorphism between $Q$
and the following Borel construction. $H\times H$ acts on $H$ on the
left by 
$$
\begin{aligned}
(H\times H)\times H &\overset{\mu}\lra H\\
((\lambda_1,\lambda_2),h) & \longmapsto\lambda_2h\lambda_1^{-1}
\end{aligned}
$$
Set $E_1:=((G\times G)\times_\mu H\overset{\pi}\lra X)$, $E_2:=Q$ and consider the
following map 
$$
\begin{aligned}
E_1 &\overset{\mathcal{F}}\lra E_2\\
[g_1,g_2,h] & \longmapsto(g_1H,g_2hg_1^{-1})
\end{aligned}
$$
To begin with $\mathcal{F}$ is well defined:
$$
g_1\lambda_1H,g_2\lambda_2,\lambda_2^{-1}h\lambda_1)\longmapsto
(g_1,\lambda_1H, g_2hg_1^{-1})=(g_1H,g_2hg_1^{-1}).
$$ 
The inverse is given by $(x,g)\overset{\mathcal{F}^{-1}}\longmapsto
[g_1,gg_1,1]$, where $g_1H=x$. If $g_1\lambda$ is chosen instead
with $\lambda\in H$ then $[g_1\lambda ,gg_1\lambda
,\lambda^{-1}1\lambda]=[g_1,gg_1,1]$ so $\mathcal{F}^{-1}$ is
well-defined. It is easy to see that it is indeed the inverse to
$\mathcal{F}$.

We claim that both diagrams \eqref{iso} with $\pi_1,\pi_2$ replaced by $\pi$, $\alpha$
respectively commute. For instance, 
$$
(\alpha\circ\mathcal{F})([g_1,g_2,h])=\alpha
(g_1H,g_2Hg_1^{-1})=(g_1H,g_2hH)=(g_1H,g_2H)=\pi([g_1,g_2,h]).
$$
Therefore the bundle of shifts $Q=E_2$ is indeed a fiber bundle and $\mathcal{F}$ is a bundle isomorphism.
\end{proof}

Given a pair of maps $M\xra{\pfi ,\psi}X$ one
obtains a single map $M\overset{(\pfi ,\psi)}\lra X\times X$ into
the base of the bundle of shifts. The following characterization of the homotopy type follows directly from the homotopy lifting property in the bundle of shifts.

\begin{corollary}\label{C:1.2}
Let $G$ be a compact connected Lie group, $H<G$ a closed subgroup,
$X=G/H$ and $M$ a $CW$--complex. Then two continuous maps
$M\xra{\pfi ,\psi}X$ are homotopic if and only
if there exists a nullhomotopic $M\overset{u_0} \lra G$ such that
$\psi =u_0\pfi$. Given an arbitrary map $M\overset{u}\lra G$ maps
$\pfi$, $u\pfi$ are homotopic if and only if $u=u_0w$, where $u_0$
is nullhomotopic and $w\pfi =\pfi$.
\end{corollary}
\begin{proof}
If $u_0^t$ is a homotopy that translates $u_0$ into constant
$1$ map then $\psi_t:=u_0^t\pfi$ translates $u_0\pfi$ into $\pfi$ and $\Phi(m,t):=(\pfi (m),\psi_t (m))$ translates $(\pfi,\pfi)$ into
$(\pfi,\psi)$. The former admits a lift $(\pfi,1)$ into $Q$, indeed $\alpha\circ(\pfi,1)=(\pfi ,\pfi)$. 
Since $Q$ is a fiber bundle by \refL{1.1} the homotopy lifting property implies that 
the following diagram can be completed as indicated: 
\be
\begin{diagram}
M\times\{0\}&  \rTo{(\pfi,1)}  & X\times G\\
\dInto    &  \ruDotsto^{\widetilde{\Phi}} & \dTo{\alpha}\\
M\times I &  \rTo^{\Phi} & X\times X
\end{diagram}
\ee
By the upper triangle $\widetilde\Phi_2 (m,0)=1$ and by the lower one
$\widetilde\Phi_1 (m,t)=\Phi_1 (m,t)=\pfi (m)$, $\widetilde\Phi_2
(m,t)\widetilde\Phi_1 (m,t)=\widetilde\Phi_2(m,t)\pfi (m)=\psi_t (m)$. Set
$u_0(m):=\widetilde\Phi_2 (m,1)$ then $u_0\pfi =\psi$ and $\widetilde\Phi_2
(\cdot ,t)$ is a homotopy that translates the constant map $1$ into
$u_0$ as required.

For the second claim note that $u=u_0w$ implies $u\pfi
=u_0w\pfi=u_0\pfi$ and is homotopic to $\pfi$. Conversely, if
$u\pfi$ is homotopic to $\pfi$ then by the first claim there is also
a second nullhomotopic $u_0$ such that $u_\pfi =u_0 \pfi$. It
suffices to set $w:=u_0^{-1}u$.
\end{proof}

\begin{remark}
Note that $\pfi$, $u\pfi$ homotopic does not imply that $u$ is nullhomotopic. Characterization of such $u$ as products given in \refC{1.2} is rather indirect and we will give a more explicit one in \refT{1.2}.
\end{remark}

\section{Characterization of the $2$-homotopy type}\label{S:1.3}

We established above that if $\psi =u\pfi$ and $u$ has a special
form $u=u_0w$ then $\pfi$ and $\psi$ are homotopic. If no
restriction is imposed on $u$ it is not necessarily so but the
restrictions of $\pfi$, $\psi$ to the 2-skeleton of $M$ are
homotopic at least if $m$ is a 3-dimensional $CW$ complex. This is
in turn sufficient for the existence of such $u$. This fact is much
more complicated than \refC{1.2}. We will prove it by reducing both
the lifting problem and the $2$-homotopy problem to problems in the
obstruction theory \cite{Brd,DK,Sp,St} and then showing that the obtained obstructions 
are essentially the same.

Let us start with the lifting problem. As before given two maps
$M\xra{\pfi ,\psi}X$ define $M\overset{(\pfi,\psi)}\lra X\times X$ 
and consider the {\it ratio bundle}:
\begin{equation}\label{e0.23}
\bal
Q_{\pfi ,\psi}:=(\pfi ,\psi)^*Q &=\{(m,x,g)\in M \times X\times G|(\pfi (m),\psi(m)=(x,gx)\}\\
&=\{(m,g)\in M\times G|\psi(m)=g\pfi(m)\}
\eal
\end{equation}
As is obvious from the second representation sections of this bundle
$M\overset{\sigma} \lra Q_{\pfi ,\psi}\subset M\times G$ have the form $\sigma (m)=(m,u(m))$,
where $\psi =u\pfi$. In other words they play the role of non-existent 'ratios' $\psi/\pfi$.
Hence the problem of finding a lift $u$ is equivalent to constructing a section of the bundle $Q_{\pfi ,\psi}$,
which is a standard problem in the obstruction theory.

Let us recall some basic notation following N.Steenrod \cite{St}. Assume
that in a fiber bundle $F\overset{\i} \hra E\overset{\pi}\lra B$
 the base $B$ is a $CW$--complex and the fiber $F$ is {\it homotopy simple} up to dimension $n$
 (i.e. $\pi_1(F)$ acts trivially on $\pi_k(B)$ for $1\leq k\leq n$),
 where $n$ is {\it the lowest homotopy non-trivial dimension}
 (i.e. $\pi_k(F)=0$ for $1\leq k\leq n-1$ but $\pi_n (F)\neq 0$).
 This means that there is no obstruction to constructing a section up to
 dimension $n$ and we may assume that $B^{(n)}\overset{\sigma}\lra E$
 is already constructed, here $B^{(n)}$ is the $n$-skeleton of $B$.
  Let $\Delta \subset B$ be an $(n+1)$ cell of $B$ which we may assume
  to be contractible (or even a simplex). Then the restriction $E\mid_\Delta$
  is a trivial bundle and we have a trivialization
  $\Delta \times F \overset{\Phi_\Delta}\lra E\mid_\Delta$. Let $\pi_1$,$\pi_2$
   denote the projections to the first and the second factor of $\Delta\times F$.
    Then the map $\pi_2\circ \Phi_\Delta^{-1}\circ\sigma :\partial\Delta\lra F$
    defines an element of $\pi_n (F)$. It turns out that this element does not depend on a choice of trivialization and
\bee\label{e0.24}
c_\sigma
(\Delta):=[\pi_2^{-1}\circ\Phi_\Delta^{-1}\circ\sigma
|_{\partial\Delta}]\in\pi_n(F)
\eee
is a $\pi_n (F)$-valued cochain
and in fact a cocycle. Its cohomology class
$\overline{c}_\sigma\in H^{n+1}(B,\pi_n (F))$ is called the primary obstruction
to extending $\sigma$. This cohomology class does not even depend on a choice
of $\sigma$ on the $n$-skeleton of $B$ and is an invariant of the bundle
$E\overset{\pi}\lra B$ itself. This invariant is called {\it the primary characteristic class}
of $E$ and denoted 
$$
\varkappa (E):=\overline{c}_\sigma.
$$ 
The characteristic class is natural
with respect to the pullback of bundles: 
$$
\varkappa (\pfi^*E)=\pfi^*\varkappa (E)
$$ 
and the Eilenberg extension theorem claims that a section $\sigma$ can be altered
on $B^{(n)}$ so as to be extendable to $B^{(n+1)}$ if and only if $\overline{c}_\sigma=0$.
This completely solves the sectioning problem when $\pi_k(F)=0$ for $n+1\leq k<\dim B$ (i.e. there are no further obstructions). A section exists if and only if $\varkappa(E)=0$.

In our case the bundle in question is $H\overset{\i}\hra Q_{\pfi
,\psi}\overset{\pi}\lra M$. The fiber is a Lie group so it is
homotopy simple in all dimensions. The first non-trivial dimension
is $n=1$ as $\pi_0 (H)=0$ by \refC{1.1} and $\varkappa (Q_{\pfi
,\psi})\in H^2(M,\pi_1 (H))$. Since $\pi_2(H)=0$ for all Lie groups
and $\dim M=3$ there is no further obstruction and a section exists
if and only if $\varkappa (Q_{\pfi ,\psi})=0$. Thus we want to
compute this characteristic class. By naturality $\varkappa (Q_{\pfi
,\psi})=\varkappa((\pfi ,\psi)^*Q)=(\pfi ,\psi)^*\varkappa(Q)$ and we
need to compute $\varkappa$ for the bundle of shifts.

Recall from \refL{1.1} that $Q$ is isomorphic to the following Borel construction:
$\widehat{E}:=\widehat{P}\times_{\widehat{\mu}}H$ with $\widehat{P}=G\times G$
and the action
$$
\begin{aligned}
(H\times H)\times H&\ovs{\widehat{\mu}}\lra H\\
((\lambda_1 ,\lambda_2),h)&\longmapsto \lambda_2 h\lambda_1^{-1}
\end{aligned}
$$
The form of the action suggests that we can 'decompose' $\widehat{E}$ into a
combination of two simple bundles $E$ and $E'$, namely 
\be
E:=P\times_\mu H\quad \text{with}\quad \mu(\lambda )h:=\lambda h
\ee 
and its dual 
\be 
E':=P\times_{\mu '}H\quad \text{with}\quad \mu'(\lambda)h:=h\lambda^{-1}
\ee 
(in our case $P=G$ and one can multiply on both sides). We will not explain precisely what 
the 'decomposition' means in this case but it should be clear from the proof of \refL{1.2}(ii). Note that $E$ is
bundle isomorphic to $P$ itself by $\bal P & \lra E\\ p & \longmapsto [p,1] \eal$ so we write $\varkappa(P)$ for $\varkappa(E)$.

\begin{lemma}\label{L:1.2}
Let $P\ovs{\pi}\lra X$ be a principal bundle with
the structure group $H$. Define $\widehat{P}:=(P\times P\lra X\times
X)$, $E$, $E'$, $\widehat{E}$ as above and let $\pi_1$, $\pi_2$
denote the projections from $X\times X$ to the first and the second
components. Then

{\rm(i)} $\varkappa (P)=\varkappa (E)=-\varkappa (E')$.

{\rm (ii)} If also $H^k(X,\mathbb{Z})=0$ for $0\leq k\leq n$ then
$$
\varkappa (\widehat{E})=\pi_2^*\varkappa (P)-\pi_1^*\varkappa (P).
$$
\end{lemma}
\begin{proof}
{\rm(i)} Note that if $\sigma (x)=[p,h]$ gives a section of $E$
then $\sigma '(x)=[p,h^{-1}]$ gives a section of $E'$. Also if
$\Delta\ovs{S_\Delta}\lra P|_\Delta$ is a local section of $P$ then
$$
\bal
\Delta\times F & \ovs{\Phi|_\Delta}\lra(P\times_\mu F)|_\Delta\\
(x,f) & \longmapsto [S_\Delta (x),f]\\
(\pi (p),\mu(\lambda^{-1})f) & \longmapsfrom [p,f],\quad \text{with}\ S_\Delta(\pi(p))=p\lambda,
\eal 
$$
is a local trivialization of the associated bundle.

We choose a section $S_\Delta$ of $P$ and denote $\Phi_\Delta$,
$\Phi^\prime_\Delta$ the corresponding trivializations of $E$, $E'$.
Also if $\sigma$ is the chosen section of $E$ on $B^{(n)}$ then the
$\sigma'$ is the one we choose for $E'$. By definition:
\begin{align*}
\pi_2\circ\Phi_\Delta^{-1}\circ\sigma'(x) &=\pi_2\circ\Phi_\Delta^{-1}([p,h^{-1}]), && \pi (p)=x=\pi_2\\
&=(\pi(p),\mu'(\lambda^{-1})h^{-1}), && S_\Delta (\pi (p))=S_\Delta(x)=p\lambda\\
&=h^{-1}(\lambda^{-1})^{-1}=(\lambda ^{-1}h)^{-1}=(\mu(\lambda ^{-1})h)^{-1} &&\\
&=(\pi_2 \circ\Phi_\Delta^{-1}([p,h])^{-1})=(\pi_2\circ\Phi_\Delta^{-1}\circ\sigma(x))^{-1}. &&
\end{align*}
In other words, $c_{\sigma'}(\Delta)=[o^{-1}]$ if $c_{\sigma}(\Delta)=[o]$,
with $o$ being a map $\partial\Delta\lra H$ and $[\cdot]$ denoting a
class in $\pi_n (H)$. But in $\pi_n(H)$ one has $[o^{-1}]=-[o]$ (see e.g. \cite{Dy}) for any $o$ and $\varkappa(E')=\overline{c}_{\sigma'}=-\overline{c}_\sigma =-\varkappa(E)$.

{\rm(ii)}  Under our assumptions the K\"unneth formula and the universal coefficients theorem \cite{Brd} imply that
\begin{align*}
H^{n+1}(X\times X,\pi_n(H)) &\simeq H^{n+1}(X, \pi_n (H))\oplus H^{n+1}(X,\pi_n(H)),\\
\omega &\longmapsto (\i_1^*\omega ,\i_2^*\omega)\\
\pi_1^*\omega^*+\pi_2^*\omega_2 &\longmapsfrom (\omega_1,\omega_2),
\end{align*}
where $x\ovs{\i_1} \longmapsto (x,x_0)$, $x\ovs{\i_2}\longmapsto (x_0,x)$
for some fixed point $x_0\in X$. Let $p_0\in P$ be any point with $\pi (p_0)=x_0$, then
\begin{align*} 
\i^*_1\widehat{E} &=\{(x,[p,p_0,h])\in X\times \widehat{E}|\ (x,x_0)=(\pi(p),\pi (p_0))\}\\ 
&\simeq \{(x,[p,h])\in X\times E|\ \pi (p)=x\}\simeq E'
\end{align*} 
since $p_0$ is fixed and $\widehat{\mu}$ reduces to $\mu '$ on the
first component. Analogously, $\i^*_2\widehat{E}\simeq E$. Therefore
from naturality and {\rm(i)}
\begin{multline*}
\varkappa
(\widehat{E})=\pi_1^*\i_1^*\varkappa (\widehat{E})+ \pi_2^*\i_2^*\varkappa
(\widehat{E})=\pi_1^*\varkappa (\i_1^*\widehat{E}) +\pi_2^*\varkappa
(\i_2^*\widehat{E})\\
=\pi_1^*\varkappa (E')+\pi_2^*\varkappa (E)
=\pi_2^*\varkappa (P)-\pi_1^*\varkappa (P)
\end{multline*}
\end{proof}
The next example gives an application of the primary characteristic class.
\begin{example}\label{E:Chern}
Let $P$ be a principal $U_n=U_n(\mathbb{C})$
bundle and $U_k<U_n$ sit in  it block diagonally. Then $U_n$ acts on
$U_n/U_k$ on the left and we have an associated bundle
$E_k:=P\times_\mu(U_n/U_k)$. N.Steenrod \cite{St} defines the {\it
$k$-th Chern class} of $P$ as 
$$
c_k (P):=\varkappa(E_{k-1}).
$$
Equivalence to other definitions is proved in \cite{BH} (Appendix 1). For $k=1$
this is exactly the bundle $E$ from \refL{1.2}. Hence in this case
$\varkappa (P)=c_1 (P)\in H^2(X,\pi_1 (U_n))\simeq H^2 (X,\mathbb{Z})$.
\end{example}

In our case $P$ is the quotient bundle $G\lra X$ and we write
$\varkappa (G)$ with the usual abuse of notation (of course
$\varkappa (G)$ also   depends on $H<G$). It is easy to compute
$\varkappa (Q_{\pfi ,\psi})$ now since $Q_{\pfi ,\psi}=(\pfi
,\psi)^*Q$ and $Q=\widehat{E}$ for the quotient bundle $G\lra X$:
\begin{align*}
\varkappa (Q_{\pfi ,\psi}) &=\varkappa ((\pfi ,\psi)^*Q)=(\pfi ,\psi)^*\varkappa (Q)) && \text{by  naturality}\\
&=(\pfi ,\psi)^*(\pi_2^*\varkappa (G)-\pi_1^*\varkappa (G)) && \text{by \refL{1.2}}\\
&=(\pi_2\circ (\pfi ,\psi))^*\varkappa (G)-(\pi_1\circ (\pfi,\psi))^*\varkappa (G) &&\\
&=\psi^*\varkappa (G)-\pfi^*\varkappa(G).
\end{align*}

\begin{corollary}\label{C:1.3}
Let $X=G/H$ be a simply connected homogeneous
space presented as in \refC{1.1}, $M$ be a $3$-dimensional $CW$--complex 
and $M \xra{\psi,\pfi}X$ continuous maps. Then a continuous $M\ovs{u}\lra G$
with $\psi=u\pfi$ exists if and only if 
$$
\psi^*\varkappa(G)=\pfi^*\varkappa (G),
$$
where $\varkappa (G)$ is the primary characteristic class of the quotient bundle $G\to X$.
\end{corollary}

\begin{remark}
In fact the conditions of \refL{1.2} are satisfied with
$n=1$ if $H$ is connected and $X$ is simply connected (simple
connectedness of $G$ is not necessary). Hence \refC{1.3} can be
applied directly to $U_n$ homogeneous spaces without reducing them
to $SU_n$ ones as long as the subgroup $H<U_n$ is already connected.
\end{remark}

Now we also want to reduce characterization of $2$-homotopy type of maps
$M\lra X$ to computing an obstruction. This requires more data
from the obstruction theory. Let $B$ be a $CW$--complex and
$B\ovs{\psi,\pfi}\lra F$ be two maps homotopic on $B^{(n-1)}$ by $\Phi
:B^{(n-1)}\times I\lra F$. If $\Delta\subset B^{(n)}$ is an $n$-cell
then 
$$
\partial (\Delta\times I)\subset(B\times \{0\})\bigcup
(B^{(n-1)}\times I)\bigcup (B\times \{1\})
$$ 
so $\Phi$ is defined on it and $\partial(\Delta\times I)\simeq S^n$. 
Therefore we can set
\be 
d_\Phi (\pfi ,\psi)(\Delta):=[\Phi(\partial(\Delta\times
I))]\in\pi_n (F) 
\ee 
and this defines a $\pi_n (F)$--valued cochain
on $B$ called {\it the difference cochain} \cite{St}. It turns out to
be a cocycle and its cohomology class 
$$
\overline{d}(\pfi,\psi):=\overline{d_\Phi (\pfi ,\psi)}
$$ 
does not depend on a choice
of homotopy on $B^{(n-1)}$. Obviously $\overline{d_\Phi (\pfi
,\psi)}\in H^n (B,\pi_n (F))$. The homotopy $\Phi$ can be extended
from $B^{(n-2)}$ to $B^{(n)}$ (it may have to be altered on
$B^{(n-1)}$) if and only if $\overline{d}(\pfi ,\psi)=0$. The difference
is natural
\be
\overline{d} (\pfi\circ f,\psi\circ f)=f^*\overline{d}(\pfi
,\psi)
\ee
and additive
\be
\overline{d}(\pfi ,\chi)=\overline{d}(\pfi
,\psi)+\overline{d}(\psi ,\chi)
\ee
Since $\pfi$ is always homotopic to
itself $\overline{d} (\pfi ,\pfi)=0$ and additivity implies 
$$
\overline{d}(\psi,\pfi)=-\overline{d}(\pfi ,\psi).
$$ 

Now let $n$ be the lowest homotopy
non-trivial dimension of $F$ and $F$ be homotopy simple up to this
dimension. Then any two maps into $F$ are homotopic on $B^{(n-1)}$
and ${\overline{d}(\pfi ,\psi)}$ is defined for any pair. It is called
{\it the primary difference} between $\pfi$ and $\psi$ \cite{St}. 
\begin{theorem*}[Eilenberg classification theorem] 
If the primary difference is the only obstruction to homotopy, i.e. 
$$
\pi_k(F)=0\quad \text{for}\quad n+1\leq k\leq\dim B
$$
then $\pfi,\psi$ are homotopic if and only if\ $\overline{d}(\pfi ,\psi)=0$. 
Moreover, for any $\omega\in H^n (B,\pi_n
(F))$ and a given $B\ovs{\pfi}\lra F$ there is $B\ovs{\psi}\lra F$
such that\ $\overline{d}(\pfi,\psi)=\omega$. 
\end{theorem*}
In other words, in conditions of the theorem maps are classified up to homotopy 
by their primary differences with a fixed map $\pfi$ and their is a one-to-one correspondence 
between homotopy classes and $H^n (B,\pi_n(F))$. In general one can only
claim that $\pfi$, $ \psi$ are $(n+q-1)$-homotopic, where $(n+q)$ is
the next after $n$ homotopy non-trivial dimension of $F$. In our
case $B=M$, $F=X$, $n=2$ since $X$ is simply connected and $q=1$
since generally speaking $\pi_3 (X)\neq 0$. 
So $M\ovs{\psi,\pfi}\lra X$ are $2$-homotopic if and only if $\overline{d}(\pfi
,\psi)=0$.

We can do a little better. For any connected space $F$ there are two
special maps $F\lra F$: the identity $\id_F$ and the constant map
$\mbox{pt}_F (x)=x_0\in F$. The primary difference $\overline{d}(\id_F,
pt_F)$ only depends on $F$
itself (since all constant maps into a connected space are homotopic to each other). 
This class can also be described more explicitly. If $\pi_0
 (F)=...=\pi_{n-1}(F)=0$ then by the Hurewicz theorem $H_0 (F,\mathbb{Z})=...=H_{n-1}(F,\mathbb{Z})=0$
, $H_n (F,\mathbb{Z})\simeq\pi_n (F)$ and by the universal
coefficients theorem $H^n(F,\pi_n(F))\simeq
\mbox{Hom}(H_n(F,\mathbb{Z}),\pi_n(F))$. Let $\pi_n
(F)\ovs{\mathcal{H}}\lra H_n(F,\mathbb{Z})$ be the Hurewicz
isomorphism. The basic class $\b_F\in H^n(F,\pi_n(F))$
is the class that corresponds to the homomorphism
$H_n(F,\mathbb{Z})\ovs{\mathcal{H}^{-1}}\lra \pi_n(F)$ under the
above isomorphism. 

\begin{definition}[The basic class]\label{D:basic}
The basic class $\b_F\in H^n(F,\pi_n(F))$ is the cohomology class that maps every homology class in $H_n(F,\Z)$ 
into its image in $\pi_n(F)$ under the Hurewicz isomorphism ($\b_F$ is also called fundamental or
characteristic class of $F$ by some authors \cite{DK,MT,St}).
\end{definition}

Note that $\overline{d}(\mbox{id}_F,\mbox{pt}_F)\in H^n(F,\pi_n(F))$ as
well and one can show \cite{St} that
\be
\overline{d}(\id_F,\pt_F)=\b_F
\ee
Now let $H\ovs{\psi, \pfi}\lra X$ be any continuous maps and
$M\ovs{\pt_{M,X}}\lra X$ be a constant map. Then by naturality and
additivity
\begin{equation}\label{e0.25}
\bal
\overline{d}(\pfi,\psi) &=\overline{d}(\pfi,\mbox{pt}_{M,X})+\overline{d}(\mbox{pt}_{M,X},\psi)\\
&=\overline{d}(\pfi,\mbox{pt}_{M,X})-\overline{d}(\psi,{pt}_{M,X})\\
&=\overline{d}(\id_X\circ\pfi ,\pt_X\circ\pfi)-\overline{d}(\id_X\circ\psi,\pt_{X}\circ\psi)\\
&= \pfi^*\overline{d}(\id_X,\pt_X)+\psi^*\overline{d}(\id_X,\pt_X)=\pfi^*\b_X-\psi^*\b_X.
\eal
\end{equation}

\begin{corollary}\label{C:1.4} In the conditions of Corollary 2 the maps $\pfi$, $\psi$ are
$2$-homotopic if and only if $\psi^*\b_X=\pfi^*\b_X$.
\end{corollary}

This condition has the same form as in \refC{1.3} with $\varkappa(G)$ replaced by $\b_X$. The next example demonstrates a relation between the two classes in a simple case.

\begin{example}\label{E:basic}
The complex projective space $\CP^n$ can
be represented as $SU_{n+1}/U_n$. Since $\pi_2(\CP^n)\simeq
\mathbb{Z}$ the basic class $\b_{\CP^n}\in
H^2(\CP^n,\pi_2(\CP^n))\simeq H^2(\CP^n,\mathbb{Z})$ is just the
generator of the second cohomology under this identification -- the
Poincare dual of the hyperplane class. On the other
hand, by \refE{Chern}: $\varkappa (SU_{n+1})=c_1(SU_{n+1})$ and the
first Chern class of this bundle is also known to be the generator
(under the identification $\pi_1(U_n)\simeq \mathbb{Z}$) \cite{BT}.
Hence with the above identifications we must have $\varkappa
(SU_{n+1})=\pm \b_{\CP^n}$.
\end{example}

In general, $\varkappa (G)\in H^2(X,\pi_1 (H))$  and $\b_X\in H^2(X,\pi_2(X))$ but from \eqref{e0.22} we have $\pi_1(H)\simeq\pi_2(X)$ under the connecting homomorphism. 
The rest of this section is denoted to establishing
that $\varkappa (G)=-\partial\circ\b_X$.\ Since the connecting homomorphism in this case is an isomorphism once the relation is established Corollaries \ref{C:1.3},\ref{C:1.4} directly imply

\begin{theorem}\label{T:1.1}
Let $X$ be a compact simply connected homogeneous space and $M$ a
$3$-dimensional $CW$ complex. Then three conditions are equivalent
for continuous $M\ovs{\psi ,\pfi} \lra X$:

{\rm(i)}\ $\pfi$, $\psi$ are $2$-homotopic (i.e. homotopic on the
$2$-skeleton of $M$);

{\rm(ii)}\ $\psi^*\b_X=\pfi^*\b_X\in H^2(M,\pi_2(X))$, $\b_X$ is the
basic class of $X$;

{\rm (iii)}\ There exists a continuous $M\ovs{u}\lra G$ such that
$\psi =u\pfi$, where $X=G/H$ as in Corollary 1.
\end{theorem}
Note that equivalence of the first two conditions is just a particular case of the Eilenberg classification theorem. An additional notion we need to tie $\varkappa (G)$ to $\b_X$ is the transgression \cite{DK,HW,MT,Sp,St}.
\begin{definition}[Transgression]
Let $F\ovs{\i}\hra E\ovs{\pi}\lra B$ be a fiber bundle and
$\mathbb{A}$ an Abelian group. An element $\alpha\in
H^n(F,\mathbb{A})$ is called transgressive if there are cochains
$\xi\in C^n(E,\mathbb{A})$ and $\eta\in C^{n+1}(B,\mathbb{A})$ such
that
\begin{equation}\label{e0.26}
\begin{aligned}
\overline{\i^*\xi} &=\alpha\\
\delta\xi &=\pi^*\eta,
\end{aligned}
\end{equation}
where the bar denotes the corresponding cohomology
class and $\delta$ is the cohomology differential. When $\alpha$ is
transgressive classes $\tau ^{\#}\alpha :=\overline{\eta}\in
H^{n+1}(B,\mathbb{A})$ are called its (cohomology) transgressions.

Dually, an element $a\in H_{n+1}(B,\mathbb{A})$ is transgressive if
there exist chains $w \in C_{n+1}(E,\mathbb{A})$ and $\upsilon\in
C_n(F,\mathbb{A})$ such that
\begin{equation}\label{e0.27}
\begin{aligned}
\overline{\pi_*\omega} &=a\\
\eth w &=\i_*\upsilon,
\end{aligned}
\end{equation}
with $\eth$ denoting the homology differential. Any
$\tau_{\#}a:=\overline{\upsilon}\in H_n(F,\mathbb{A})$ is called
a (homology) transgression of $a$.
\end{definition}

Note that $\pi^* (\delta\eta)=\delta (\pi^*\eta)=\delta ^2\xi=0$ and
$\delta\eta =0$ since $\pi^*$ is injective on cochains. Analogously,
$\partial\upsilon =0$ so taking $\overline{\eta}$, $\overline{\upsilon}$ makes
sense. Also note that $\xi$, $\eta$ (respectively $w$, $\upsilon$)
when they exist may not be unique and hence $\tau^\#$, $\tau_\#$
really map into a quotient of the cohomology (homology) group. For
the case of homology we are only interested in the case
$\mathbb{A}=\mathbb{Z}$. There is an $\mathbb{A}$-valued pairing
(the Kronecker pairing \cite{DK}) between $H^*(Y,\mathbb{A})$ and
$H_*(Y,\mathbb{Z})$ given by evaluation of cochains on chains,
$\tau^\#$ and $\tau_\#$ are dual to each other with respect to this
pairing. Indeed, when $\alpha$, $a$ are transgressive
\bee\label{e0.28}
\tau^\#\alpha
(a)=\overline{\eta}(\overline{\pi_*w})=\pi^*\eta
(w)=\delta\xi (w)=\xi (\partial w)\\
=\xi (\i_*\upsilon)=\i^*\xi (\upsilon)=\overline{\i^* \xi}
(\overline{\upsilon})=\alpha (\tau_\# a)
\eee
One has to be careful with
the ambiguity in $\tau^\#$ and $\tau_\#$ in \eqref{e0.28}, in general it
only says that $\tau^\#\alpha$, $\tau_\# a$ can be adjusted so that
the equality holds.

Unlike the connecting homomorphism
$\pi_{n+1}(B)\ovs{\partial}\lra\pi_n (F)$ which is everywhere
defined and unambiguous the homology transgression $\tau_\#$ in general
maps from a subgroup of $H_{n+1}(B,\mathbb{Z})$ to a quotient of
$H_n(F,\mathbb{Z})$. In a sense it 'imitates' the non-existent
connecting homomorphism in homology \cite{DK}. More precisely,
spherical classes in $H_{n+1}(B,\mathbb{Z})$ are always
transgressive and the diagram
\bee\label{e0.29}
\begin{diagram}
\pi_{n+1}(B)           & \rTo{\partial}   &\pi_n(F)            \\
\dTo{\mathcal{H}_B}      &                  &\dTo{\mathcal{H}_F} \\
H_{n+1}(B,\mathbb{Z})  & \rTo{\tau_\#}     &H_n(F,\mathbb{Z})
\end{diagram}
\eee
commutes. Here $\mathcal{H}_B$, $\mathcal{H}_F$ are Hurewicz
homomorphisms and it is understood that $\mathcal{H}_F(\partial
(z))$ is just one of transgressions of $\mathcal{H}_B(z)$.
Commutativity can be established by inspecting the definitions of
$\tau_\#$ and $\partial$ (see \cite{Hu}).

There is a case when the transgression is unambiguous. When $H^i
(B,\mathbb{A})=0$ for $0<i <k$ and $H^j(F,\mathbb{A})=0$ for $0<j<l$
a result of J.-P. Serre says that $H^m(F,\mathbb{A})\ovs{\tau^\#}\lra
H^{m+1}(B,\mathbb{A})$ is well-defined and one has the {\it Serre exact
sequence} \cite{HW,MT}:
\bee\label{e0.30}
0\lra H^1(B,\mathbb{A})\ovs{\pi^*}\lra
H^1(E,\mathbb{A})\ovs{\i^*}\lra H^1(F,\mathbb{A})\ovs{\tau^\#}\lra
H^2(B,\mathbb{A})\ovs{\pi^*}\lra ...\ovs{\i^*}\lra
H^{k+l-1}(F,\mathbb{A}).
\eee
Analogous statement is also true for the
homology transgression. Conditions of the Serre exact sequence are
satisfied in particular if $n$, $n+1$ are the lowest homotopy
non-trivial dimensions for $F$ and $B$ respectively and $k=n+1$,
$l=n$. In this case one has the following \cite{St} (see also \cite{BH}, Appendix 1):

\begin{theorem*}[Whitehead transgression theorem] Let $F\ovs{\i}\hra
E\ovs{\pi}\lra B$ be a fiber bundle with the fiber $F$ being
homotopy simple up to dimension $n$  and let $n$, $n+1$ be the
lowest homotopy non-trivial dimensions of $F$ and $B$ respectively.
Then the primary characteristic class of $E$ is transgressed from
the minus basic class of $F$, i.e.
\bee\label{e0.31}
\varkappa
(E)=-\tau^\# \b_F\in H^{n+1}(B,\pi_n(F))
\eee
\end{theorem*}
Using \eqref{e0.31} it is not difficult now to relate $\varkappa (E)$ also to the basic class of
$B$.

\begin{corollary}
In conditions of the Whitehead transgression theorem
\bee\label{e0.32}
\varkappa (E)=-\partial\circ \b_B,
\eee
where
$\pi_{n+1}(B)\ovs{\partial}\lra\pi_n (F)$ is the connecting
homomorphism (cf. \cite{Nk2}).
\end{corollary}

\begin{proof}
By the universal coefficients theorem \cite{Brd}:
$$
0\lra \Ext(H_n(B,\mathbb{Z}),\pi_n(F))\lra H^{n+1}(B,\pi_n(F))\lra
\Hom (H_{n+1}(B,\mathbb{Z}),\pi_n(F))\lra 0
$$
is exact and since $n+1$ is the lowest homotopy non-trivial
dimensional of $B$ the group $H_n(B,\mathbb{Z})=0$ and the $\Ext$
term vanishes. Hence the elements of $H^{n+1}(B,\pi_n(F))$ are
completely determined by their pairing with integral homology
classes. By the Serre exact sequences both transgressions
$H^n(F,\pi_n(F))\ovs{\tau^\#}\lra H^{n+1}(B,\pi_n(F))$ and
$H_{n+1}(B,\mathbb{Z})\ovs{\tau_\#}\lra H_n(F,\mathbb{Z})$ are
unambiguous. Thus using \eqref{e0.28},\eqref{e0.29} and \eqref{e0.31} we have
\be
\bal \varkappa
(E)(a)&=-\tau^\# \b_F(a)=-\b_F(\tau_\# a)=-\mathcal{H}_F^{-1}(\tau_\#
a)=-\partial
(\mathcal{H}_B^{-1}(a))\\
&=-\partial (\b_B(a))=-\partial\circ \b_B (a).\eal
\ee
Since $a\in
H_{n+1}(B,\mathbb{Z})$ is arbitrary (all elements are spherical by the Hurewicz theorem and
hence transgressive) we get \eqref{e0.32}.
\end{proof}

In our application the bundle is $H\hra G\lra X=G/H$ and $n=1$ since
$H$ is connected. Therefore, 
$$
\varkappa(G)=-\partial\circ \b_X\in
H^2 (X,\pi_1(H))
$$ 
as required for \refT{1.1}.

\section{Secondary invariants and the homotopy type}\label{S:1.4}

By the Eilenberg classification theorem  maps $M\to X$ are $2$--homotopic if and only if they have the same pullbacks of the basic class $\b_X$. This pullback $\pfi^*\b_X$ is known as the {\it primary invariant} of a map $\pfi$. If $\pi_3(X)=0$ then $2$-homotopy type gives the entire homotopy type (recall that we only consider a $3$--dimensional $M$), otherwise some {\it secondary
invariants} have to be specified. Unlike in the case of the primary invariant these classical secondary invariants require a pair of maps to be defined and the definiton is not constructive \cite{Bo,MT}. This is inconvenient for our purposes so we use the following bypass. As was proved in \refS{1.2} a continuous map $\psi=u\pfi$ is homotopic to $\pfi$ if and only if $u=u_0\omega$
with a nullhomotopic $u_0$ and $w\pfi=\pfi$. In this section we derive an explicit characterization for such $u$ in terms of $u^*\b_G$, where $\b_G$ is the basic class of $G$. In other words, we are using $u^*\b_G$ as a secondary invariant of a pair $\psi,\pfi$ while for the lift $u$ it is a primary invariant and is defined straightforwardly.

Let $(M,G)$ denote the space of continuous maps $M\to G$ and $(M,G)\pfi$ the space of maps $M\to X$ that have the form $u\pfi$ for $u\in(M,G)$. We denote further
\begin{equation}\label{stab}
\Stab_\pfi:=\{w\in (M,G)|w\pfi=\pfi\}
\end{equation}
and call it the {\it stabilizer} of $\pfi$. Then one has the following fibration
\begin{align*}
(M,G)&\ovs{\Pi}\lra (M,G)\pfi\\
u&\mapsto u\pfi.
\end{align*}
If $v\pfi =u\pfi$ then $w:=u^{-1}v \in\Stab_\pfi$ and the
fiber of this fibration is exactly the $\Stab_\pfi$. To show that this is indeed a fibration we follow an idea from \cite{AS}. By definition \cite{Brd} we need to complete the diagram as indicated
\bee\label{e0.33}
\begin{diagram}
A\times\{0\}& \rTo{F_0}          &(M,G)  \\
\dInto     &\ruDotsto         &\dTo_{\Pi}   \\
A\times I   &\rTo{f}          &(M,G)\pfi
\end{diagram}
\eee
where $I:=[0,1]$. Set $\overline{F}_{0}(m,a):=F_0(a)(m)$ and $\overline{f}(m,a,t):=f(a,t)(m)$. 
Recall from \refL{1.1} that the bundle of shifts \eqref{shifts} is a fiber bundle and therefore
a fibration so the following diagram can be completed as indicated:
\be
\begin{diagram}
(M\times A)\times {0}&  \rTo{(\overline{F}_0,\pfi )}  & G\times X\\
\dInto       &  \ruDotsto^{\overline{\Phi}}    & \dTo_{\alpha}        \\
(M\times A)\times I      &  \rTo{(\overline{f},\pfi )}      &X\times X
\end{diagram}
\ee
Inspecting the definitions of $\overline{F}_{0}$, $\overline{f}$ one concludes that the original diagram can be completed as well using $\overline{\Phi}$.  

Denote
\begin{align*}
[M,G] &:=\pi_0((M,G)),\\
[(M,G)\pfi] &:=\pi_0((M,G)\pfi).
\end{align*}
Using the homotopy exact sequence of the fibration 
$$
\Stab_\pfi\ovs{\i}\hra(M,G)\ovs{\pi}\lra (M,G)\pfi
$$ 
which is
$$
\pi_0(\Stab_\pfi)\ovs{\i_*}\lra\pi_0((M,G))\pi_*\lra\pi_0((M,G)\pfi)\lra 0.
$$ 
one gets
\begin{align}\label{secon1}
[(M,G)\pfi]\simeq \frac{[M,G]}{\i_*\pi_0(\mbox{Stab}_\pfi)} &&(\simeq\text{means bijection}).
\end{align}
Note that $[M,G]$ is the set of homotopy classes of continuous maps $M\lra G$ and $[(M,G)\pfi]$ is the set of homotopy classes of
continuous maps into $X=G/H$ $2$--homotopic to $\pfi$ by \refT{1.1}. 

If $G$ is compact simply connected $\pi_1(G)=\pi_2(G)=0$ and it follows from the Eilenberg classification theorem that
\begin{align*}
[M,G] &\simeq H^3(M,\pi_3(G))\\
[u] &\longmapsto u^*\b_G
\end{align*}
is a group isomorphism. Under this isomorphism the subgroup
$\i_*\pi_0(\Stab_\varphi)=\pi_0(\i(\Stab_\varphi))$ is mapped
into a subgroup of $H^3(M,\pi_3(G))$ that we denote $\O_\varphi$, i.e
\begin{equation}\label{Ofi}
\O_\pfi:=\{w^*\b_G\mid w\in\Stab_\pfi\}<H^3(M,\pi_3(G)).
\end{equation}
With this notation \eqref{secon1} becomes
\bee\label{secon2} 
[(M,G)\pfi]\simeq H^3(M,\pi_3(G))/\O_\pfi. 
\eee

Although the definition \eqref{Ofi} uses the map $\pfi$ explicitly we will show that in fact this subgroup only depends on its $2$--homotopy type. To this end we need the following Lemma which essentially follows from the Hopf-Samelson theorem \cite{Dy,WG}:
\begin{lemma}\label{L:1.3}
Let $\pi_1,\pi_2$ be the natural projections from $G\times G$ to the first and the second factor
and $\begin{aligned}G\times G &\ovs{m}\lra G\\ (g_1,g_2) &\longmapsto g_1g_2\end{aligned}$ be the multiplication map. Then 
\bee\label{HopSam1} 
m^*\b_G=\pi_1^*\b_G +\pi_2^*\b_G, 
\eee 
and given two maps $M\ovs{u,v}\lra G$ 
\bee\label{HopSam2} 
(u\cdot v)^*\b_G =u^*\b_G+v^*\b_G. 
\eee
\end{lemma}
\begin{proof}
Since $G$ is simply connected by the K\"unneth theorem 
\be
H_3(G\times G,\Z)=H_3(G,\Z)\times 1+1\times H_3(G,\Z),
\ee 
where $1\in H_0(G,\Z)$ is the class of a point (and one can take ${1\in G}$)
and $\times$ is the cross-product of homology classes \cite{Brd,Dy}. 
By the universal coefficients 
\be 
0 \to\Ext(H_2(G\times G,\Z),\pi_3(G))\to H^3(G\times G,\pi_3(G)) \to\Hom
(H_3(G\times G,\Z),\pi_3(G)) \to 0 
\ee 
and the first term vanishes since $H_2(G\times G,\Z)=0$. Thus elements of $H^3(G\times G,\Z)$
are determined by evaluation on homology classes and 
\begin{align*}
m^*\b_G(z\times 1+1\times w) &=\b_G(m_*(z\times1+1\times w))=\b_G(z+w)\\
&=\b_G(\pi_{1*}(z\times1)+\pi_{2*}(1\times w))\\
&=\b_G(\pi_{1*}(z\times1+1\times w)+\pi_{2*}(z\times 1+1\times w))\\
&\qquad\text{since $\pi_{1*}(1\times w)=\pi_{2*}(z\times1)=0$}\\
&=(\pi_1^*\b_G+\pi_2^*\b_G)(z\times 1+1\times w)\ \text{as claimed in \eqref{HopSam1}.}
\end{align*}
Furthermore, 
$$
(u\cdot v)^*\b_G=(m\circ(u,v))^*\b_G=(u,v)^*(\pi_1^*\b_G+\pi_2^*\b_G)=u^*\b_G+v^*\b_G
$$
as claimed in \eqref{HopSam2}.
\end{proof}

\begin{corollary}\label{C:1.5}
$\O_\pfi$ only depends on the $2$-homotopy type of $\pfi$ or
equivalently on $\pfi^*\b_X$ and not on $\pfi$ itself.
\end{corollary}

\begin{proof}
By \refT{1.1}  $\psi$ is $2$--homotopic to $\pfi$ if there is $M\ovs{u}\lra G$ such that $\psi=u\pfi$. 
Therefore
\be
\Stab_{\psi}=\{w| w\psi=\psi\}=\{w|wu\pfi=u\pfi\}=\{w|u^{-1}wu\in\Stab_\pfi\}=u(\Stab_\pfi)u^{-1}
\ee
Therefore by the definition \eqref{Ofi}
\begin{align*}
\O_\psi &=\{w^*\b_G| w\in\Stab_\psi\}=\{(uw'u^{-1})^*\b_G| w'\in\Stab_\pfi\} &&\\
&=\{u^*\b_G+ (w')^*\b_G-u^*\b_G| w'\in\Stab_\pfi\}= \O_\pfi && \text{by \refL{1.3}}
\end{align*}
\end{proof}
Hence $ \O_\pfi = \O_{\pfi^*\b_X}$ and since every $\varkappa\in
H^2(M,\pi_2(X))$ is presentable by a $\pfi$ one can talk about
$\O_\varkappa$. 

Summarizing the above discussion we conclude:
\begin{theorem}\label{T:1.2}
Two continuous maps $M\ovs{\psi,\pfi}\lra X$ are homotopic if and only if 
$\psi=u\pfi$ and $u^*\b_G\in\O_\pfi$ for some $M\xra{u}G$. 
\end{theorem}
It is instructive to compare this characterization to the classical one
given by the Postnikov classification theorem \cite{Bo,Ps,WJ}. Its formulation uses a homotopic operation known as the Whitehead product $\pi_2(X)\times\pi_2(X)\lra\pi_3(X)$ \cite{Brd,WG} to define a cup product of $\pi_2(X)$--valued cohomology classes and a cohomology operation known as the Postnikov square ${\rm Ps}:H^1(M,\pi_2(X))\to H^3(M,\pi_3(X))$ \cite{Bo,Nk1,Ps,WJ}.
\begin{theorem*}[Postnikov classification theorem] Let $M$ be a $3$--dimensional CW--complex and $X$ a connected simply connected complex of any dimension. The two continuous maps $M\ovs{\psi,\pfi}\lra X$ are $2$--homotopic if and only if $\psi^*\b_X=\pfi^*\b_X=:\varkappa$. There exists $\widetilde{\psi}$ homotopic to $\psi$ on $M$ and equal to $\pfi$ on the $2$--skeleton. The primary difference $\overline{d}(\pfi,\widetilde{\psi})\in H^3(M,\pi_3(X))$ is then defined and independent of a choice of such $\widetilde{\psi}$. The maps $\psi,\pfi$ are homotopic if and only if there exists $\alpha\in H^1(M,\pi_2(X))$ such that $\overline{d}(\pfi,\widetilde{\psi})=\varkappa\smile\alpha+{\rm Ps}(\alpha)$. In particular,
\begin{equation}\label{Postnikov}
[(M,G)\pfi]\simeq\frac{H^3(M,\pi_3(X))}{\pfi^*\b_X\smile H^1(M,\pi_2(X))+{\rm Ps}(H^1(M,\pi_2(X)))}
\end{equation}
\end{theorem*}
As one can see from the Postnikov theorem the definition of the classical secondary invariant requires the map $\psi$ to be 'preconditioned' by a $2$--homotopy and we are not aware of a more explicit procedure for defining it. Also notice that the classical invariant takes values in $H^3(M,\pi_3(X))$ whereas $u^*\b_G\in H^3(M,\pi_3(G))$. The relation between the two can be derived using
that $\pi_3(X)\simeq \pi_3(G)/\i_*\pi_3(H)$ by \eqref{e0.22}, in particular if $\i_*\pi_3(H)=0$ as in the case of $U_1$ in $SU_2$ our invariant can be identified with the classical one. For $M=S^3$ all maps are $2$--homotopic to the constant map and the secondary invariants can be given for a single map rather than a pair by fixing $\pfi$ to be the constant map. If also $X=SU_2/U_1$ both definitions give the classical Hopf invariant.

For applications in \refCh{3} it is convenient to reinterpret the basic class and the secondary invariant 
in terms of the deRham cohomology. 
Let us start with the group $H^3(G,\pi_3(G))$. Recall that we assume that $G$ is compact connected and simply connected. By the universal coefficients theorem \cite{Brd,DK} the following sequence is exact:
\be
0\lra \mathop{\rm Tor}\nolimits(H^2(G,\Z),\pi_3(G))\lra
H^3(G,\pi_3(G))\lra H^3(G,\Z)\otimes\pi_3(G)\lra 0.
\ee
Since $G$ is a simply connected Lie group $H^2(G,\Z)=0$ and the torsion term vanishes so
\be
H^3(G,\pi_3(G))\simeq H^3(G,\Z)\otimes\pi_3(G).
\ee
Since $G$ is also compact it is a direct product
of simple components $G=G_1\times\dots\times G_N$ and therefore
$$
\pi_3(G)\simeq\pi_3(G_1)\oplus\dots\oplus\pi_3(G_N).
$$
The sum on the right $\simeq\Z^N$ because $\pi_3(\Gamma)\simeq\Z$ for any simple Lie group $\Gamma$ \cite{BtD}. 
Thus
\be
H^3(G,\pi_3(G))\simeq H^3(G,\Z)\otimes\Z^N
\ee
Both third cohomology groups $H^3(G,\Z)$, $H^3(M,\Z)$ are free Abelian, 
the first one by the Hurewicz theorem and the second by Poincare duality since $M$ is a closed connected $3$--manifold (if $M$ is not orientable $H^3(M,\Z)=0$).
This means that not only are elements of $H^3(G,\Z)\otimes\Z^N$ completely 
represented by integral classes in $H^3(G,\R)\otimes\R^N$ but also that their 
pullbacks are completely characterized as integral classes in $H^3(M,\R)\otimes\R^N$. But real cohomology classes from 
$H^3(G,\R)\otimes\R^N$ are represented by $\R^N$--valued differential $3$--forms by the deRham theorem \cite{GHV}. 

Let $\Theta$ be a differential form that represents $\b_G$. Being $\R^N$--valued it is a collection
$\Theta=(\Theta_1,\dots,\Theta_N)$ of $N$ scalar $3$--forms and the pullback
$$
u^*\Theta:=(u^*\Theta_1,\dots,u^*\Theta_N)
$$
is defined as a vector--valued $3$--form. We can go one step further. 
Assuming $M$ orientable $H^3(M,\Z)\simeq\Z$ and again by the universal coefficients:
$$
H^3(M,\pi_3(G))\simeq H^3(M,\Z)\otimes\pi_3(G)\simeq H^3(M,\Z)\otimes\Z^N\simeq\Z^N.
$$
The last isomorphism is given by evaluation of cohomology classes on the fundamental class of $M$  
or in terms of differential forms by integration over $M$ \cite{GHV}. Thus we get a combined isomorphism
\bee\label{deRham}
\bal
H^3(M,\pi_3(G)) &\ovs{\sim}\lra \Z^N\\
u^*\b_G &\longmapsto \int\limits_Mu^*\Theta:=(\int\limits_Mu^*\Theta_1,\dots,\int\limits_Mu^*\Theta_N).
\eal
\eee
Under this isomorphism the subgroup $\O_\pfi<H^3(M,\pi_3(G))$ is transformed into a subgroup of $\Z^N$ 
and we denote its image by the same symbol, explicitly
\begin{equation}\label{I-Ofi}
\O_\pfi:=\{\int\limits_Mw^*\Theta\mid w\in\Stab_\pfi\}<\Z^N.
\end{equation}
Now \refT{1.2} can be restated as
\begin{corollary}\label{C:secinv}
Two continuous maps $M\ovs{\psi,\pfi}\lra X$ are homotopic if and only if 
$\psi=u\pfi$ and $\int\limits_Mu^*\Theta\in\O_\pfi$ for some $M\xra{u}G$. 
\end{corollary}
If $M$ is not orientable then the secondary invariant is always $0$.

\chapter{Gauge theory on coset bundles}\label{S:2}

As we know from the previous chapter $2$-homotopic maps $M\ovs{\psi, \pfi}\lra X$ 
into a homogeneous space $X=G/H$ are related by a
'lift' $M\ovs{u}\lra G$, namely $\psi=u\pfi$. Therefore if we want to
minimize a functional within a given homotopy type we can fix a {\it
reference map} $\pfi$ to fix a $2$-homotopy type and consider all 
maps of the form $u\pfi$.

As was demonstrated in \cite{AK1} it is even more convenient to work
with the $1$-form $a=u^{-1}du$ instead of $u$ for a number of
reasons. First, unlike maps $M\lra G$ differential forms form a
linear space that allows straightforward definition of Sobolev
spaces. Second, $a$ can be interpreted as a gauge potential of a
pure-gauge connection on $M\times G$ and the Faddeev-Skyrme functional
admits a very nice representation in these terms that allows use of
technics from the gauge theory. Finally, the stabilizer subgroup of
the reference map $\Stab_\pfi :=\{w :M\lra G|w\pfi =\pfi\}$ turns
out to be isomorphic to the group of gauge transformations of the
quotient bundle $H\hra G\to G/H$ pulled back by 
$\pfi$ (\refL{2.5})\footnote[1]{By abuse of notation we use the symbol for the total space to denote a bundle, 
for instance $\pfi^*G$ denotes a pullback of a bundle $H\hra G\to G/H$ regardless of what $H$ is.}. 
This fact is very useful in the description of secondary invariants. All of
this leads to the idea of restating the whole problem in terms of
gauge theory on the quotient bundles and their pullbacks that we call {\it coset bundles}. 

Note that trivial bundles can be seen as pullbacks of the one-point quotient bundle $G\hra G\to\pt$.
Quotient bundles are distinguished from general principal bundles by
having a group as the total space. This leads to many nice
constructions on their pullbacks analogous to constructions on trivial bundles and not available in the general case. 
Two examples are invariant connections and {\it untwisting} of gauge potentials (see
\refD{cospot}).
 
After introducing some general notions and fixing the
notation we devote the rest of this chapter to description of these
constructions. It turns out that gauge potentials on $\pfi^*G$ can
be realized as $\g$-valued (not just bundle-valued) forms on $M$ and we
derive formulas for the gauge action and curvature in this representation (\refT{2.1}). 
As a consequence we can interpret the coisotropy form
$\omega^\bot$ in terms of the simplest invariant connection $\pr_\h (g^{-1}dg)$ on $\pfi^*G$ (\refC{refcurv}).

\section{Connections on principal bundles}\label{S:2.1}

In this section we fix the notation and list some basic facts and formulas
about Lie algebra valued, matrix valued and connection forms for future reference (see
\cite{BM,FU,GHV,MM} among others).

Let $\mathbb{E}$ be a Euclidian space and $\mbox{End}(\mathbb{E})$
the algebra of linear operators on it. Exterior $k$-form is a
multilinear antisymmetric map $\mathbb{E}^k\ovs{\alpha}\lra
\mathbb{R}$, the space of such forms is denoted
$\Lambda^k\mathbb{E}$ and $|\alpha |:=k$ is called the degree of
$\alpha$. If $M$ is a smooth manifold then a differential $k$-form
$\alpha$ is an assignment of an exterior $k$-form $\alpha_m$ to each
tangent space $T_m M$, that varies smoothly with $m\in M$, the notation is $\Gamma
(\Lambda^k M)$. Smooth here means $C^\infty$, however we will later use Sobolev spaces of
forms such as $L^2$ and $W^{1,2}$. If $\mathbb{R}$ is replaced by End$(\mathbb{E})$ as
a set of values we talk about matrix valued forms instead of scalar
ones and denote their space $\Gamma
(\Lambda^kM\otimes\mbox{End}(\mathbb{E}))$. For each form $\alpha$
we define its {\it differential} $\Gamma
(\Lambda^kM\otimes\mbox{End}(\mathbb{E}))\ovs{d}\lra\Gamma
(\Lambda^{k+1}M\otimes\mbox{End}(\mathbb{E}))$ by
\begin{multline}\label{dif}
d\alpha (X_1,\dots ,X_{k+1}):=\sum\limits^{k+1}_{i=1}
(-1)^{i+1}X_i\alpha (X_1,\dots ,\widehat{X}_i ,\dots ,X_{k+1})+ \\
+\sum\limits^{k+1}_{i<j=1}(-1)^{i+j}\alpha ([X_i,X_j],X_1,\dots
,\widehat{X}_i,\dots ,\widehat{X}_j, \dots ,X_{k+1})
\end{multline}
where as usual $X_i$ are vector fields, $Xf$ is the derivative in
the direction of $X$ of a function $f$, $[X,Y]$ is a bracket of
vector fields and $\widehat{X}$ means omission. For each pair of
forms we define the {\it wedge product} $\Gamma
(\Lambda^kM\otimes\mbox{End}(\mathbb{E}))\times \Gamma
(\Lambda^lM\otimes\mbox{End}(\mathbb{E}))\ovs{\wedge}\lra\Gamma
(\Lambda^{k+l}M\otimes\mbox{End}(\mathbb{E}))$
\bee\label{e0.34}
\alpha\wedge\beta (X_1,\dots
,X_{k+l}):=\frac{1}{k!l!}\sum\limits_{\sigma\in
S_{k+l}}\mbox{sgn}(\sigma)\alpha (X_{\sigma (1)},\dots ,X_{\sigma
(k)})\beta(X_{\sigma (k+1)},\dots ,X_{\sigma (k+l)})
\eee
where
$S_n$ is the group of permutations of $n$ elements and sgn$(\sigma)$
is the sign of a permutation, and the {\it graded commutator}
\bee\label{e0.35}
[\alpha ,\beta](X_1,\dots
,X_{k+l}):=\frac{1}{k!l!}\sum\limits_{\sigma\in
S_{k+l}}\mbox{sgn}(\sigma)[\alpha (X_{\sigma (1)},\dots ,X_{\sigma
(k)}),\beta (X_{\sigma (k+1)},\dots ,X_{\sigma (k+l)})]
\eee
with
$[\xi ,\eta]:=\xi\eta -\eta\xi$ in End$(\mathbb{E})$. Note that for
scalar forms always $[\alpha ,\beta]=0$. If $\g$ is a Lie algebra
then one may consider $\g$--valued forms $\alpha\in \Gamma
(\Lambda^kM\otimes \g)$ with the differential \eqref{dif} and the graded
commutator \eqref{e0.35}, where $[\xi ,\eta]$ means the Lie bracket in $\g$
but the wedge product is no longer defined. To help this note that
by the Ado theorem there is always a faithful representation
$\g\hra\mbox{End}(\mathbb{E})$ \cite{BtD} that can be used to define
\eqref{e0.34}. Of course in general $\alpha\wedge\beta$ is no
longer $\g$-- but only $\mbox{End}(\mathbb{E})$--valued. 

If $G$ is a Lie group with the Lie algebra $\g$ then the {\it adjoint action} of
$G$ on $\g$ extends to forms
\bee\label{e0.36}
(\Ad_*(g)\alpha)(X_1,\dots ,X_k):=\Ad_*(g)(\alpha (X_1,\dots ,X_k)).
\eee
If $G$ is a compact Lie group then it also has a faithful
representation $G\hra\mbox{End}(\mathbb{E})$ which induces the
corresponding representation $\g\hra\mbox{End}(\mathbb{E})$ and this is
always the one we use. 
The following properties are more or less straightforward from the
definitions (see \cite{BM}):
\bee\label{e0.37}
\bal
&\quad \text{\bf Wedge-commutator relations} \\
&{\rm (i)}\ (\alpha\wedge\beta)\wedge\gamma=\alpha\wedge(\beta\wedge\gamma)\\
&{\rm (ii)}\ [\alpha ,\beta]=\alpha\wedge\beta-(-1)^{|\alpha||\beta|}\beta\wedge\alpha\\
&{\rm (iii)}\ [\beta ,\alpha]= -(-1)^{|\alpha||\beta|}[\beta ,\alpha]\\
&{\rm (iv)}\ \alpha\wedge\alpha=1/2[\alpha ,\alpha] \\
&{\rm (v)}\ [\alpha\wedge\alpha ,\alpha]=[\alpha ,\alpha\wedge\alpha]=0\\
&\quad \text{\bf Cancellation formula} \\
&{\rm (vi)}\ [[\alpha ,\beta], \beta]=[\alpha ,\beta\wedge\beta],\quad |\beta|\ \text{odd}\\
&\quad \text{\bf Adjoint action} \\
&{\rm (vii)}\ \Ad_*(g)(\alpha\wedge\beta)=(\Ad_*(g)\alpha)\wedge (\Ad_*(g)\beta)\\
&{\rm (viii)}\ d(\Ad_*(g)\alpha)=\Ad_*(g)(d\alpha +[g^{-1}dg,\alpha]) \\ 
&\quad \text{\bf Product rules} \\
&{\rm (ix)}\ d(\alpha\wedge\beta)=d\alpha\wedge\beta+(-1)^{|\alpha|}\alpha\wedge d\beta\\
&{\rm (x)}\ d[\alpha ,\beta]=[d\alpha ,\beta]+(-1)^{|\alpha|}[\alpha ,d\beta]\\
&{\rm (xi)}\ d(\alpha\wedge\alpha)=[d\alpha ,\alpha]=-[\alpha ,d\alpha],\quad |\alpha|\ \text{odd.}
\eal
\eee
It is worth pointing out that by \eqref{e0.37}(iii) for forms of odd degree one has contrary to the intuition
$$
[\alpha ,\beta]=\alpha\wedge\beta+\beta\wedge\alpha.
$$
Again for forms of odd degree the wedge square $\alpha\wedge\alpha$ is invariantly defined 
by \eqref{e0.37}(iv) although in general $\alpha\wedge\beta$ is not $\g$--valued 
and depends on a choice of representation for $\g$. If $\alpha$ is a $\g$--valued form and $\Phi$ is
an End$(\g)$--valued one then $\Phi\wedge\alpha$ can be defined by
the expression \eqref{e0.34} if 'multiplication' there is interpreted as application of an
operator from End$(\g)$ to an element of $\g$. The product rule
\eqref{e0.37}(ix) still applies but generally speaking \eqref{e0.37}(i) fails:
$(\Phi\wedge\alpha )\wedge\beta\neq\Phi\wedge (\alpha\wedge\beta)$.

Given a Lie group $G$ its Lie algebra $\g$ is canonically identified
with the tangent space at identity $T_1G$. Left action of $G$ on
itself also gives a canonical isomorphism between $\g=T_1G$ and
$T_g G$ for any $g\in G$. Namely, if $L_\gamma g:=\gamma g$ then
the isomorphism is $T_1G\ovs{L_{g*}}\lra T_g G$ and we write
abusively $g\xi :=L_{g*}\xi$ for $\xi\in \g$, $g\in G$. One then
gets a tautological $\g$--valued $1$-form $\theta_ L$ on $G$: 
$\theta_L(g\xi):=\xi$. It is traditionally denoted $g^{-1}dg$ and
called the (left-invariant) Maurer-Cartan form for it satisfies
$L_g^*\theta_L=\theta_L$ for any $g\in G$. Analogously one can
define the right-invariant form $\theta_R=dgg^{-1}$ using the
right action of $G$ on itself.

Let $H\ovs{\i}\hra P\ovs{\pi}\lra M$ be a smooth principal bundle
with the structure group $H$ (see \refD{princp}), $\h$ be the Lie algebra
of $H$ and $R_h$ denote the right action of $H$ on $P$ ($G$ is
reserved since for the quotient bundles we have $P=G$).

\begin{definition}[Connection forms]\label{D:2.1}
An $\h$--valued $1$--form $A\in\Gamma^1(\Lambda^1 P\otimes\h)$ on $P$ is called a {\it connection form} if
\bee\label{conn}
\bal
&1) \i^* A=h^{-1}dh\\
&2) R_h^*A=\Ad_*(h^{-1})A
\eal
\eee
\end{definition}

According to this definition the Maurer-Cartan form $g^{-1}dg$ is a
connection form on the principal bundle $G\hra G\lra \pt$
over one point. If $H\hra Q\lra N$ is another $H$-bundle and
$Q\ovs{f}\lra P$ is a morphism of $H$-bundles, i.e. $f(qh)=f(q)h$
then a connection form $A$ on $P$ pulls back to a connection form
$f^*A$ on $Q$. Automorphisms of $P$ that cover the identity
\bee\label{e0.38}
\begin{diagram}
P     & \rTo{f}          &P        \\
\dTo{\pi}   &            &\dTo{\pi}      \\
M      &   \rTo{id}    & M
\end{diagram}
\eee
are called {\it gauge transformations} and form a group denoted
$\Aut(P)$ or $\mathcal{G} (P)$. The above remark gives an action of
gauge transformations on the space of connection forms.

\begin{definition}[Curvature forms]\label{D:2.2}
Given a connection form $A$ on $P$ the $\h$--valued $2$-form $F(A):=dA+A\wedge A$
is called the {\it curvature form} of the connection $A$. Any curvature from satisfies
\bee\label{curv}
\bal
&1) \i^*F(A)=0 && \text{($F(A)$ is horizontal)}\\
&2) R_h^*F(A)=\Ad_*(h^{-1})F(A) && \text{($F(A)$ is equivariant)}
\eal
\eee
\end{definition}
Let $A_0$ be a fixed {\it reference connection} on $P$ and $A$ be an
arbitrary one then the difference $A-A_0$ is also horizontal and equivariant. 
Unlike the connection forms themselves that only form an affine space the differences $A-A_0$ form a linear one. 
Horizontality implies that if $\overline{X}\in T_pP$ is a lift of $X\in T_mM$ with $\pi (p)=m$,
i.e. $\pi_*\overline{X}=X$ the value $(A-A_0)(\overline{X})$ only
depends on $X$ and not on a choice of the lift. This property allows
one to descend forms on $P$ to forms on $M$. If the value were also independent of a choice of $p$ in the fiber over $m$ (invariance) we could obtain an $\h$--valued form on $M$. As it is however, one has to deal with bundle--valued forms. 

Consider the following Borel construction (see \eqref{Borel} or \cite{Hus}). The structure group $H$ acts on $\h$ on the left by
the adjoint representation $\Ad_*$ and we set
$\Ad_*(P):=P\times_{\Ad_*}\h$. This is a vector bundle over $M$ with
fibers isomorphic to the Lie algebra $\h$. Bundle--valued forms are
defined in the same way as $\End(\mathbb{E})$--valued ones in the
beginning of this section except $\alpha_m$ now takes values in the
corresponding fiber $\Ad_* (P)_m$ of the bundle. The notation is
$\alpha\in\Gamma (\Lambda^kM\otimes \Ad_*(P))$.
\begin{definition}[Gauge potentials]\label{D:pot}
The gauge potential $\alpha$ of a connection $A$ on $P$ with respect to
a reference connection $A_0$ is an $Ad_*(P)$--valued $1$-form on
$M$ given by $\alpha_m(X):=[p,(A-A_0)(\overline{X})]$, where $\pi (p)=m$ and
$\overline{X}$ is an arbitrary lift of $X$ to $T_pP$.
\end{definition}
One can check that taking $ph$ instead of $p$ gives the same value.
There is one-to-one correspondence between the gauge potentials and the connection forms  
but the fact that they are bundle--valued is a nuisance. For coset bundles we will give a different presentation of connections by 'untwisted' gauge potentials that are $\g$--valued forms ({\it not} $\h$--valued forms) in \refS{2.3}.

The gauge transformations can also be described in a similar fashion.
Since $H$ acts on the left on itself by the adjoint action $\bal
H\times H &\ovs{\Ad}\lra H\\ (\lambda ,h)&\longmapsto \lambda
h\lambda^{-1}\eal$ we can set $\Ad P:=P\times_{\Ad}H$. Elements of a fixed fiber 
form a group under the multiplication $[p,h_1]\cdot[p,h_2]:=[p,h_1h_2]$ and 
therefore so do the sections of $\Ad P$. There is an isomorphism Aut$(P)\simeq\Gamma (\mbox{Ad}P)$ and both
groups are known as {\it the gauge group} of $P$ \cite{MM}.

\section{The coisotropy form}\label{S:2.2}

In this section we introduce the coisotropy form of a homogeneous space that plays a central role in our formulation of the Faddeev-Skyrme energy. On a Lie group one has two canonical forms, the left--invariant one $g^{-1}dg$ and the right-invariant one $dg\,g^{-1}$. Note that the latter although not invariant under the left action is however left $\Ad_*$--equivariant, i.e. $L_{\gamma*}(dg\,g^{-1})=\Ad_*(\gamma)dg\,g^{-1}$. On a homogeneous space $G/H$ we only have left action of the group $G$ and although it is impossible to define a meaningful $\g$--valued left--invariant form on $G/H$ it is possible to define a left--equivariant one at least when $G$ admits a bi-invariant Riemannian metric (e.g. when $G$ is compact \cite{BtD}). 
This form is our coisotropy form and it reduces to $dg\,g^{-1}$ when $H$ is trivial.

We start by fixing a bi-invariant Riemannian metric on $G$. On the Lie algebra $\g$ of $G$ identified with the tangent space $T_1G$ this metric induces the {\it isotropy decomposition} 
$$
\g=\h\oplus\h^\bot 
$$
with $\h$ being the Lie algebra of $H$. Since the metric is
bi-invariant 
$$
(\Ad_*(g)\xi ,\Ad_*(g)\eta)=(\xi ,\eta).
$$ 
If $A^*$ denotes the adjoint of $A$ with respect to the metric then
$$
(\Ad_*(g))^*=\Ad_*(g)^{-1}=\Ad_*(g^{-1}).
$$ 
In other words, $\Ad_*(g)$ is an isometry on $\g$ for any
$g\in G$. If $h\in H$ then $\Ad(h)H\subset H$ and by differentiation
$\Ad_*(h)\h\subset\h$. With $\Ad_*(h)$ being an isometry it also yields
$\Ad_*(h)\h^\bot\subset\h^\bot$ and both subspaces of the isotropy decomposition are
$\Ad_*(H)$-invariant. If $\pr_{\h}$, $\pr_{\h^\bot}$ denote the corresponding orthogonal projections then the invariance implies that $\Ad_*(h)$ commutes with both of them.

Now the adjoint action $\ad$ of $\g$ on itself is the derivative of the $\Ad_*$ action of $G$ on $\g$ and the isometric property translates into
$$
\ad^*_\xi =-\ad_\xi
$$ 
for any $\xi\in\g$ and we also have that $\h,\h^\bot$ are invariant under $\ad_\xi$. Since $\ad_\xi\eta =[\xi,\eta]$ 
the invariance can be expressed as
\begin{gather}\label{hh}
[\h ,\h]\subset\h ,\qquad [\h ,\h^\bot]\subset\h^\bot.
\end{gather}

The coset space $X=G/H$ inherits a metric from $G$ by the Riemann
quotient construction \cite{Pe}. If $G\xra{\pi}X$ is the quotient map set
$(S,T)_X:=(\overline{S},\overline{T})_G$, where
$\overline{S}$, $\overline{T}$ are the unique lifts of $S$, $T$ to $TG$
that are orthogonal to the kernel of the projection $TG\ovs{\pi_*}\lra TX$.
Bi-invariance implies that $(\cdot ,\cdot)_X$ is invariant under the
left action of $G$ on $X$: $(L_{g*}S,L_{g*}T)_X=(S,T)_X$. Moreover, $G$ 
is the {\it isometry group} of the Riemannian manifold $X$ \cite{Ar,Pe}. 

Let $x_0:=1H=\pi(1)\in X$ then the projection
$\g=T_1G\ovs{\pi_*}\lra T_{x_0}X$ identifies $\h^\bot$ with the
tangent space to $X$ at $x_0$. Left action of $G$ on $X$ allows one
to extend the isomorphism of $\h^\bot$ to an arbitrary $T_xX$ but
since there is more than one way to present $x$ as $gx_0$ this
isomorphism is not canonical. Note that every vector in $T_xX$ has the form
$g(\pi_*\xi)$ for $\xi\in\g=T_1G$ (we take the liberty of writing $gT$ instead of $L_{g*}T$).
\begin{definition}[The coisotropy form]\label{D:coisot}
The coisotropy form $\omega^\bot\in\Gamma(\Lambda^1X\otimes\g)$ of $X$ is
\bee\label{e0.41}
\omega^\bot
(g(\pi_*\xi)):=\Ad_*(g)\pr_{\h\bot}(\xi)
\eee
or equivalently
\bee\label{coisot}
\pi^*\omega^\bot:=\Ad_*(g)\pr_{\h\bot} (g^{-1}dg).
\eee
\end{definition}

There is another description of the coisotropy form that makes it
more transparent that it is well-defined (does not depend on a
choice of $g$ in $x=gx_0$). It uses the isotropy subalgebra of a point. 
Recall that the isotropy subgroup of a point $x\in X$ is 
$$
H_x:=\{\gamma\in G|\gamma x=x\}.
$$
If $x=gx_0=gH$ then $\gamma gH=gH$ is equivalent to $\gamma\in\Ad(g)H$ and
$$
H_x=\mbox{Ad}(g)H,\qquad x=gH.
$$ 
By analogy we define the isotropy subalgebra $\h_x$ of $x\in X$ and the {\it coisotropy subspace} $\h^\bot_x$:
\begin{equation*}
\bal
\h_x &:=\mbox{Ad}_*(g)\h,\quad x=gH\\
\h_x^\bot &:=\mbox{Ad}_*(g)\h^\bot 
\eal
\end{equation*}
This is well-defined since $\mbox{Ad}_*(gh)=\mbox{Ad}_*(g)\mbox{Ad}_*(h)$ and both $\h,\h^\bot$ are $\mbox{Ad}_*(H)$--invariant.
Moreover, \eqref{hh} implies
\bee\label{e0.42}
[\h_x ,\h_x]\subset\h_x,\qquad [\h_x,\h_x^\bot]\subset\h_x^\bot.
\eee
Here is a more geometric interpretation. In the same spirit as $gT$ denotes the action of an element of $G$ on a tangent vector in $X$ we can write $\xi x$ for the action of a vector in $\g$ on a point in $X$. Both are induced by the left action of $G$ on $X$ and rigorously
$$
\xi x:=\frac{d}{dt}\exp(t\xi)x|_{t=0}.
$$
Since $G$ acts transitively, for each $x\in X$ the map $\xi\mapsto\xi x$ is
onto $T_xX$ and its kernel is exactly the isotropy subalgebra $\h_x$. The next lemma establishes some important properties of the coisotropy form.
\begin{lemma}\label{L:2.1}
{\rm(i)} $ \omega^\bot$ is well-defined and
\bee\label{e0.43}
\omega^\bot (\xi x)=\pr_{\h_x^\bot}(\xi).
\eee
{\rm (ii)} $L^*_\gamma\omega^\bot=\mbox{Ad}_*(\gamma)\omega^\bot$, i.e. $\omega^\bot$ is
left--equivariant.

\noindent {\rm (iii)}$|\omega^\bot(S)|=|S|$ for any $S\in TX$.
\end{lemma}
\begin{proof}
{\rm (i)} Since $\xi x\in T_xX$ it has the form
$$
g(\pi_*\widetilde{\xi})=\xi x=\xi gH=g\widetilde{\xi}H,
$$
where $x=gH$. Thus one can take
$\widetilde{\xi}=\Ad_*(g^{-1})\xi)$. Now by \eqref{e0.41}
\bee\label{e0.44}
\omega^\bot (\xi x)=\omega^\bot
(g(\pi_*\widetilde{\xi}))=\Ad_*(g)\pr_{\h^\bot}
(\widetilde{\xi})=\Ad_*(g)\pr_{\h^\bot}(\Ad_*(g^{-1})\xi)
\eee
By linear algebra if $\m$ is a subspace of a Euclidian space and ${\rm U}$ is an
isometry then
\be
\pr_{{\rm U}\m}={\rm U}\pr_{\m}{\rm U}^*={\rm U}\pr_{\m}{\rm U}^{-1}
\ee
and since $\Ad_*(g)$ is an isometry we obtain from \eqref{e0.44}
\be
\omega^\bot(\xi x)=\pr_{\Ad_*(g)\h^\bot}(\xi)=\pr_{\h_x^\bot}(\xi)
\ee
as required. Since the last expression depends only on $x\in X$
and not on $g\in G$ we conclude that $\omega^\bot$ is well defined.

{\rm (ii)} Since in our notation $L_{\gamma *}S=\gamma S$:
\begin{multline*}
L_\gamma^*\omega^\bot(g(\pi_*\xi))=\omega^\bot(\gamma
g(\pi_*\xi))=\Ad_*(\gamma g)\pr_{\h^\bot}(\xi)\\
=\Ad_*(\gamma)(\Ad_*(g)\pr_{\h^\bot}(\xi ))=\Ad_*(\gamma)\omega^\bot(g(\pi_*\xi)).
\end{multline*}

{\rm (iii)} Since $\Ad_*(\gamma)$ is an isometry and the metric on
$X$ is left-invariant it suffices to check the equality for $x=x_0$,
$g=1$. But there the lift of $S=\pi_*\xi$ is exactly
$\overline{S}=\pr_{\h^\bot}(\xi)$ since $\Ker \pi_*=\h$. Therefore
by definition of the Riemann quotient: $|S|:=|\overline{S}|=|\omega^\bot(S)|$.
\end{proof}
\begin{remark}
For most of the above it would have been sufficient to fix an
$\Ad_*(H)$--invariant $\m \subset \g$ which is coisotropic, i.e. $\g=\h\overset{.}+ \m$ (direct sum). 
With a bi-invariant metric one just takes $\m=\h^\bot$. In general, closed subgroups of Lie groups that admit
existence of such a subspace are called reductive \cite{Ar,BC,KN}. Obviously, in a compact Lie group every closed subgroup is
reductive. We chose to use a metric since we need it to define the Faddeev-Skyrme functional anyway.
\end{remark}

As one can see from \refL{2.1} the coisotropy form is just a way to rewrite tangent vectors on $X$ as vectors in $\g$ in an algebraically nice way (the Maurer-Cartan form $dg\,g^{-1}$ plays the same role on $G$). The next example gives a more explicit description in the case of $\CP^1=SU_2/U_1$.

\begin{example}[The coisotropy form of $\CP^1$]\label{E:CP1coisot}
Recall that $SU_2$ is represented by
\be
\begin{aligned}
SU_2&=\left\{\left.\begin{pmatrix} z &w\\
-\overline{w}&\overline{z}\end{pmatrix}\right|\ z, w\in\C,
|z|^2+|w^2|=1\right\}\\
U_1&=\left\{\left.\begin{pmatrix} z&0\\ 0&\overline{z}\end{pmatrix}\right|\ z\in\C,
|z|=1\right\}<SU_2
\end{aligned}
\ee
It is convenient to use the isomorphism $\begin{pmatrix}
z&w\\ -\overline{w}&\overline{z}\end{pmatrix}\longmapsto z+w
j\in\H$ with  the algebra of quaternions and use the quaternionic
notation. In this notation
\bee\label{quat}
\begin{aligned} 
G &=SU_2=\{q\in\H|\ |q|=1\}\\
H &=U_1=\{q\in\C|\ |q|=1\}\\
\g &=\su_2=\{\ q\in\H|\Re(q)=0\}=\Im\H\\
\h &=\uu_1=\{\ q\in\C |\Re(q)=0\}=\Im\C =i\R
\end{aligned}
\eee
There is a useful embedding 
\begin{align*} 
\CP^1 &\ovs{\tau}\hra \H\\
qU_1 &\mapsto qiq^{-1}=\Ad_*(q)i
\end{align*} 
with the image
\be
\tau (\C P^1)=S^2=\{q\in\Im\H|\ |q|=1\}\subset\Im\H =\g
\ee
and it is convenient to identify $\CP^1$ with this image. 

We will now compute the coisotropy form under this identification. Since by \refL{2.1}(ii) $\omega^\bot$ is
left-equivariant it suffices to compute it for $x_0=\pi(1)$ that is mapped into $i$ under $\tau$. 
Differentiating $\tau$ one gets 
\begin{align*} 
T_{x_0}\C P^1&\ovs{\tau_*}\lra T_iS^2\\
\xi x_0&\longmapsto [\xi ,i]
\end{align*} 
where as usual $T_iS^2$ is identified with a subspace in $\Im\H$. 
Therefore by \refL{2.1}(i)
$$
\omega_{x_0}^\bot (\xi
x_0)=\pr_{\h_{x_0}^\bot}(\xi)=\pr_{\h^\bot}(\xi)=\frac{1}{2}i[\xi
,i]=\frac{1}{2}i(\tau_*(\xi x_0)).
$$
Hence if we identify $T_{x_0}\CP^1$ with $T_iS^2$ and write $\omega_i^\bot$
as a form on $\Im\H$ it becomes $\omega_i^\bot
(\eta)=\frac{1}{2}i\eta$. Analogously identifying $T_x\CP^1$ with $T_{\tau(x)}S^2\subset\Im\H$ and using the left equivariance 
we get $\omega_x^\bot(\xi x)=\frac{1}{2}\tau (x)(\tau_*(\xi x))$. Thus 
\bee\label{e0.54}
\omega_q^\bot (\eta)=\frac{1}{2}q\eta,\quad q\in
S^2,\quad \eta\in T_qS^2.
\eee
Geometrically this means that $\omega^\bot$ takes half of a vector in a tangent plane to $S^2$ and rotates it by $90^0$
counterclockwise in that plane. The resulting vector is interpreted as an element of $\g=\Im\H=\R^3$.
\end{example}

Although we do not reflect it in the notation $\omega^\bot$ depends
on a choice of presentation $X=G/H$ and a choice of a bi-invariant
metric on $G$. We want to investigate this dependence now. Recall
that in \refS{1.1} assuming $X$ simply connected we used the following
operations on $G$ to obtain the presentation of \refC{1.1}:
\begin{itemize}
\item[1.] Taking the identity component $G_0$ and replacing $G/H$ by $G_0/({G_0}\cap H)$;
\item[2.] Taking the universal cover $\widetilde{G}\ovs{\pi}\lra G$ and
replacing $G/H$ by $\widetilde{G}/{\pi^{-1}(H)}$; 
\item[3.] Taking a maximal compact subgroup $K(G)$ and replacing $G/H$ by\\ 
$K(G)/({K(G)\cap H})$.
\end{itemize}
Since $G$ is compact its universal cover decomposes \cite{BtD}:
\be
\widetilde{G}=\widetilde{G}_1\times\dots\times\widetilde{G}_k\times\R^n
\ee
with $K(\widetilde{G})=\widetilde{G}_1\times\dots\times\widetilde{G}_k\triangleleft\widetilde{G}
$ being a normal subgroup of $\widetilde{G}$. Moreover, by
the Montgomery theorem \cite{Mg} $K(\widetilde{G})$ still acts transitively on $X$
and therefore $\widetilde{G}=K(\widetilde{G})\cdot\pi^{-1}(H)$.
Analogously, $G_0\triangleleft G$ and since $X$ is connected one also has
$G=G_0\cdot H$. In other words, the above three operations are particular
cases of the following two:
\begin{itemize}
\item[(R1)] $G$ is replaced by a cover $\overline{G}\ovs{\pi}\lra G$
and $X=G/H=\overline{G}/{\pi^{-1}(H)}$;
\item[(R2)] $G$ is replaced by a normal subgroup $N\triangleleft G$
such that $G=N\cdot H$ and\\ $X=G/H=N/({N\cap H})$.
\end{itemize}
It turns out that given a bi--invariant metric on $G$ the metrics on
$\overline{G}$ and $N$ can be chosen so that the metric on $X$ and
the coisotropy form $\omega^\bot$ stay the same. This means that the presentation of \refC{1.1} can be used without loss of generality even assuming that a homogeneous space $X$ comes equipped with a metric and a coisotropy form.

\begin{lemma}\label{L:2.2}
Given a bi-invariant metric on $G$ define a metric on $\overline{G}$ from {\rm (R1)} by pullback 
and on $N$ from {\rm (R2)} by restriction. Then the Riemann quotient metric on $X$ and $\omega^\bot$ 
are the same in the new presentation of $X$.
\end{lemma}
\begin{proof}
Note that in both cases we obtain a bi-invariant metric on
$\overline{G}$ and $N$ respectively (for $\overline{G}$ since $\pi$
is a group homomorphism). We will give a proof for {\rm (R2)}, for {\rm (R1)} it
is analogous and simpler. 

If $S,T\in T_xX$ we can choose their lifts
$\overline{S},\overline{T}\in T_nG$ with $n\in N$, $x=nH$ since
$G=NH$. Then
$(S,T)_x=(\overline{S},\overline{T})_G=(\overline{S},\overline{T})_N$
by definition of the Riemann quotient \cite{Pe}. Let $\omega_N^\bot$,
$\omega_G^\bot$ denote the coisotropy forms induced from $N$, $G$
respectively. We have the diagram:
\be
\begin{diagram}
N           & \rTo{\i}  & G \\
\dTo{\pi_N} &           &\dTo{\pi_G}  \\
X           &\rTo{\id}  & X
\end{diagram}
\ee
By \eqref{coisot}
$$
\pi_N^*\omega_G^\bot=\i^*\pi_G^*\omega_G^\bot=\i^*(\Ad_*(g)\pr_{\h^\bot}(g^{-1}dg))=\Ad_*(n)\pr_{\h^\bot}(n^{-1}dn)
$$
But $G=N\cdot H$ implies $\g=\nn+\h$, where $\nn$ is the Lie algebra of
$N$ and hence $\h^\bot\subset\nn$. The orthogonal complement to $\nn\cap\h$ in $\nn$ is $\nn\cap(\nn\cap\h)^\bot$ and therefore
by \refD{coisot}
$$
\pi_N^*\omega_N^\bot=\Ad_*(n)\pr_{\nn\cap(\nn\cap\h)^\bot}(n^{-1}dn).
$$
But since $\h^\bot\subset\nn\subset\g$ we have 
$$
\nn\cap(\nn\cap\h)^\bot=\nn\cap\h^\bot=\h^\bot\subset\g.
$$
Therefore $\pi_N^*\omega_G^\bot=\pi_N^*\omega_N^\bot$ and
$\omega_G^\bot=\omega_N^\bot$ since $\pi_N^*$ is mono. 
\end{proof}

\begin{remark}
Both operations {\rm (R1)},{\rm (R2)} can be interpreted in terms of principal
bundles. In {\rm (R1)} we have a bundle $P=(\overline{G}\lra
X)$ and a normal subgroup $\pi^{-1}(1)\triangleleft\pi^{-1}(H)$ of the
structure group $\pi^{-1}(H)$ and pass to the quotient bundle
$P/\pi^{-1}(1)$ \cite{Hus}. In {\rm (R2)} the structure group
$H\triangleleft G$ reduces to $(N\cap H)\triangleleft H$ and $N$
is the total space of the bundle obtained from the original one by the reduction of the structure group \cite{Hus,KN}. 
Thus our 'changes of presentation' are just quotients and reductions of the bundle $H\hra G\lra X$.
\end{remark}

\section{Trivial bundles and coset bundles}\label{S:2.3}

This section provides the main technical tools needed for application of the gauge theory to the Faddeev-Skyrme models in the next chapter. After reviewing briefly some special constructions that are available on trivial principal bundles (trivial connection, pure-gauge connections, global gauges, etc.) we proceed to generalize them to coset bundles and develop some necessary 'calculus' for such bundles.

Trivial bundles are the simplest principal bundles and their total spaces are 
products $P=M\times G$. The principal action is
multiplication by $G$ on the right in the second component 
$$
\bal
(M\times G)\times G &\lra M\times G\\
((m,g),\gamma)&\mapsto (m,g\gamma)
\eal
$$
and the projection is the projection $M\times G\ovs{\pi_1}\lra M$ to the first component. 
Trivial bundles and only those can be obtained by pullback from the bundle over one point $G\lra\pt$.
Indeed, in general pullback of a principal bundle $P\ovs{\pi}\lra X$ by a map $M\ovs{\pfi}\lra X$ is
\be
\pfi^* P:=\{(m,p)\in M\times P|\pfi (m)=\pi (p)\}
\ee
and for $P_{\pt}:=(G\to\pt)$ the defining condition trivializes leaving just $M\times
G$. 

For each pullback bundle there is a canonical bundle morphism
$M\times P \supset \pfi^* P\ovs{\pi_2}\lra P$ that allows 'transfer'
of connection forms: every connection $A$ on $P$ induces a
connection $\pi_2^*A$ on $\pfi^*P$. For $P_{\pt}$ the
left-invariant Maurer-Cartan form $\theta_L=g^{-1}dg$ gives a
canonical connection and $\pi_2^*\theta_L$ (also denoted $g^{-1}dg$ when no confusion can result)
is called the {\it trivial connection} on $M\times G$. More
connections can be obtained by using gauge transformations (bundle automorphisms) $f$ of $M\times G$. 
Since $f(m,g\gamma)=(m,f_2(m,g)\gamma)$ we
have $f_2(m,g)=f_2(m,1)g=u(m)g$, where $M\ovs{u}\lra G$ and
$f(m,g)=(m,u(m)g)$. Conversely, any map into $G$ induces a gauge
transformation and we have a one-to-one correspondence between maps
$M\to G$ and $\Aut(M\times G)$. Applying them to the trivial
connection we get new ones:
\begin{multline}
f^*\pi_2^*(g^{-1}dg)=(\pi_2\circ f)^*(g^{-1}dg)=(ug)^{-1}d(ug)\\
=g^{-1}u^{-1}(dug+udg)=\Ad_*(g^{-1})(u^{-1}du)+g^{-1}dg.
\end{multline}
Such connections are called {\it pure-gauge} since they are trivial up to gauge equivalence (one could define
'pure-gauge' connections on any principal bundle with a reference
connection $A_0$ as those of the form $f^*A_0$ but this is not
common). Thus we have a canonical choice of a reference connection
$A_0:=\pi_2^*(g^{-1}dg)=g^{-1}dg$ (by our abuse of notation) and may consider differences $A-A_0$. The reason
that the gauge potentials from \refD{pot} had to be bundle--valued was
that the differences $A-A_0$ although horizontal are not
invariant under the right action of the structure group. We only have $\Ad_*$-equivariance:
$$
R_g^*(A-A_0)=Ad_*(g^{-1})(A-A_0).
$$ 
On a trivial bundle (and, as we will see shortly on a coset bundle) this can be fixed by a correction
factor $Ad_*(g)$. Indeed, the form $Ad_*(g)(A-A_0)$ is horizontal, invariant
and therefore descends to a $\g$--valued form on $M$.
\begin{definition}[Untwisted gauge potentials on trivial bundles]\label{D:trivpot}
The untwisted gauge potential of a connection $A$ on $M\times G$ is
the form $a\in\Gamma(\Lambda^1\otimes\g)$ satisfying
\bee\label{trivpot}
\pi_1^*a=\Ad_*(g)(A-g^{-1}dg).
\eee
\end{definition}
It is immediate from \eqref{trivpot} that for pure-gauge connections
$A=f^*(g^{-1}dg)$ one gets $a=u^{-1}du$. Note that conventionally $a$ is introduced via 'local gauges' and is called 'connection in a local gauge' rather than gauge potential \cite{MM,DFN} (of course, on trivial bundles local gauges happen to be global). It is in this sense that $a$ is a 'pure-gauge connection' in \cite{AK1,AK2}. We use the above construction instead because it conveniently generalizes to coset bundles while global gauges do not.

The 'untwisting' can also be applied to curvature forms. For the untwisted gauge potential 
$a$ of a connection $A$ define $F(a)$ by
\bee\label{e0.39}
\pi_1^*F(a)=\Ad_*(g)F(A).
\eee
Then a simple computation shows that
\bee\label{e0.40}
F(a)=da+a\wedge a.
\eee
Connections (potentials) with $F(A)=0$ $(F(a)=0)$ are
called {\it flat}. Every pure-gauge connection is flat as can be
seen directly from the expression $a=u^{-1}du$ for the potential.
The converse is true if $\pi_1(M)=0$, otherwise there is a topological
obstruction to constructing a {\it developing map} $u$ called
the holonomy \cite{AK1,KN}.

Now let us replace the one-point bundle $G\ovs{\i}\hra
G\ovs{\pi}\lra\pt$ by a quotient bundle $H\ovs{\i}\lra G\ovs{\pi}\lra
G/H=:X$. Most of the above generalizes to pullbacks of these bundles
under maps $M\ovs{\pfi}\lra X$ that generalize trivial bundles
$M\times G$ and are next to them in complexity. Formally:
\begin{definition}[Coset bundles]\label{D:coset}
A principal bundle is called a coset bundle if it is isomorphic to a pullback of a quotient bundle
$H\hra G\to G/H=X$, where $H<G$ is a closed subgroup of a Lie group $G$. 
\end{definition}
Given $M\ovs{\pfi}\lra X$ we have
$$
\pfi^* G:=\{(m,g)\in M\times G|\ \pfi(m)=gH\}\subset M\times G
$$ 
and any connection form $A$ on the
trivial bundle $M\times G$ restricted to $\pfi^* G$ has the {\it
isotropy decomposition}:
\bee\label{e0.45}
A=\pr_\h A+\pr_{\h^\bot}A=:A^\vert+A^\bot
\eee
Since $\Ad_*(h)$ commutes with $\pr_\h$ it
follows immediately from \eqref{conn} that $A^\vert$ is a connection form on
$\pfi^*G$. Therefore the reference connection $A_0=g^{-1}dg$ on
$M\times G$ gives us a natural choice of a reference connection on
$\pfi^*G$:
\bee\label{e0.46}
B_0:=A_0^\vert=(g^{-1}dg)^\vert=\pr_\h(g^{-1}dg)
\eee
Since $\pfi^*G\subset M\times G$ the correction
factor $\Ad_*(g)$ is still available and we can copy \refD{trivpot} to
set
\begin{definition}[Untwisted gauge potentials on coset bundles]\label{D:cospot}
The untwisted gauge potential of a connection $B$ on $\pfi^*G$ is
the form $b\in \Gamma (\Lambda^1M\otimes\g)$ satisfying
\bee\label{e0.47}
\Ad_*(g)(B-(g^{-1}dg)^\vert)=\pi_1^*b.
\eee
\end{definition}
Note that $B$ can also be represented by a usual gauge potential $\beta$ from
\refD{pot} which is an $\Ad_*(\pfi^*G)$--valued $1$-form. This
bundle has fiber $\h$ while $b$ is a $\g$-- but not $\h$--valued form.
Thus to 'untwist' the potentials we have to pay by enlarging the target algebra. 

To establish a relation between twisted and untwisted potentials we need the following notion.
\begin{definition}[Isotropy subbundle of algebras]
Given $M\ovs{\pfi}\lra X=G/H$ define the isotropy subbundle of
$M\times\g$:
\begin{align*} \h_\pfi &:=\{(m,\xi)\in M\times \g
|\ \xi\in\h_{\pfi (m)}\}\\
&=\{(m,\xi)\in M\times \g |\ \xi\in\Ad_*(g)\h,\ \pfi (m)=gH\}.
\end{align*}
\end{definition}
The isotropy subbundle is clearly a vector bundle with fiber $\h$. Each fiber has a Riemannian metric by restriction from $M\times\g$. The following Lemma explains why the 'untwisting' is possible.
\begin{lemma}\label{L:2.3}
There is an isometric isomorphism of vector bundles
$$
\begin{aligned}
\Ad_*(\pfi^*G)&\ovs{\sim}\lra\h_\pfi\\
(m,[g,\xi])&\longmapsto (m,\Ad_*(g)\xi)
\end{aligned}
$$ 
that induces isomorphisms on differential forms 
$$
\Gamma(\Lambda^kM\otimes\Ad_*(\pfi^*G))\simeq\Gamma
(\Lambda^kM\otimes\h_\pfi)\subset\Gamma (\Lambda^kM\otimes\g).
$$
The gauge potential $\beta$ of a connection $B$ is transformed by this isomorphism into its untwisted gauge potential $b$.
\end{lemma}
\begin{proof}
It is easy to see that the above map as well as the map
$$
\begin{aligned} 
\h_\pfi&\lra\Ad_*(\pfi^*G)\\
(m,\eta)&\longmapsto (m,[g,\Ad_*(g^{-1})\eta])
\end{aligned}
$$ 
are both well-defined and inverses of each other. Therefore they are both isomorphisms and they are isometric
because $\Ad_*(g)$ is an isometry. Now let $S\in
T_mM,\pfi (m)=gH$. Then by Definitions \ref{D:pot},\ref{D:cospot}
\begin{multline*}
\beta(S)=(m,[g,(B-(g^{-1}dg)^\vert)(\overline{S})])\\
\longmapsto\Ad_*(g)(B-(g^{-1}dg)^\vert)(\overline{S})=\pi_1^*b(\overline{S})=b(\pi_{1*}\overline{S})=b(S)
\end{multline*}
and $\beta\mapsto b$ as claimed.
\end{proof}

\noindent {\bf Notational convention}: Since we have little use for the
(twisted) gauge potentials of \refD{pot} from now on expressions 'gauge
potential' or 'potential' will refer to the untwisted ones of Definitions \ref{D:trivpot},\ref{D:cospot}
unless otherwise stated. Since the isomorphism of \refL{2.3} is
isometric results stated in the literature for twisted potentials
(such as the Uhlenbeck compactness theorem \cite{Ul1,We} that we use in
\refS{3.2}) are trivially rephrased in terms of our untwisted ones. We utilize such rephrasings 
without special notice.

The isotropy decomposition of connection forms has a parallel for the gauge potentials.
\begin{definition}[Isotropy decomposition of gauge potentials]\label{D:isotrop}
Let $A$ be a connection on $M\times G$ with potential $a$ of \refD{cospot}
then $a^\vert$, $a^\bot$ are defined by
\bee\label{e0.48}
\begin{aligned}
\pi_1^*a^\vert&:=\Ad_*(g)(A^\vert-(g^{-1}dg)^\vert),\\
\pi_1^*a^\bot &:=\Ad_*(g)(A^\bot -(g^{-1}dg)^\bot).
\end{aligned}
\eee
\end{definition}
Obviously, $\pi_1^*(a^\vert+a^\bot)=\Ad_*(g)(A-g^{-1}dg)$ so
$a=a^\vert+a^\bot$. 

We now want to compute the isotropic and coisotropic
components explicitly. 
\begin{lemma}\label{L:2.4}
Components of $a$ are given by
\bee\label{e0.49}
a^\vert=\pr_{\h_\pfi}(a),\quad a^\bot=\pr_{\h_\pfi^\bot}(a).
\eee 
If $A_u$ is a pure-gauge connection with the potential $a_u=u^{-1}du$ then
\bee\label{mapcon}
 a_u^\bot=\Ad_*(u^{-1})(u\pfi)^*\omega^\bot -\pfi^*\omega^\bot.
\eee
\end{lemma}
\begin{proof}
Let $\pfi (m)=gH$ then $\pr_{\h_{\pfi
(m)}}=\pr_{\Ad_*(g)\h}=\Ad_*(g)\pr_\h\Ad_*(g^{-1})$ and since
$(m,g)\in\pfi^*G$ always satisfies $\pfi (m)=gH$ we have
\be
\begin{aligned}
\pi_1^*(\pr_{\h_\pfi}(a))&=\pr_{\Ad_*(g)\h}(\pi_1^*a)=\Ad_*(g)\pr_\h\Ad_*(g^{-1})(\Ad_*(g)(A-g^{-1}dg))\\
&=\Ad_*(g)\pr_\h (A-g^{-1}dg)=\Ad_*(g)(A^\vert-(g^{-1}dg)^\vert)=\pi_1^*a^\vert
\end{aligned}
\ee
Since $\pi_1^*$ is mono we have the first formula. The second formula follows from $a^\bot
=a-a^\vert$. For the third one recall that $A_u=f_2^*(g^{-1}dg)$, where
$f_2$ is the second component of the gauge transformation
$$
\begin{aligned} 
M\times G&\ovs{f} \lra M\times G\\
(m,g)&\longmapsto (m,u(m)g)
\end{aligned}
$$
It is easy to see by inspection that the following diagram commutes:
$$
\begin{diagram}
\pfi^* G & \rTo{f_2}  &G       \\
 \dTo{\pi_1}   &         &\dTo{\pi} \\
M           & \rTo{u\pfi}     & X
\end{diagram}
$$
Therefore, 
\bee\label{e0.50}
\begin{aligned}
\pi_1^*(\Ad_*(u^{-1})&(u\pfi)^*\omega^\bot)=\Ad_*((u\circ\pi_1)^{-1})\pi_1^*(u\pfi)^*\omega^\bot &&\\
&=\Ad_*((u\circ\pi_1)^{-1})f_2^*\pi^*\omega^\bot &&\\
&=\Ad_*((u\circ\pi_1)^{-1})f_2^*\Ad_*(g)(g^{-1}dg)^\bot && \text{by \eqref{coisot}}\\
&=\Ad_*((u\circ\pi_1)^{-1})\Ad_*((u\circ\pi_1)g)(f_2^*(g^{-1}dg)^\bot) && \text{since $f_2={(u\circ\pi_1)g}$}\\
&=\Ad_*(g)A_u^\bot &&
\end{aligned}
\eee
When $u$ is the constant $1$ map this equality turns into
\bee\label{e0.51}
\pi_1^*(\pfi^*\omega^\bot)=\Ad_*(g)(g^{-1}dg)^\bot
\eee
Subtracting \eqref{e0.51} from \eqref{e0.50} and using the definition \eqref{e0.48} we get
the desired formula.
\end{proof}

\begin{example}[Isotropy decomposition on $\CP^1$]\label{E:CP1isotrop}
Recall from \refE{CP1coisot} that on $\CP^1=SU_2/U_1$ we can identify $\su_2$ with the space $\Im\H$ of
purely imaginary quaternions and $\uu_1$ with $i\R\subset\Im\H$. Therefore
\be
\begin{aligned}
\pr_\h (\xi)&=(\xi ,i)i=\Re (\xi
\overline{i})i=\frac{\xi\overline{i}+\overline{\xi}i}{2}i=-\frac{\xi
i+i\xi}{2}i=\frac{1}{2}(\xi -i\xi i)\\
\pr_{\h^\bot}(\xi)&=\xi -\frac{1}{2}(\xi -i\xi i)=\frac{1}{2}(\xi
+i\xi i)=\frac{1}{2}i(-i\xi +\xi i)=\frac{1}{2}i[\xi ,i],
\end{aligned}
\ee
where $\h^\bot =\uu_1^\bot$ is the linear span of\ $j$, $k$. Also recall that 
we can identify $\CP^1$ itself with the unit sphere $S^2$ in $\Im\H$. Under this
identification a map $M\ovs{\pfi}\lra\CP^1$ turns into a map $M\ovs{\phi}\lra S^2$ with
\be
\phi(m):=qiq^{-1}=qi\overline{q},\quad \text{if $\pfi (m)=qU_1$.}
\ee
With this notation:
\begin{multline*}\pr_{\h_\pfi}
(\xi)=\Ad_*(q)\pr_\h(\Ad_*(q^{-1})\xi)=\Ad_*(q)(\Ad_*(q^{-1})\xi,i)i\\
=(\xi ,\Ad_*(q)i)\Ad_*(q)i=(\xi,\phi)\phi
\end{multline*}
since $\Ad_*(q)$ is an isometry and $(\xi ,\eta)\in\R$ and
therefore commutes with all quaternions. Analogously
$$
\pr_{\h_\phi^\bot}(\xi)=\frac{1}{2}\phi[\xi,\phi].
$$ 
Thus by \eqref{e0.49} we get in terms of $\phi$:
\bee\label{e0.53}
a^\vert=(a,\phi)\phi,\quad a^\bot =\frac{1}{2}\phi
[a,\phi]. \eee
These are the expressions used in \cite{AK2}.
\end{example}

Gauge transformations on coset bundles can also be
'untwisted' into $G$--valued maps. Recall from the end of \refS{2.1} that for general principal bundles gauge transformations can be described as sections of the bundle $\Ad(P)=P\times_{\Ad} H$. Just as we described $\Ad_*(P)$ as isomorphic to a subbundle of $M\times\g$ in \refL{2.4} we can describe $\Ad(P)$ as isomorphic to a subbundle of $M\times G$.
\begin{definition}[Isotropy subbundle of groups]\label{isot2}
Given $M\ovs{\pfi}\lra X=G/H$ the isotropy subbundle of $M\times G$
relative to a closed subgroup $H<G$ is
 \be
 \begin{aligned}
 H_\pfi &:=\{(m,\gamma)\in M\times G|\gamma\in H_{\pfi (m)}\}\\
 &=\{ (m,\gamma )\in M\times G|\gamma\in\Ad_*(g)H,\quad \pfi
 (m)=gH\}.
 \end{aligned}
 \ee
\end{definition}
This is a fiber bundle with fiber $H$. Sections of this bundle are
maps $M\ovs{w}\lra G$ that satisfy $w (m)\in H_{\pfi (m)}$
for all $m\in M$. Recall that we denoted 
$$
\Stab_\pfi :=\{M\ovs{w}\lra G|w \pfi =\pfi\}.
$$
By analogy to \refL{2.4} one obtains
\begin{lemma}\label{L:2.5}
There is an isomorphism 
$$
\bal 
\Ad(\pfi^*G)&\ovs{\sim}\lra H_\pfi\\
(m,[g,\lambda])&\longmapsto (m,\Ad (g)\lambda)
\eal
$$ 
that induces isomorphism of the gauge group $\Gamma (\Ad
(\pfi^*G))\ovs{\sim}\lra\Gamma (H_\pfi)$, i.e
\bee\label{e0.55}
\Gamma (H_\pfi)=\Stab_\pfi\simeq\Gamma (\Ad (\pfi^*G))
\eee
\end{lemma}
\begin{proof}
The isomorphism is proved word to word as in \refL{2.4} with $\Ad$ in
place of $\Ad_*$. For the second claim note that $w
(m)=ghg^{-1}$ for some $h\in H$ and $w (m)\pfi (m)=w
(m)gH=ghg^{-1}gH=gH=\pfi (m)$, the converse follows similarly.
\end{proof}
Thus instead of being represented by sections of a twisted bundle with fiber $H$ gauge
transformations are represented by $G$--valued maps. As in the case of gauge potentials the 
untwisting comes at a price of extending the target space. Also note that \refL{2.5} gives a gauge description of the stabilizer of a reference map. This description will play a crucial role in applications to minimization in the next chapter.

Recall that the main function of gauge transformations is their
action on connection forms -- the gauge action. As connections are
now presented by (untwisted) gauge potentials $b\in\Gamma (\Lambda^1M\otimes\g)$
(\refD{cospot}) and gauge transformations by maps $M\ovs{w}\lra G$ we
would like to have an explicit expression for the action of $w$
on $b$. Similarly, curvature of a connection $B$ on $\pfi^*G$ is a
horizontal equivariant $2$-form on $\pfi^*G$ and after applying the
correction factor $\Ad_*(g)$ we can make it invariant and descend it to
$M$. Again we would like an explicit expression for the result in terms of
the potential $b$. This prompts the following definition.
\begin{definition}[{\bf Gauge action and curvature for gauge potentials}]\label{D:bwF}
Let $f_w$ be the gauge transformation corresponding to the map
$M\ovs{w}\lra G$, $w\in\Gamma (H_\pfi)$ and $b$ be the potential of a connection $B$. Then $b^w$
denotes the gauge potential of the transformed connection $f_w^*B$. The curvature potential $F(b)$ is defined by
\bee\label{e0.56}
\pi_1^*F(b)=\Ad_*(g)F(B)=\Ad_*(g)(dB+B\wedge B).
\eee
\end{definition}
Obviously, $F(b)\in\Gamma(\Lambda^2 M\otimes\g)$, moreover $F(b)\in\Gamma
(\Lambda^2M\otimes\h_\pfi)$ since $dB+B\wedge B$ is $\h$--valued.
Note that usually $F(\beta)$ is defined for a twisted potential $\beta$ from \refD{pot}
and is an $\Ad_*(P)$--valued $2$-form descended from $F(B)$. This $F(\beta)$ corresponds to our $F(b)$ 
under the induced isomorphism of \refL{2.4}.

Before we derive explicit expressions for $b^w$, $F(b)$ let us make several preparations. 
First, it is convenient to extend the notation $\vert$, $\bot$ to all $\g$--valued forms on
$\pfi^*G$ and $M$:
\bee\label{e0.57}
\bal 
R^\vert &:=\pr_\h (R)  && \text{for $R\in\Gamma(\Lambda^\bullet(\pfi^*G)\otimes\g)$}\\
R^\bot &:=\pr_{\h^\bot}(R) &&\\
r^\vert &:=\pr_{\h_\pfi}(r)  && \text{for $r\in\Gamma(\Lambda^\bullet M\otimes\g)$.}\\
r^{\bot} &:=\pr_{\h_\pfi^\bot}(r) && 
\eal
\eee
By \eqref{e0.45}, \eqref{e0.49} this agrees with the previous notation for $A$ and $a$. 

Second, note that every connection form $B$ on $\pfi^*G$ is the isotropic part of a
(non-unique) connection $A$ on $M\times G$. It is easy
to see using the gauge potentials $b$. By \refD{cospot} one has that $b$ is an $\h_\pfi$--valued $1$--form. 
But $\h_\pfi\subset\g$ and it can also be treated as a $\g$--valued one. By \refD{trivpot} any 
$\g$--valued $1$--form represents a connection on $M\times G$. Let $A$ denote this connection for 
$b$ treated as a $\g$--valued form then $B=A^\vert$ as required. More explicitly we have
\be
\bal
\pi_1^*b &=\Ad_*(g)(B-(g^{-1}dg)^\vert) && \text{on $\pfi^*G\subset M\times G$}\\ 
\pi_1^*a &=\Ad_*(g)(A-g^{-1}dg)  && \text{on  $M\times G$} 
\eal
\ee
and therefore 
\bee\label{minext}
A=B+(g^{-1}dg)^\bot
\eee
on $\pfi^*G$. It can be uniquely extended to the entire $M\times G$ by equivariance.
This is the {\it minimal extension} of $B$. More generaly we could take any $\h_\pfi^\bot$--valued $1$-form $\delta$ on $M$, set
$a=b+\delta$ and take $A$ on $M\times G$ that corresponds to $a$.
 
Third, the gauge transformation $f_w$ from \refD{bwF} can be found explicitly. By \refL{2.5} $w$
corresponds to a section $\sigma$ of $\Ad(\pfi^*G)$ given by
$$
\sigma(m):=(m,[g,\Ad_*(g^{-1})w (m)])
$$
In its turn by the isomorphism between $\Gamma (\Ad (\pfi^*G))$ and $\Aut(\pfi^*G)$ (see e.g. \cite{MM}) 
this section corresponds to
$$
f_w(m,g)=(m,g\Ad_*(g^{-1})w(m))=(m,w(m)g).
$$ 
Although we obtained it as a gauge transformation of $\pfi^*G$ only, 
it obviously extends to a gauge transformation of $M\times G$ that we denote
by the same symbol. If $A$ is a connection on $M\times G$ with the gauge potential $a$ then 
the gauge potential $a^w$ of $f_w^*A$ is easily found to be \cite{DFN,MM}:
\bee\label{e0.59}
a^w =\Ad_*(w^{-1})a+w^{-1}dw .
\eee

Now we are ready to derive the promised formulas. The coisotropy form $\omega^\bot$ makes an important appearence here. The idea of the proof is to extend a connection on $\pfi^*G$ to a connection on $M\times G$, use the well-known formulas for potentials on a trivial bundle and then 'project' to the potentials on a coset bundle. 
\begin{theorem}\label{T:2.1}
Let $B$ be a connection on $\pfi^*G$, $b$ be its (untwisted) gauge potential and
$w$ be a section of $H_\pfi\subset M\times G$. Then
\bee\label{e0.60}
\bal
&{\rm (i)}\ \ \ \ \ \ b^w=\Ad_*(w^{-1})b+w^{-1}dw-(\Ad_*(w^{-1})-I)\pfi^*\omega^\bot\\
&{\rm (ii)}\ F(b^w)=\Ad_*(w^{-1})F(b)\\
&{\rm (iii)}\ \ F(b)=db+b\wedge b-[b,\pfi^*\omega^\bot]-(\pfi^*\omega^\bot\wedge\pfi^*\omega^\bot)^\vert.
\eal
\eee
\end{theorem}
\begin{proof}
{\rm (i)} Let $A$ be the minimal extension of $B$ to $M\times G$ then we have for the potentials $a,b$
then
\begin{align*}
\pi_1^*a^w &=\Ad_*(g)(f_w^*A-g^{-1}dg) && \text{by \refD{trivpot}}\\
\intertext{and}
\pi_1^*b^w &=\pi_1^*(a^\vert)^w=\Ad_*(g)(f_w^*A^\vert-(g^{-1}gd)^\vert) && \text{by \refD{cospot}}\\
&=\Ad_*(g)((f_w^*A)^\vert-(g^{-1}dg)^\vert) && \text{since $\pr_\h$ commutes with $f_w^*$}\\
&=\pi_1^*(a^w)^\vert && \text{by \refD{isotrop}.}
\end{align*}
Therefore $b^w=(a^w)^\vert$. Since $\pr_{\h_\pfi}$ commutes with $\Ad_*(w^{-1})$ 
for $w\in\Gamma (H_\pfi)$ we have further
$$
b^w=(a^w)^\vert=(\Ad_*(w^{-1})a+w^{-1}dw)^\vert=\Ad_*(w^{-1})a^\vert+(w^{-1}dw)^\vert.
$$
But by definition of the minimal extension $b=a^\vert=a$ and 
\bee\label{e0.63}
b^w=\Ad_*(w^{-1})b+(w^{-1}dw)^\vert.
\eee
When $w\pfi =\pfi$ the equality \eqref{mapcon} becomes
\bee\label{e0.62}
(w^{-1}dw)^\bot =\Ad_*(w^{-1})\pfi^*\omega^\bot-\pfi^*\omega^\bot
\eee
and therefore
$$
(w^{-1}dw)^\vert=w^{-1}dw-(w^{-1}dw)^\bot=w^{-1}dw-(\Ad_*(w^{-1})-I)\pfi^*\omega^\bot
$$
Substituting this into \eqref{e0.63} we get the required formula.

{\rm (ii)} For any horizontal equivariant form $R$ on $\pfi^*G$ one has 
$\Ad_*(g)R=\pi_1^*r$ with a unique form $r$ on $M$. We claim that then
\bee\label{e0.61}
\Ad_*(g)(f_w^*R)=\pi_1^*(\Ad_*(w^{-1})r).
\eee
Indeed,
$$
f_w^*(\Ad_*(g)R)=\Ad_*((w\circ\pi_1)g)f_w^*R=\Ad_*(w\circ\pi_1)(\Ad_*(g)f_w^*R)
$$
and
\begin{multline*}
\Ad_*(g)(f_w^*R)=\Ad_*((w\circ\pi_1)^{-1})f_w^*(\pi_1^*r)=\Ad_*((w\circ\pi_1)^{-1})(\pi_1\circ f_w)^*r\\
=\Ad_*((w\circ\pi_1)^{-1})\pi_1^*r=\pi_1^*(\Ad_*(w^{-1})r).
\end{multline*}
Applying \eqref{e0.61} to $R=F(B)=dB+B\wedge B$ one obtains
\be
\Ad_*(g)F(f_w^*B)=\Ad_*(g)(f_w^*F(B))=\pi_1^*(\Ad_*(w^{-1})F(b))=\pi_1^*F(b^w),
\ee
which implies {\rm (ii)} since $\pi_1^*$ is mono. 

{\rm (iii)} Again let $A$ be the minimal extension of $B$. Even though for potentials $a=b$ 
we now have two different curvatures: one induced from the curvature of $A$ by \eqref{e0.39}, 
the other induced from the curvature of $B$ by \eqref{e0.56} and they are not equal. 
To avoid confusion we denote the former $\widehat{F}(a)$ for the duration of this proof only. 
Thus
\begin{align*}
\pi_1^*F(b)=\pi_1^*F(a) &=\Ad_*(g)(dB+B\wedge B)\\
\pi_1^*\widehat{F}(a) &=\Ad_*(g)(dA+A\wedge A)
\end{align*}
Since $A$ is the minimal extension by \eqref{minext}
\be
\bal
dA &=dB+d(g^{-1}dg)^\bot\\
A\wedge A &=B\wedge B+[B,(g^{-1}dg)^{\bot}]+(g^{-1}dg)^\bot\wedge (g^{-1}dg)^\bot
\eal
\ee
Since $g^{-1}dg$ is flat it satisfies
$$
d(g^{-1}dg)=-(g^{-1}dg)\wedge(g^{-1}dg)
$$
Decomposing $g^{-1}dg=(g^{-1}dg)^\vert+(g^{-1}dg)^\bot$ and 
taking into account \eqref{hh} we get
\be
d(g^{-1}dg)^\bot=-(g^{-1}dg\wedge g^{-1}dg)^\bot
=-[(g^{-1}dg)^\vert,(g^{-1}dg)^\bot]-((g^{-1}dg)^\bot\wedge (g^{-1}dg)^\bot)^\bot.
\ee
Putting it together:
\begin{multline*}
dA+A\wedge A=dB+d(g^{-1}dg)^\bot+B\wedge B+[B,(g^{-1}dg)^{\bot}]+(g^{-1}dg)^\bot\wedge (g^{-1}dg)^\bot\\
=dB+B\wedge B+[B,(g^{-1}dg)^\bot]+(g^{-1}dg)^\bot\wedge (g^{-1}dg)^{\bot}
-[(g^{-1}dg)^\vert,(g^{-1}dg)^\bot]-((g^{-1}dg)^\bot\wedge
(g^{-1}dg)^\bot)^\bot\\
=dB+B\wedge B+[(B-(g^{-1}dg)^\vert),(g^{-1}dg)^\bot]+((g^{-1}dg)^\bot\wedge (g^{-1}dg)^\bot)^\vert.
\end{multline*}
Now apply $\Ad_*(g)$ to both sides and distribute it under $\wedge$ and $[\cdot ,\cdot]$ using
\eqref{e0.37}(vii) and under the $\vert,\perp$ signs using that $\Ad_*(g)\pr_\h =\pr_{\h_\pfi}\Ad_*(g)$.
Since 
$$
\Ad_*(g)(B-(g^{-1}dg)^\vert)=\pi_1^*b
$$
by \eqref{e0.48} and
$$
\Ad_*(g)(g^{-1}dg)^\bot =\pi_1^*(\pfi^*\omega^\bot)
$$ 
by \eqref{e0.51} the equality turns into
\be
\pi_1^*\widehat{F}(a)=\pi_1^*F(b)+\pi_1^*[b,\pfi^*\omega^\bot]+\pi_1^*(\pfi^*\omega^\bot\wedge\pfi^*\omega^\bot)^\vert
\ee
Removing $\pi_1^*$ and recalling that $\widehat{F}(a)=da+a\wedge a=db+b\wedge b$ by \eqref{e0.40} we get the required formula.
\end{proof}

Note that the formulas from \refT{2.1} look like their analogs for trivial bundles with 'correction terms' depending on the pullback of the coisotropy form $\pfi^*\omega^\bot$. If $\pfi$ is a constant map and the bundle $\pfi^*G$ is trivial then $\pfi^*\omega^\bot=0$ and we recover the formulas for trivial bundles.

An interesting consequence of \refT{2.1} is
\begin{corollary}\label{C:Dafi}
\bee\label{Dafi} 
(a^w)^\bot +\pfi^*\omega^\bot
=\Ad_*(w^{-1})(a^\bot +\pfi^*\omega^\bot) 
\eee
\end{corollary}
\begin{proof}
By direct computation from \eqref{e0.63}
\be
\bal
(a^w)^\bot 
&=a^w-(a^w)^\vert=\Ad_*(w^{-1})a +w^{-1}dw-\Ad_*(w^{-1})a^\vert -(w^{-1}dw)^\vert\\
&=\Ad_*(w^{-1})a^\bot +(w^{-1}dw)^\bot\\ 
&=\Ad_*(w^{-1})a^\bot+\Ad_*(w^{-1})\pfi^*\omega^\bot-\pfi^*\omega^\bot\\ 
&=\Ad_*(w^{-1})(a^\bot+\pfi^*\omega^\bot)-\pfi^*\omega^\bot.
\eal
\ee
\end{proof}
Comparing \eqref{Dafi} to \eqref{e0.60}(ii) we see that the quantity $a^\bot
+\pfi^*\omega^\bot$ 'transforms as curvature'. This reflects the following situation for connections.
In a principal bundle the only local gauge-equivariant functional of a connection $A$ is its curvature $F(A)$ 
but only if we consider the gauge action induced by that {\it same bundle}. If on the other hand, 
we consider the gauge action induced by a {\it subbundle}
the curvature is joined by the coisotropic part $A^\bot$ with respect to this subbundle. It follows from \eqref{e0.48} and \eqref{e0.51} that
\bee\label{e0.65}
\Ad_*(g)A^\bot =\pi_1^*(a^\bot
+\pfi^*\omega^\bot).
\eee
Such 'partial gauge equivalence' arises in {\it coset models} of quantum physics \cite{BMSS}. 
The gauge principle implies in this situation that physical Lagrangians should be functions of $a^\bot +\pfi^*\omega^\bot$,
$F(a^\vert)$. The Faddeev-Skyrme functional (reformulated for potentials) will depend on the first quantity only (see \refD{FadSkyrp}).

Taking $\pfi=\id_X$ and $b=0$ in \eqref{e0.60}(iii) corresponds to computing the curvature potential of the
reference connection $(g^{-1}dg)^\vert$ on the quotient bundle $H\hra G\to G/H=X$.
\begin{corollary}\label{C:refcurv}
The curvature potential of the reference connection $(g^{-1}dg)^\vert$
on $G\ovs{\pi}\lra X$ is
\bee\label{e0.66}
F(0)=-(\omega^\bot\wedge\omega^\bot)^\vert.
\eee
\end{corollary}
This is another indication of a role the coisotropy form plays
in geometry of homogeneous spaces. It becomes especially nice for 
symmetric spaces \cite{Ar,Br2,Hl}. There in addition to relations \eqref{hh} 
one also has 
$$
[\h^\bot ,\h^\bot]\subset\h
$$ 
and \eqref{e0.66} becomes 
$$
F(0)=-\omega^\bot\wedge\omega^\bot.
$$
Finally projecting \eqref{e0.60} to $\h_\pfi$, $\h_\pfi^\bot$ and taking into account \eqref{e0.42} we get
\begin{corollary} 
For any gauge potential on a coset bundle $\pfi^*G$ one has
\bee\label{e0.67} 
\bal 
F(b) &=(db)^\vert+b\wedge b-(\pfi^*\omega^\bot\wedge\pfi^*\omega^\bot)^\vert\\
(db)^\bot &=[\pfi^*\omega^\bot ,b]
\eal 
\eee
\end{corollary}

\chapter{Faddeev-Skyrme models and minimization}\label{S:3}

After topological and gauge-theoretic preliminaries in the first two
chapters we are ready to introduce our version of the Faddeev-Skyrme
functional for general homogeneous targets and prove existence of
minimizers of different topological types. Since regularity theory
for Skyrmions is lacking the main difficulty is to balance
regularity requirements that allow a notion of 'topological type' to
be introduced against compactness properties that lead to
consideration of more singular maps. We end up with admissible maps
of \refD{admis} that have meaningful '$2$--homotopy sectors' (generalizing
$2$-homotopy type) and strongly admissible maps of \refD{sadmis} that have meaningful 'homotopy sectors'.
Both spaces are 'large enough' to ensure compactness. We analyze the {\it primary minimization} (within each $2$--homotopy sector) for arbitrary homogeneous spaces in \refS{3.2} and the {\it secondary minimization} (within each homotopy sector) for symmetric spaces in \refS{3.3}.

It turns out to be technically convenient (and perhaps even
more 'natural') to write the Faddeev-Skyrme functional for (pure-gauge)
potentials rather than maps. In particular our main tool in the
the proof of existence of minimizers is based on the K.Uhlenbeck's
gauge-fixing procedure for connections \cite{Ul1,We}. The reason
the Faddeev-Skyrme functional is so interesting analytically is that additional
regularity comes not from control over higher derivatives of maps but
$2$-determinants of the first derivatives. This type of conditions
has been intensively studied recently (see \cite{GMS4} and references
therein) and can be utilized using the fact that wedge products of
Sobolev forms are 'better' than predicted by the Sobolev multiplication
theorems \cite{IV} due to 'cancellation of singularities' in
determinants. In our case this works especially well for symmetric
spaces (\refS{3.3}), where additional symmetry leads to more
cancellations.

\section{Faddeev-Skyrme energy}\label{S:3.1}

In this section we introduce our version of the Faddeev-Skyrme functional,
and compare it to other 'Skyrme functionals' found in the literature. Then we
introduce admissible maps and their $2$--homotopy sectors and give a precise 
formulation of the primary minimization problem.

Let $X=G/H$ be a homogeneous space with $G$ a compact Lie group and
$H<G$ a closed subgroup. $G$ is equipped with a bi-invariant metric
that induces an inner product on $\g$ and a Riemann quotient metric on
$X$ (see \refS{2.2}). Recall that the coisotropy form $\omega^\bot$ on
$X$ is defined by:
$$
\pi^*\omega^\bot =\Ad_*(g)(g^{-1}dg)^\bot,
$$
where $G\ovs{\pi}\lra X$ (\refD{coisot}). If $E\lra M$ is a
Riemannian vector bundle over $M$ and $\omega\in\Gamma(\Lambda^kM\otimes E)$ 
we set as usual
\bee\label{e0.82}
|\omega_m|:=\mbox{sup}\{|\omega_m(S_1,\dots ,S_k)|_E|\ S_i\in
T_mM,\ |S_i|_{TM}=1\},\quad m\in M.
\eee
for the norm of the form at the point $m$.
Given a map $M\ovs{\psi}\lra
X$ we have $d\psi\in\Gamma (\Lambda^1M\otimes\psi^*TX)$ and
$\psi^*\omega^\bot\in\Gamma (\Lambda^1M\otimes\g)$ so $|d\psi|$,
$|\psi^*\omega^\bot|$ are defined at each point. By \refL{2.1}(iii) 
$$
|\psi^*\omega^\bot(s)|=|\omega^\bot(\psi_*S)|=|\psi_*S|=|d\psi(S)|
$$ 
for any $S\in TM$ and hence $|d\psi|=|\psi^*\omega^\bot|$.

\begin{definition}[Faddeev-Skyrme functional]\label{D:FadSkyr}
The Faddeev-Skyrme energy of a map $M\ovs{\psi}\lra X$ is
\bee\label{FadSkyr}
\bal
E(\psi) &:=\int\limits_M\frac{1}{2}|d\psi|^2+\frac{1}{4}|\psi^*\omega^\bot\wedge\psi^*\omega^\bot|^2\,dm\\
        &=\int\limits_M\frac{1}{2}|\psi^*\omega^\bot|^2+\frac{1}{4}|\psi^*\omega^\bot\wedge\psi^*\omega^\bot|^2\,dm.
\eal
\eee
\end{definition}
The second expression does not require using the bundle $\psi^*TX$ that is only
defined for smooth $\psi$. By \eqref{e0.66}
$$
-(\psi^*\omega^\bot\wedge\psi^*\omega^\bot)^\vert=-\psi^*(\omega^\bot\wedge\omega^\bot)^\vert
$$ 
is the pullback of the curvature potential of the reference
connection on $G\lra X$. Perhaps in view of this 
it would have been more natural to replace $\psi^*\omega^\bot\wedge\psi^*\omega^\bot$ in 
\eqref{FadSkyr} by its parallel component and in fact all our arguments 
for primary minimization still go through if this is done. As for the secondary 
minimization we only consider symmetric spaces so 
$(\omega^\bot\wedge\omega^\bot)^\vert=\omega^\bot\wedge\omega^\bot$ anyway.

Now let us compare our definition to some similar ones.
\begin{example}
{\rm\bf (Lie groups)} Here $H=\{1\}$ and $X=G$ is a compact Lie
group. The Skyrme functional is usually written as \cite{AK1}:
\bee\label{e0.85}
E(u)=\int\limits_M\frac{1}{2}|du|^2+\frac{1}{4}|u^{-1}du\wedge
u^{-1}du|^2\,dm.
\eee
The coisotropy form becomes
$$
\omega^\bot=\Ad_*(g)g^{-1}dg=dgg^{-1}.
$$ 
Since the metric is bi-invariant $|dgg^{-1}|=|g^{-1}dg|$ and moreover since $\Ad_*(g)$ is an isometry
$$
|dgg^{-1}\wedge dgg^{-1}|=|\Ad_*(g)(g^{-1}dg\wedge g^{-1}dg)|=|g^{-1}dg\wedge g^{-1}dg|.
$$ 
Hence our functional \eqref{FadSkyr} coincides with \eqref{e0.85} as
$u^*(dgg^{-1})=duu^{-1}$.
\bigskip

\noindent {\rm\bf (The $2$--sphere)} The functional of the Faddeev model can
be written as \cite{AK2}:
\bee\label{e0.86}
E(\psi)=\int\limits_M\frac{1}{2}|d\psi|^2+\frac{1}{4}|d\psi\times
d\psi|^2\,dm
\eee
where $M\ovs{\psi}\lra S^2$ and 
$$
(d\psi\times d\psi)(S,T):=d\psi(S)\times d\psi(T)
$$
Here on the right $\xi\times\eta$ is the cross-product of two vectors in $\R^3$
(we assume $S^2\hra\R^3$ as the unit sphere so $d\psi$ is $\R^3$--valued). As in \refE{CP1coisot} identify
$\R^3\simeq\Im\H$, the space of purely imaginary quaternions and use the quaternionic notation. 
Then we have 
$$
\xi\times\eta=[\xi,\eta]=\xi\eta-\eta\xi
$$ 
and as we computed in \refE{CP1coisot} $\omega_{\pi(q)}^\bot(\xi)=\frac{1}{2}\pi(q)\xi$ with
$\pi(q)=qiq^{-1}$. Since vectors in $\Im\H$ are orthogonal if and
only if they anticommute, $\xi\in T_{\pi(q)}S^2\bot\pi(q)$ and $\pi(q)^2=i^2=-1$ we  have:
\begin{multline*}
\omega_{\pi(q)}^\bot\wedge\omega_{\pi(q)}^\bot(\xi,\eta)
=\frac{1}{2}\pi(q)\xi\frac{1}{2}\pi(q)\eta-\frac{1}{2}\pi(q)\eta\pi(q)\xi
=\frac{1}{4}(-\pi(q)^2\xi\eta+\pi(q)^2\eta\xi)\\
=\frac{1}{4}(\xi\eta-\eta\xi)=\frac{1}{4}[\xi ,\eta].
\end{multline*}  
Therefore $\psi^*\omega^\bot\wedge\psi^*\omega^\bot=\frac{1}{4}d\psi\times d\psi$ and 
up to constant multiples \eqref{e0.86} is the same as \eqref{FadSkyr}. 
\bigskip

\noindent {\rm\bf (Riemannian manifolds)} N.Manton \cite{Mn} suggested a definition of a
'Faddeev-Skyrme functional' that works for maps $M\ovs{\psi}\lra N$ between
arbitrary Riemannian manifolds:
\bee\label{Mn}
E_M(\psi):=\int\limits_M\frac{1}{2}|d\psi|^2+\frac{1}{4}|d\psi\wedge
d\psi|^2\,dm,
\eee
where $d\psi\wedge d\psi\in\Gamma(\Lambda^2M\otimes\psi^*(TN)^{\wedge 2})$ is defined by
$$
d\psi\wedge d\psi(S,T)=d\psi(S)\otimes d\psi(T)-d\psi(T)\otimes
d\psi(S).
$$
Since $d\psi\wedge d\psi$ is universal any quadratic antisymmetric expression in components of $d\psi$ factors through it. In particular there is a smooth section $L$ of $(\psi^*(TN)^{\wedge 2})^*$ such that 
$$
\psi^*(\omega^\bot\wedge\omega^\bot)=<L,d\psi\wedge d\psi>.
$$ 
Therefore $E(\psi)\leq CE_M(\psi)$ for some $C>0$. 

However even for Lie groups \eqref{Mn} is strictly stronger than the usual one \eqref{e0.85}.
Indeed, $\psi^*(\omega^\bot\wedge\omega^\bot)$ takes values in $\g$
of dimension say $n$ and the fiber of $(TG)^{\wedge 2}$ has dimension
$\frac{n(n-1)}{2}$ strictly greater than $n$ for $n>3$. Therefore \eqref{Mn} controls all components of
$d\psi\wedge d\psi$ while \eqref{e0.85} only controls some linear
combinations. Nonetheless, for $X=SU_2\simeq S^3$ Manton's
functional coincides with \eqref{e0.85} and for $X=S^2$ it coincides with \eqref{e0.86}.
\bigskip

\noindent {\rm\bf (Symplectic manifolds)} In the original formulation of
the Faddeev model the functional \eqref{e0.86} was written differently:
\bee\label{e0.88}
E_{Sp}(\psi)=\int\limits_M\frac{1}{2}|d\psi|^2+\frac{1}{4}|\psi^*\Omega|^2\,dm,
\eee
where $\Omega$ is the volume form of $S^2$. Since $S^2$ is
$2$-dimensional its volume form is also a symplectic form and \eqref{e0.88}
can be generalized to $M\ovs{\psi}\lra N$ with any symplectic target
manifold $N$ (see \cite{Ar} for definitions). In contrast to \eqref{Mn}
which is stronger than our functional \eqref{FadSkyr} $E_{Sp}$ is in fact
much weaker for $\psi^*\Omega$ only controls one linear combination
of components in $d\psi\wedge d\psi$. In fact, the symplectic form can be
chosen so that $E_{Sp}(\psi)\leq CE(\psi)$. 

It can be shown that the curvature potential $(-\omega^\bot\wedge\omega^\bot)^\vert$ 'contains' all possible invariant symplectic forms on $G/H$ \cite{Ar} (i.e. all those if they exist can be recovered by contracting it with some $\g^*$--valued functions). In other words, \eqref{FadSkyr} with $\psi^*\omega^\bot\wedge\psi^*\omega^\bot$ replaced by $(\psi^*\omega^\bot\wedge\psi^*\omega^\bot)^\vert$ can be obtained as a sum of functionals \eqref{e0.88} with $\Omega$--s forming a basis in the space of invariant symplectic forms. This is essentially how L.Faddeev and A.Niemi introduce their 'Skyrme functional' for complex flag manifolds \cite{FN2}.
\end{example}

So far we wrote the Faddeev-Skyrme functional \eqref{FadSkyr} having in mind only
smooth (or at least $C^1$) maps $\psi$. But it is well-known that spaces of such maps lack necessary weak compactness
properties for solving minimization problems \cite{GMS4} and we need to use Sobolev maps. 

A traditional way of defining Sobolev maps between Riemannian manifolds is the following
(see e.g. \cite{Wh,HL1,HL2}). Let $N$ be a Riemannian manifold and $N\hra\R^n$
an isometric embedding into a Euclidian space of large dimension.
Then the spaces $W^{k,p}(M,\R^n)$ are defined in the usual way and one sets
\bee\label{e0.89}
W^{k,p}(M,N):=\{\psi\in W^{k,p}(M,\R^n)|\psi(m)\in N\ a.e.\}
\eee
Note that the Faddeev-Skyrme energy density in \eqref{FadSkyr}
\bee\label{e0.90}
e(\psi):=\frac{1}{2}|\psi^*\omega^\bot|^2+\frac{1}{4}|\psi^*\omega^\bot\wedge\psi^*\omega^\bot|^2
\eee
is defined almost everywhere for any $\psi\in W^{1,2}(M,X)$. Of
course it does not have to be integrable and we define the 'space' of {\it finite energy maps}:
\bee\label{e0.91}
\bal
W_E^{1,2}(M,X):&=\{\psi\in
W^{1,2}(M,X)|e(\psi)\in L^1(M,\R)\}\\
&=\{\psi\in W^{1,2}(M,X)|E(\psi)<\infty\}.
\eal
\eee
Note that neither $W^{1,2}(M,X)$ nor $W_E^{1,2}(M,X)$ are Banach spaces or
even convex subsets of a Banach space and the word 'space' can only
mean metric or topological space. 

Since $\pi_2(G)=0$ smooth maps are dense in $W^{1,2}(M,G)$ \cite{HL2} but
not in $W^{1,2}(M,X)$ because $\pi_2(X)\neq0$. This means in particular that formulas
derived for smooth maps can not be extended to Sobolev maps into $X$ simply
by smooth approximation. For instance we can extend the formula
\eqref{mapcon} to $u\in W^{1,2}(M,G)$ but we have to keep $\pfi$ smooth (or at least $C^1$).

We now want a notion of $2$-homotopy type for maps in
$W_E^{1,2}(M,X)$. In general for $W^{1,p}(M,N)$ maps such a notion was
introduced by B.White \cite{Wh} but his $n$-homotopy type is defined
only for $[p]>n$ ($[\cdot]$ is the integral part). In our case this only yields $1$-homotopy type
which is not very interesting since $\pi_1(X)=0$ by assumption. In the case of the Faddeev-Skyrme functional additional regularity comes not from integrability of higher derivatives but from integrability of $2$--determinants of the first derivatives. One needs a version of $n$--homotopy type that takes advantage of this regularity information. Our alternative is motivated by \refT{2.1} 
which claims that two continuous maps $M\ovs{\psi ,\pfi}\lra
X$ are $2$--homotopic if and only if there is a continuous 'lift'
$M\ovs{u}\lra G$ with $\psi =u\pfi$. 
\begin{definition}[$2$--homotopy sector]
We say that $\pfi ,\psi\in W_E^{1,2}(M,X)$ are in the same $2$--homotopy sector
if there is a map $u\in W^{1,2}(M,G)$ such that $\psi=u\pfi$ a.e.
\end{definition}
Note that if $N$ is compact $W^{1,2}(M,N)\subset L^\infty(M,N)$.
Therefore the product rule and the Sobolev multiplication theorems
\cite{Pl} imply that $W^{1,2}(M,G)$ is a group that acts on
$W^{1,2}(M,X)$. In particular, $W_E^{1,2}(M,X)$ is divided into
disjoint $2$--homotopy sectors. However, $W^{1,2}(M,G)$ no longer
acts on $W_E^{1,2}(M,X)$. In fact, even if $\pfi$ is smooth and $u\in
W^{1,2}(M,G)$ the product $\psi=u\pfi$ may not have finite Faddeev-Skyrme
energy. Indeed, by \eqref{mapcon}
\bee\label{e0.92}
\bal
\psi^*\omega^\bot&=\Ad_*(u)((u^{-1}du)^\bot+\pfi^*\omega^\bot)\\
\psi^*\omega^\bot\wedge\psi^*\omega^\bot&=\Ad_*(u)((u^{-1}du)^\bot\wedge(u^{-1}du)^\bot+[(u^{-1}du)^\bot
,\pfi^*\omega^\bot]+\pfi^*\omega^\bot\wedge\pfi^*\omega^\bot)
\eal
\eee
and $E(\psi)<\infty$ is equivalent to
$$
(u^{-1}du)^\bot\wedge(u^{-1}du)^\bot\in L^2(\Lambda^2M\otimes\g),
$$
which does not hold for an arbitrary $a\in W^{1,2}(M,G)$.
Despite the appearence this condition still depends on $\pfi$ since $\bot$ stands for $\pr_{\h_\pfi^\bot}$. To avoid cumbersome symbols we often do not reflect dependence on $\pfi$ in the notation assuming that a reference map is fixed once and for all.

\begin{definition}[Finite energy lifts]\label{D:finlift}
We say that $u\in W^{1,2}(M,G)$ has finite energy if
$E(u\pfi)<\infty$ or equivalently $((u^{-1}du)^\bot)^{\wedge 2}\in
L^2(\Lambda^2M\otimes\g)$. The notation is $W_E^{1,2}(M,G)$.
\end{definition}

We can fix a $2$--homotopy sector in $W_E^{1,2}(M,X)$ by choosing a smooth
reference map $\pfi\in C^\infty(M,X)$ and considering all maps in $W_E^{1,2}(M,G)\pfi$. Since 
$$
W^{1,2}(M,G)\pfi\cap W^{1,2}(M,X)=W_E^{1,2}(M,G)\pfi
$$ 
by \eqref{e0.92} these maps exhaust the entire $2$--homotopy sector of $\pfi$. Note however that it is unclear if
$$
\widetilde{W}_E^{1,2}(M,X):=\bigcup_{\pfi\in C^\infty}W_E^{1,2}(M,G)\pfi
$$ 
contains all finite energy maps. In this respect we can only guess:
\begin{conjecture}\label{Cj:1}
Every $2$--homotopy sector of finite energy maps contains a smooth
representative, i.e. $\widetilde{W}_E^{1,2}(M,X)=W_E^{1,2}(M,X)$.
\end{conjecture}
Although $C^\infty(M,X)$ is not dense in $W^{1,2}(M,X)$ it is dense
in $\widetilde{W}_E^{1,2}(M,X)$ (in the $W^{1,2}$ norm) since all
such maps are of the form $u\pfi$ and $u\in W^{1,2}(M,G)$ can be
approximated by smooth maps. In other words, if this conjecture is
true it implies that $W_E^{1,2}(M,X)$ is essentially 'smaller' than
$W^{1,2}(M,X)$. For $X=S^2$ this conjecture is proved in \cite{AK3} but
the proof relies heavily on the fact that in $U_1\hra SU_2\to S^2$ the
subgroup $H=U_1$ is Abelian.

Appearence of $(u^{-1}du)^\bot$ in \eqref{e0.92} suggests a formulation of
the Faddeev-Skyrme energy in terms of gauge potentials. Denote $a:=u^{-1}du$ then
since $\Ad_*(u)$ is an isometry \eqref{e0.92} yields
\bee\label{e0.93}
\bal
|\psi^*\omega^\bot|&=|\pfi^*\omega^\bot+a^\bot|\\
|\psi^*\omega^\bot\wedge\psi^*\omega^\bot|&=|(\pfi^*\omega^\bot+a^\bot)\wedge(\pfi^*\omega^\bot+a^\bot)|.
\eal
\eee
\begin{definition}[Faddeev-Skyrme functional for potentials]\label{D:FadSkyrp}
Denote 
$$
D_\pfi a:=\pfi^*\omega^\bot+a^\bot,
$$ 
where $a\in L^2(\Lambda^1M\otimes\g)$ is a gauge potential.
Then for a fixed reference map $M\xra{\pfi}X$ the Faddeev-Skyrme energy of $a$ is
\bee\label{FadSkyrp}
E_\pfi(a):=\int\limits_M\frac{1}{2}|D_\pfi a|^2+\frac{1}{4}|D_\pfi
a\wedge D_\pfi a|^2\,dm.
\eee
\end{definition}
By \eqref{e0.92}, \eqref{e0.93} for $u\in W^{1,2}(M,G)$ one has
$$
E(u\pfi)=E_\pfi(u^{-1}du),
$$ 
where $E$ is the Faddeev-Skyrme functional \eqref{FadSkyr} for maps. By analogy to 
\refD{finlift} we now define
\begin{definition}[Finite energy potentials]\label{D:finpot}
A gauge potential $a\in L^2(\Lambda^1M\otimes\g)$ has finite energy
if $E_\pfi(a)<\infty$ or equivalently $a^\bot\wedge a^\bot\in L^2(\Lambda^2M\otimes\g)$. We
denote this space $L_E^2(\Lambda^1M\otimes\g)$.
\end{definition}

The presentation $\psi=u\pfi$ when it exists is not unique. Any
$w\in W^{1,2}(M,G)$ satisfying $w\pfi =\pfi$ a.e. produces another
lift $\widetilde{u}=uw$ with $\psi=\widetilde{u}\pfi$. In terms of
potentials this manifests as gauge freedom: we established in \refL{2.5}
that such $w$ are sections of the isotropy subbundle $H_\pfi\subset
M\times G$ isomorphic to $\Ad(\pfi^*G)$ whose sections
are gauge transformations. Changing $u$ to $uw$ corresponds to
changing  $a$ to  $a^w=\Ad_*(w^{-1})a+w^{-1}dw$ and by \refC{Dafi}
\bee\label{e0.95}
D_\pfi(a^w)=\Ad_*(w^{-1})D_\pfi(a).
\eee
Therefore $E_\pfi(a^w)=E_\pfi(a)$ as expected. By the way, this holds for any
gauge potential $a$, not just pure-gauge potentials $a=u^{-1}du$.  If one wants 
to consider non-flat potentials $a$ the functional \eqref{FadSkyrp} should be augmented 
by the Yang-Mills term $|F(a^\vert)|^2$:
\bee\label{e0.96}
E_\pfi^{YM}(a):=\int\limits_M\frac{1}{2}|D_\pfi
a|^2+\frac{1}{4}|D_\pfi a\wedge D_\pfi
a|^2+\frac{1}{2}|F(a^{\|})|^2\,dm.
\eee
We will only consider
pure-gauge potentials and functionals \eqref{FadSkyrp} but our results
trivially extend to arbitrary potentials with the functional \eqref{e0.96}.

The definition of space $L_E^2(\Lambda^1M\otimes\g)$ imposes no
additional restriction on $a^\vert$. Since we consider only
pure-gauge potentials $a=u^{-1}du$ there is however a hidden restriction. It
follows by smooth approximation in $W^{1,2}(M,G)$ that such $a$
satisfy 
$$
da+a\wedge a=0\quad \text{(equality in $W^{-1,2}(\Lambda^2M\otimes\g)),$}
$$
i.e. are distributionally flat. Projecting the flatness condition to $\h_\pfi$ 
one finds that $F(a^{\|})\in L^2(\Lambda^2M\otimes\g)$ (see \refL{dcurv}). In addition to that by
\refL{2.5} stabilizing maps $w\pfi =\pfi$ represent gauge transormations
exactly on the bundle where $a^{\|}$ is a gauge potential. In other
words, {\it the Faddeev-Skyrme functional \eqref{FadSkyrp} allows gauge-fixing of $a^{\|}$
without changing its value}. Along with the bound on $F(a^{\|})$ this
gives us control over the isotropic component while the coisotropic one $a^\bot$ 
is controlled directly by the functional. For technical reasons explained in the next section (see the discussion after \eqref{e0.106}) to use the gauge-fixing we need to restrict the class of finite energy maps.
\begin{definition}[Admissible maps, lifts and potentials]\label{D:admis}
A gauge potential $a$ is admissible if
\bee\label{admis}
\bal
&{\rm
1)}\ a^\bot\in
L^2(\Lambda^1M\otimes\g),\\
&{\rm 2)}\ a^\bot\wedge a^\bot\in L^2(\Lambda^2M\otimes\g),\\
&{\rm 3)}\ a^{\|}\in W^{1,2}(\Lambda^1M\otimes\g).
\eal
\eee
The space of admissible potentials is denoted
$\mathcal{E}(\Lambda^1M\otimes\g)$. A lift $M\ovs{u}\lra G$ is
admissible if $u^{-1}du\in\mathcal{E}(\Lambda^1M\otimes\g)$, a map
$M\ovs{\psi}\lra X$ is admissible if $\psi =u\pfi$ for a smooth $\pfi$
and an admissible $u$. We write $\mathcal{E}(M,G)$, $\mathcal{E}(M,X)$
for admissible lifts and maps respectively and often shortly
$\mathcal{E}\pfi$ instead of $\mathcal{E}(M,G)\pfi$ for the
admissible $2$--homotopy sector of $\pfi$.
\end{definition}
Note that conditions {\rm 1)}, {\rm 2)} of \eqref{admis} simply mean $a\in
L_E^2(\Lambda^1M\otimes\g)$ and hence $u\in W_E^{1,2}(M,G)$, whereas
{\rm 3)} is stronger since generally one only has $a^{\|}\in
L^2(\Lambda^1M\otimes\g)$. Obviously,
$$
\mathcal{E}(M,X)=\bigcup_{\pfi\in C^\infty}\mathcal{E}\pfi
$$
is analogy to $\widetilde{W}_E^{1,2}(M,X)$ and of course
$$
\mathcal{E}(\Lambda^1M\otimes\g)\varsubsetneqq L_E^2(\Lambda^1M\otimes\g),\quad
\mathcal{E}(M,G)\varsubsetneqq W_E^{1,2}(M,G).
$$
Nonetheless we believe in
\begin{conjecture}\label{Cj:2}
For any smooth $\pfi$ and finite energy $u$ there is an admissible $\widetilde{u}\in\mathcal{E}(M,G)$ with $u\pfi
=\widetilde{u}\pfi$ (every finite energy lift is equivalent to an admissible one). Equivalently, 
$$
\mathcal{E}(M,X)=\widetilde{W}_E^{1,2}(M,X).
$$ 
\end{conjecture}
Of course, $\widetilde{u}$ may and will depend on
$\pfi$. Together Conjectures \ref{Cj:1}, \ref{Cj:2} imply that any finite energy map
has the form $\psi=u\pfi$ with $\pfi$ smooth and $u$ admissible. In
the next section we will prove \refCj{2} for the case when $H$ is
a torus (\refC{Cj2}). Along with the result of \cite{AK3} on \refCj{1}
for $X=S^2$ this implies 
$$
W_E^{1,2}(M,S^2)=\mathcal{E}(M,S^2).
$$
In terms of potentials \refCj{2} means that every finite
energy pure-gauge potential is gauge equivalent to an admissible one
and hence the latter are sufficient for minimization.

We already mentioned that unlike $W^{1,2}(M,G)$ the space
$W_E^{1,2}(M,G)$ is not a group. Neither is $\mathcal{E}(M,G)$. In
fact, even if $v\in W^{2,2}(M,G)$ the product $uv$ may not have
finite energy. This is because
\be
(uv)^{-1}d(uv)^\bot=(\Ad_*(v^{-1})u^{-1}du)^\bot+(v^{-1}dv)^\bot
\ee
and $\Ad_*(v^{-1})$ does not commute with $\bot$ so even the term
$((\Ad_*(v^{-1})u^{-1}du)^\bot)^{\wedge 2}$ may not be in $L^2$.

However, if $w\in W^{2,2}(H_\pfi)$, i.e. {\it if in addition to $W^{2,2}$ regularity 
$w$ stabilizes $\pfi$ then $uw$ is again admissible}. Indeed, $E(uw\pfi)=E(u\pfi)<\infty$ guarantees
conditions {\rm 1)}, {\rm 2)} in \eqref{admis} and {\rm 3)} holds because
$$
(\Ad_*(w^{-1})u^{-1}du)^{\|}=\Ad_*(w^{-1})(u^{-1}du)^{\|}
$$ 
and $(w^{-1}dw)^{\|}\in W^{1,2}(\Lambda^1M\otimes\g)$. In other words,
gauge-fixing by a $W^{2,2}$ transformation leaves us within the
class of admissible potentials. This will be crucial in the proof of \refT{3.1}.

We can now state our primary minimization problems for both maps and
potentials.

\noindent {\bf Minimization problem for maps} 
Find a minimizer of the Faddeev-Skyrme
energy \eqref{FadSkyr} in every $2$--homotopy sector of admissible maps:
\bee\label{minmap}
E(\psi)\lra\mbox{min},\quad \psi\in\mathcal{E}\pfi
\eee

\noindent {\bf Minimization problem for potentials} 
Find a minimizer of the Faddeev-Skyrme energy \eqref{FadSkyrp} among all flat
admissible potentials
\bee\label{minpot}
E_\pfi(a)\lra\mbox{min},\quad
a\in\mathcal{E}(\Lambda^1M\otimes\g),\quad da+a\wedge a=0 
\eee
Note that the above two problems are equivalent only if $\pi_1(M)=0$. In
general, if one wants an exact reformulation of the minimization problem for
maps in terms of potentials one has to introduce generalized
holonomy for Sobolev connections and require ${\rm Hol}(a)=1$
instead of flatness. This is indeed done in \cite{AK1}. However using
the fact that gauge-fixing does not spoil admissibility and keeping
track of lifts $u$ directly along with their potentials we can and
will when solving \eqref{minmap} avoid the use of holonomy altogether.

Another remark concerns the fact that the $2$--homotopy sector even for
continuous maps characterizes only the $2$--homotopy type but not
the homotopy type. Of course if $\pi_3(X)=0$, e.g. $X=\C P^n$,
$n\geq 2$ there are no additional invariants and the two notions are equivalent. 
In general, however the $2$--homotopy sector $\mathcal{E}\pfi$ should be subdivided into subsectors by
secondary homotopy invariants and more subtle {\it secondary
minimization} should be carried out within each subsector. When $X$ is
a symmetric space this will be done in \refS{3.3} (see also \cite{AK2,AK3} for the case of the Faddeev model).

\section{Primary minimization}\label{S:3.2}

In this section we first establish some analytic relations between
isotropic and coisotropic parts of flat potentials. A simple application of these relations 
is a proof of \refCj{2} for Abelian $H$. Then we discuss the Uhlenbeck compactness theorem 
and the Wedge product theorem in our context and prove the main
result (\refT{3.1}) on the existence of minimizers in the problem (\ref{minmap}). Unlike in
the case of maps problems with smooth approximation do not arise for differential forms
since the relevant spaces are linear. Hence we derive formulas for $C^\infty$ forms and 
use them for Sobolev ones assuming extension by smooth approximation wherever necessary.

In this section and the next it will be convenient to denote
$\Phi:=\pr_{\h_\pfi}$ and treat it as an $\mbox{End}(\g)$--valued
function with $d\Phi\in\Gamma(\Lambda^1M\otimes\mbox{End}(\g))$.
Differentiating the obvious relation $\Phi a^{\|}=a^{\|}$ we get
\bee\label{e0.100}
d\Phi\wedge a^{\|}=(I-\Phi)da^{\|}=(da^{\|})^\bot.
\eee
Analogously differentiating $(I-\Phi)a^\bot=a^\bot$ yields
\bee\label{e0.101}
d\Phi\wedge a^\bot =-\Phi(da^\bot)=-(da^\bot)^{\|}.
\eee
In the proof of \refT{3.1} we will need Sobolev estimates on $F(a^{\|})$ and
$da^\bot$ in terms of the Faddeev-Skyrme functional.
The next Lemma will be used to obtain such estimates for distributionally flat
gauge potentials.
\begin{lemma}\label{L:dcurv}
Let $a\in L^2(\Lambda^1M\otimes\g)$ be a distributionally flat gauge
potential, i.e. 
$$
da+a\wedge a=0\quad \text{in $W^{-1,2}(\Lambda^2M\otimes\g)$.}
$$
Then
\bee\label{dcurv}
\bal
&{\rm (i)}\ F(a^{\|})=d\Phi\wedge
a^\bot-\Phi(a^\bot\wedge
a^\bot)-\Phi(\pfi^*\omega^\bot\wedge\pfi^*\omega^\bot)\\
&{\rm (ii)}\ da^\bot=-d\Phi\wedge a^{\|}-d\Phi\wedge
a^\bot-[a^{\|},a^\bot]-(I-\Phi)(a^\bot\wedge a^\bot)
\eal
\eee
\end{lemma}
\begin{proof}
{\rm (i)} By the product rule and flatness:\\
\be
\bal da^{\|}&=d(\pfi a)=d\Phi\wedge a+\pfi (da)=d\Phi\wedge a-\Phi(a\wedge a)\\
&=d\Phi\wedge a-\Phi((a^{\|}+a^\bot)\wedge(a^{\|}+a^\bot))\\
&=d\Phi\wedge a-\Phi(a^{\|}\wedge a^{\|}+[a^{\|},a^\bot]+a^\bot\wedge a^\bot).
\eal
\ee 
By \eqref{e0.42} and \eqref{e0.37}(iv) the form $a^{\|}\wedge a^{\|}$ takes values in $\h_\pfi$ and
$[a^{\|},a^\bot]$ in $\h_\pfi^\bot$. Therefore 
$$
\Phi(a^{\|}\wedge a^{\|})=a^{\|}\wedge a^{\|}\quad \text{and} \quad \Phi[a^{\|},a^\bot]=0.
$$
Thus we get 
\bee\label{e0.103}
da^{\|}+a^{\|}\wedge a^{\|}=d\Phi\wedge
a^{\|}+d\Phi\wedge a^\bot-\Phi(a^\bot\wedge a^\bot).
\eee
By \eqref{e0.67}:
\be
\bal
F(a^{\|}) &=(da^{\|})^{\|}+a^{\|}\wedge a^{\|}-(\pfi^*\omega^\bot\wedge\pfi^*\omega^\bot)^{\|}\\
          &=\Phi(da^{\|}+a^{\|}\wedge a^{\|}-\pfi^*\omega^\bot\wedge\pfi^*\omega^\bot).
\eal
\ee
Subtracting $\pfi^*\omega^\bot\wedge\pfi^*\omega^\bot$ from both
sides of \eqref{e0.103}, applying $\Phi$ and taking into account that
$\Phi(d\Phi\wedge a ^{\|})=0$ by \eqref{e0.100} we get {\rm (i)}.

\noindent {\rm (ii)} Plugging $a=a^{\|}+a^\bot$ into $da+a\wedge a=0$ one gets
\be
da^\bot+a^\bot\wedge a^\bot+da^{\|}+a^{\|}\wedge a^{\|}+[a^{\|},a^\bot]=0.
\ee
Now rewriting $da^{\|}+a^{\|}\wedge
a^{\|}$ by \eqref{e0.103} and taking all terms except $da^\bot$ to the
righthand side gives {\rm (ii)}.
\end{proof}

\refL{dcurv} implies that flat potentials are better than they
'should be'. This is not surprising since for $a$ in $L^2$ the relation
$da=-a\wedge a$ implies that $da$ which is a priori only in
$W^{-1,2}$ is actually in $L^1$. If moreover $a\in
L^2_E(\Lambda^1M\otimes\g)$, then \eqref{dcurv} yields 
$$
F(a^{\|})\in L^2\quad \text{and}\quad (da^\bot)^{\|}\in L^2.
$$
The other component $(da^\bot)^\bot$ is 'spoiled' by the term $[a^{\|},a^\bot]$ which
will only be in $L^{3/2}$ even assuming that $a$ is admissible,
i.e. $a^{\|}\in W^{1,2}$.

As a first application of \refL{dcurv} we will prove \refCj{2} in the case
when $H$ is Abelian (and hence a torus \cite{BtD}). For this case it is
convenient to use the usual (twisted) gauge potentials of \refD{pot}.
In general their presentation by a differential form will depend on
a choice of local trivialization of $\Ad_*(\pfi^*G)$ bundle. 
Such a trivialization can be given by a local gauge, i.e a local section of the coset bundle $\pfi^*G$:
$$
M\supset U\ovs{\gamma}\lra\pfi^*G.
$$ 
In this gauge by \refL{2.3}
$$
\alpha=\Ad_*(\gamma^{-1})a,
$$
where $a$ is the (globally defined) untwisted gauge
potential of \refD{cospot}. Change of gauge from $\gamma$ to
$\gamma\nu$ with $U\ovs{\nu}\to H$ changes $\alpha$ 
to\footnote[1]{It may seem odd that $\alpha$ is not changed to $\Ad_*(\nu^{-1})\alpha+\nu^{-1}d\nu$ as usual. The latter gives the gauge potential with respect to a {\it new} reference connection -- the trivial connection in the trivialization given by the section $\gamma\nu$. If we keep the {\it same} reference connection and only use the new trivialization to write a bundle--valued form $\alpha$ as a Lie algebra valued one the expression is just $\Ad_*(\nu^{-1})\alpha$. The difference is that in contrast to the usual convention in gauge theory \cite{DFN,MM} {\it we are only using local gauge to trivialize the $\Ad_*$ bundle but not to simultaneously change the reference connection to the trivial one}.}
$$
\Ad_*((\gamma\nu)^{-1})a=\Ad_*(\nu^{-1})\Ad_*(\gamma^{-1})a=\Ad_*(\nu^{-1})\alpha.
$$ 
When $H$ is Abelian $\Ad_*(H)$ acts trivially on $\h$ and $\alpha$ is a globally defined section of $\Lambda^1M\otimes\h$.
Similarly, a gauge transformation $\lambda$ is a globally defined section of $M\times
H$, i.e. an $H$--valued map. 

There is nothing specific to coset bundles involved here. In any principal bundle with an Abelian structure
group $H$ the bundles
$$
\Ad_*(P)=P\times_{\Ad_*}H\quad \text{and}\quad 
\Ad(P)=P\times_{\Ad}H
$$ 
are trivial and there is no need to 'untwist' gauge transformations or potentials. Since the
relations between 'twisted' and 'untwisted' objects: 
$$
\alpha=\Ad_*(\gamma^{-1})a,\quad \lambda=\Ad(\gamma^{-1})w\quad \text{and}
\quad F(\alpha)=\Ad_*(\gamma^{-1})F(a)
$$ 
are given by multiplication by smooth maps (albeit only locally defined) Sobolev conditions imposed
on $a$, $w$, $F(a)$ are equivalent to those imposed on $\alpha$, $\lambda$, $F(\alpha)$ respectively. 

A simple computation shows that for {\it any} Abelian principal bundle the gauge action on potentials with respect to {\it any} reference connection has a very simple form: 
\bee\label{e0.104}
\alpha^\lambda=\alpha+\lambda^{-1}d\lambda,
\eee
and the curvature reduces to the differential:
\bee\label{e0.105}
F(\alpha)=d\alpha.
\eee
This is the reason we prefer $\alpha$-s to $a$-s (compare \eqref{e0.104}, \eqref{e0.105} to \eqref{e0.60}(i),(iii)).

Since $H$ is a torus the exponential map $\h\ovs{\exp}\lra H$ is globally defined
and onto. Taking $\lambda:=\exp(\xi)$ with $M\ovs{\xi}\lra \h$ we
turn \eqref{e0.104} into
\be
\alpha^\lambda=\alpha+d\xi
\ee
By a result of \cite{IV} if $\alpha,d\alpha\in L^p$ 
then there is $\xi\in W^{1,p}$ and $\widetilde{\alpha}\in W^{1,p}$ such that
\be
\alpha =\widetilde{\alpha}-d\xi.
\ee
In other words, any differential form in $L^p$ with the differential also in $L^p$
is $W^{1,p}$--cohomologous to a $W^{1,p}$ form. Since $\xi\in W^{1,p}(M,\h)$
implies $\lambda:=\exp(\xi)\in W^{1,p}(M,H)$ and
$$
\alpha^\lambda=\widetilde{\alpha}=\alpha+d\xi=\alpha+\lambda^{-1}d\lambda
$$
this result restated in terms of gauge theory reads: 
\begin{lemma}\label{L:Abfix}
In a principal bundle with an Abelian structure group every $L^p$ 
potential with $L^p$ curvature is gauge equivalent by a $W^{1,p}$ 
gauge transformation to a $W^{1,p}$ potential.
\end{lemma} 
\noindent Due to the isometric isomorphism of \refL{2.4} this lemma applies 
to the untwisted potentials and transformations $a,w$ just as it does to the twisted ones $\alpha,\lambda$.
\begin{corollary}\label{C:Cj2}
If $X=G/H$ and $H$ is a torus then Conjecture 2 holds.
\end{corollary}
\begin{proof}
We have to prove that if $\psi=u\pfi$ with $\pfi\in C^\infty(M,X)$
and $u\in W_E^{1,2}(M,G)$ then there is
$\widetilde{u}\in\mathcal{E}(M,G)$ such that $\psi=\widetilde{u}\pfi$. Let
$a=u^{-1}du$ then $a\in L^2_E$ is flat and $F(a^{\|})\in L^2$ by
\refL{dcurv}. Since \refL{Abfix} applies to untwisted 
potentials there is $w\in W^{1,2}(H_\pfi)$ such that
$$
(a^{\|})^w=(a^w)^{\|}\quad \text{with}\quad a^w=(uw)^{-1}d(uw).
$$
Set $\widetilde{u}:=uw$. Then $\widetilde{u}\pfi=u\pfi$ 
and hence $\widetilde{u}\in W_E^{1,2}(M,G)$ by \refD{finlift}.
Moreover, by construction $(\widetilde{u}^{-1}d\widetilde{u})^{\|}\in
W^{1,2}$ and $\widetilde{u}\in\mathcal{E}(M,G)$ by \refD{admis}.
\end{proof}

To extend this result to general homogeneous spaces one needs
\refL{Abfix} without the word 'Abelian'. Since a nonlinearity in
curvature $F(\alpha)$ is involved more care is required. 
For instance, by the Sobolev multiplication
theorems \cite{Pl} the product $\alpha\wedge\alpha$ with $\alpha\in W^{1,p}$ 
is in $L^p$ only for $2p\geq\dim M$. Nonetheless we still believe that the following holds.

\begin{conjecture}\label{Cj:3}
Let $P\to M$ be a smooth principal bundle and $2p\geq\dim M$. Suppose
$$
\alpha\in L^p(\Lambda^1M\otimes\Ad_*P)
$$ 
is a gauge potential on it with 
$$
F(\alpha)\in L^p(\Lambda^2M\otimes\Ad_*P).
$$ 
Then there exists a gauge transformation $\lambda\in W^{1,p}(\Ad P)$ such that
$$
\alpha^\lambda:=\Ad_*(\lambda^{-1})\alpha+\lambda^{-1}d\lambda\in W^{1,p}(\Lambda^1M\otimes\Ad_*P).
$$
\end{conjecture}
\noindent Since our $M$ is $3$--dimensional and $p=2$ \refCj{3} implies \refCj{2} for any simply connected $X$ 
(the proof is the same as in \refC{Cj2}).

The proof of \refC{Cj2} is indicative of the way we apply gauge-fixing to maps into homogeneous spaces. 
This trick will also be used to prove the main result of this section on existence of minimizers in \eqref{minmap}.
In addition we need two more results to establish weak compactness and lower semicontinuity.
First is the result of K.Uhlenbeck \cite{Ul1,We}:
\begin{theorem*}[Uhlenbeck compactness theorem] 
Let $P\to M$ be a smooth principal bundle and $2p>\mbox{dim}M$. Consider a sequence of gauge
potentials on $M$
$$
\alpha_n\in W^{1,p}(\Lambda^1M\otimes\Ad_*P)\quad \text{with}\quad ||F(\alpha_n)||_{L_p}\leq C<\infty.
$$  
Then there exists a subsequence $\alpha_{n_k}$ 
and a sequence of gauge transformations $\lambda_{n_k}\in W^{2,p}(\Ad P)$ such that
\bee\label{e0.106}
\alpha_{n_k}^{\lambda_{n_k}}\ovs{W^{1,p}}\rightharpoonup\alpha\qquad
\mbox{and}\quad ||F(\alpha)||_{L_p}\leq C.
\eee
\end{theorem*}
\noindent Note that in the Uhlenbeck compactness theorem $\alpha_n$ 
are assumed from the start to be in $W^{1,p}$ rather than just in $L^p$.
If our \refCj{3} were true one could replace this assumption with $\alpha_n\in
L^p(\Lambda^1M\otimes\Ad_*P)$ and allow $\lambda_{n_k}\in W^{1,p}(\Ad P)$. 
We will use this compactness theorem to fix the gauge for the isotropic parts $a_n^{\|}$
of potentials in a minimizing sequence. This means that we need $a_n^{\|}\in
W^{1,2}(\Lambda^1M\otimes\g)$ from the start to apply the theorem and these are the 'technical
reasons' we cited before for restricting to the admissible maps. 

The second result we need concerns weak convergence of wedge
products. Recall that even for scalar functions weak convergence of
factors to limits in $L^2$ does not imply even distributional convergence
of the product to the product of the limits. For instance,
$$
\sin(nx)\ovs{L^2}\rightharpoonup 0\quad \text{on $[0,1]$, but}\quad \sin^2(nx)=\frac{1}{2}(1-\cos(2nx))\ovs{L^2}\rightharpoonup\frac{1}{2}\neq0.
$$ 
Still the Hodge decomposition of differential forms yields \cite{RRT} (see also
\cite{IV} for a different approach):
\begin{theorem*}[Wedge product theorem] 
Assume that $\upsilon_n\ovs{L^2}\rightharpoonup\upsilon$,
$\omega_n\ovs{L^2}\rightharpoonup\omega$ are sequences of $L^2$ differential
forms on a compact manifold $M$ and $d\upsilon_n$, $d\omega_n$ are precompact in $W^{-1,2}$.
Then
$$
\upsilon_n\wedge\omega_n\ovs{\mathcal{D}'}\rightharpoonup\upsilon\wedge\omega\quad 
\text{(in the sense of distributions).}
$$
\end{theorem*}
\noindent Here as usual $\mathcal{D}(\Lambda^\bullet M\otimes\mbox{End}(\mathbb{E})^*)$
is the space of test forms ($C^\infty$ with compact support) and
$\mathcal{D}'(\Lambda^\bullet M\otimes\mbox{End}(\mathbb{E}))$ is the dual
space relative to the inner product in $L^2$ \cite{GMS4}. In the above example
the precompactness condition fails: $d\sin(nx)=n\cos(nx)$ is
unbounded even in $\mathcal{D}'$. 

It will be convenient for us to use the Wedge product theorem in a slightly weakened form. By a Sobolev embedding
theorem $L^s\hra W^{-1,p}$ compactly if
$\frac{1}{s}<\frac{1}{n}+\frac{1}{p}$ ($n:=\mbox{dim}M$). For a 
$3$-dimensional $M$ and $p=2$ this gives $s>\frac{6}{5}$. Thus
we can replace precompactness in $W^{-1,2}$ by boundedness in $L^{6/5+\ve}$ with $\ve>0$.  

\begin{theorem}\label{T:3.1}
Every $2$--homotopy sector of admissible maps has a minimizer of the Faddeev-Skyrme
energy.
\end{theorem}
\begin{proof}
We denote by $\ovs{L}\rightharpoonup$
($\ovs{L}\lra$) the weak (the strong) convergence in a Banach space $L$. 
All constants in the estimates are denoted by $C$ even though they may be different. 
Passing to subsequences is also ignored in the notation. This does not lead to any confusion. 

Recall that we assume $G\hra\mbox{End}(\mathbb{E})$ for a Euclidean space $\mathbb{E}$ and
$u\in W^{1,2}(M,G)$ means $u\in W^{1,2}(M,\mbox{End}(\mathbb{E}))$ with $u(m)\in G$ a.e. 
Let $\psi_n=u_n\pfi$ be a minimizing sequence of admissible maps in a sector $\mathcal{E}\pfi$ and 
$a_n:=u_n^{-1}du_n$. The proof is divided into several steps.

\noindent {\bf Gauge-fixing}\nopagebreak 

\noindent By definition
$$
E(u_n\pfi)=E_\pfi(a_n)\leq C<\infty.
$$
It follows by inspection from \eqref{FadSkyrp} that 
$$
||a_n^\bot||_{L^2}\leq C<\infty\quad \text{and}\quad 
||a_n^\bot\wedge a_n^\bot||_{L^2}\leq C<\infty.
$$ 
Then by \refL{dcurv}(i) also
$$
||F(a_n^{\|})||_{L^2}\leq C<\infty.
$$
Since $u_n$ are admissible $a_n^{\|}\in W^{1,2}$ 
and we may apply the Uhlenbeck compactness theorem to $a_n^{\|}$. After passing to a
subsequence we get a sequence of gauge transformations $w_n\in W^{2,2}(H_\pfi)$ such that
$$
(a_n^{\|})^{w_n}=(a_n^{w_n})^{\|}\ovs{W^{1,2}}\rightharpoonup
a^{\|}.
$$
But 
$$
a_n^{w_n}=\Ad_*(w_n^{-1})a_n+w_n^{-1}dw_n=(u_nw_n)^{-1}d(u_nw_n)
$$
and $u_nw_n$ are still admissible. Therefore we can drop $w_n$ from the notation 
and assume that $u_n$ are preselected to have the isotropic components $a_n^{\|}$ 
weakly convergent in $W^{1,2}$.

\noindent {\bf Compactness} 

\noindent Let $u_n$ be the gauge-fixed minimizing sequence from the previous step.
Since $G$ is compact it is bounded in $\End(\mathbb{E})$ and
$$
\|u_n\|_{L^\infty}\leq C<\infty.
$$
By gauge-fixing and \eqref{FadSkyrp} both
$a_n^{\|}$, $a_n^\bot$ are bounded in $L^2$. Therefore so are 
$$
a_n=a_n^{\|}+a_n^\bot=u_n^{-1}du_n\quad \text{and}\quad du_n=u_na_n.
$$ 
We conclude that
$$
\|u_n\|_{W^{1,2}}\leq C<\infty
$$ 
and after passing to a subsequence $u_n\ovs{W^{1,2}}\rightharpoonup u$. 

Since $W^{1,2}\hra L^2$ is a compact embedding we have $u_n\ovs{L^2}\lra u$ and since
$u_n$ are bounded in $L^\infty$ also $u_n^{-1}\ovs{L^2}\lra u^{-1}$. But the strong
convergence in $L^2$ implies convergence almost everywhere on a subsequence and we have $u(m)\in G$ a.e. 
so that $u\in W^{1,2}(M,G)$. 

The differential $d:W^{1,2}\to L^2$ is a bounded linear operator 
and hence it is weakly continuous. Therefore
$$
du_n\ovs{L^2}\rightharpoonup du\quad \text{and}\quad
u_n^{-1}du_n=a_n\ovs{L^2}\rightharpoonup a:=u^{-1}du.
$$ 
Moreover, by the preselection of $u_n$ we have in addition 
$$
a_n^{\|}\ovs{W^{1,2}}\rightharpoonup a^{\|}\in W^{1,2}(\Lambda^1M\otimes\g).
$$

\noindent {\bf Closure} 

\noindent In view of \eqref{FadSkyrp} 
$$
\|a_n^\bot\wedge a_n^\bot\|_{L^2}\leq C<\infty
$$ 
and (possibly after passing to another subsequence)
$$
a_n^\bot\wedge a_n^\bot\ovs{L^2}\rightharpoonup\Lambda.
$$ 
Since $a_n^\bot$ is bounded in $L^2$ and $a_n^{\|}$ is bounded in
$W^{1,2}$ we have by the Sobolev multiplication theorem \cite{Pl}:
$$
\|[a_n^{\|},a_n^\bot]\|_{L^{3/2}}\leq C<\infty
$$
and and hence by \refL{dcurv}
$$
\|da_n^\bot\|_{L^{3/2}}\leq C<\infty.
$$ 
But $3/2>6/5$ and the Wedge product theorem now implies 
$$
a_n^\bot\wedge a_n^\bot\ovs{\mathcal{D}'}\rightharpoonup a^\bot\wedge a^\bot.
$$ 
By uniqueness of the limit in $\mathcal{D}'$ one must have $\Lambda=a^\bot\wedge a^\bot$ and
$$
a_n^\bot\wedge a_n^\bot\ovs{L^2}\rightharpoonup a^\bot\wedge
a^\bot\in L^2(\Lambda^2M\otimes\g).
$$ 
Along with the previous step this yields $u\in\mathcal{E}(M,G)$ and hence
$\psi:=u\pfi\in\mathcal{E}(M,X)$. This is the map we were looking for.

\noindent {\bf Lower semicontinuity} 

\noindent $E$ in \eqref{FadSkyr} is not a weakly lower semicontinuous functional of $\psi$ 
and neither is $E_\pfi$ in \eqref{FadSkyrp} as a functional of $a$. 
However,
$$
\widehat{E}(r,\Lambda):=\frac{1}{2}\|r\|_{L^2}^2+\frac{1}{4}\|\Lambda\|_{L^2}^2
$$
{\it is} a weakly lower semicontinuous functional of a pair (see \cite{BlM}):
$$
(r,\Lambda)\in
L^2(\Lambda^1M\otimes \g)\times L^2(\Lambda^2M\otimes\g)
$$ 
But obviously,
$$
E_\pfi(a)=\widehat{E}(D_\pfi a,D_\pfi a\wedge D_\pfi a).
$$  
By the above 
$$
D_\pfi a_n=\pfi^*\omega^\bot+a_n^\bot\ovs{L^2}\rightharpoonup D_\pfi a\quad \text{and}\quad
D_\pfi a_n\wedge D_\pfi a_n\ovs{L^2}\rightharpoonup D_\pfi a\wedge D_\pfi a.
$$ 
Therefore,
\begin{multline*}
E(\psi)=E_\pfi(a)=\widehat{E}(D_\pfi a,D_\pfi a\wedge D_\pfi a)\\
\leq \liminf_{n\to\infty}E(D_\pfi a_n,D_\pfi a_n\wedge D_\pfi a_n)
=\liminf_{n\to\infty}E_\pfi(a_n)=\liminf_{n\to\infty}E(\psi_n).
\end{multline*}
Since $\psi_n$ was a minimizing sequence in $\mathcal{E}\pfi$ and $\psi=u\pfi\in\mathcal{E}\pfi$ it is a minimizer
of \eqref{FadSkyr} in the $2$--homotopy sector of $\pfi$.
\end{proof}
The minimization for flat potentials (problem \eqref{minpot}) is analogous and simpler. 
\begin{corollary}\label{C:minpot}
For every smooth $\pfi\in C^\infty(M,X)$ there exists a minimizer of
the Faddeev-Skyrme energy \eqref{FadSkyrp} among admissible flat potentials.
\end{corollary}
\begin{proof}
The proof is essentially the same as in \refT{3.1} so we only sketch
it. We gauge-fix a minimizing sequence $a_n$ to have
$a_n^{\|}\ovs{W^{1,2}}\rightharpoonup a^{\|}$ and get
$$
a_n^\bot\ovs{L^2}\rightharpoonup  a^\bot,\quad 
a_n^\bot\wedge a_n^\bot\ovs{L^2}\rightharpoonup a^\bot\wedge a^\bot
$$ 
directly from the functional $E_\pfi$ and the Wedge product theorem. Now 
$$
a_n\ovs{L^2}\rightharpoonup a,\quad
da_n\ovs{W^{-1,2}}\rightharpoonup da,
$$
and $da_n=(da_n)^{\|}+(da_n)^\bot$ is bounded in $L^{3/2}$ and hence
precompact in $W^{-1,2}$. Applying the Wedge product theorem again
we get $a_n\wedge a_n\ovs{L^2}\rightharpoonup a\wedge a$. Therefore 
$$
0=da_n+a_n\wedge a_n\ovs{W^{-1,2}}\rightharpoonup da+a\wedge a
$$ 
and $a$ is admissible and distributionally flat.
\end{proof}

\begin{remark}
If $a_n$ are not  just flat but pure-gauge it follows from a result
in \cite{AK1} that on a subsequence $a_n\ovs{L^2}\rightharpoonup a$,
where $a$ is also pure-gauge. Using this result one could prove \refT{3.1}
without introducing $u_n$ explicitly but such a proof requires a lengthy
discussion of holonomy for distributional connections. Note also
that the argument of \refC{minpot} works just as well for the
functional \eqref{e0.96} and non-flat potentials. In this case there is no
need in flatness and \refL{dcurv} since $\|F(a_n^{\|})\|_{L^2}$ are
bounded directly by the functional. Of course, the minimizers will no
longer be flat either.
\end{remark}

For $X=S^2$ \refT{3.1} is proved in \cite{AK2} (Theorem 4). In fact the
result there is stronger: $\mathcal{E}\pfi$ is subdivided into subsectors by
additional Chern-Simons invariants and there is a separate minimizer
in each subsector. This already shows that a minimizer in
$\mathcal{E}\pfi$ is not unique. But even if $\pi_3(X)=0$ and the
$2$--homotopy sectors characterize homotopy classes completely there is little
hope that the minimizers of \eqref{FadSkyr} are unique since the functional is
nowhere near being convex.

\section{Secondary minimization for symmetric spaces}\label{S:3.3}

In the primary minimization we considered minimizing \eqref{FadSkyr} over the
entire set $\mathcal{E}\pfi$. If $\mathcal{E}(M,G)$ is replaced by
$C^\infty(M,G)$ this would correspond to minimizing over all smooth maps
$2$-homotopic to $\pfi$. To minimize over maps that are {\it homotopic} to $\pfi$ 
one needs to add a constraint given by the secondary invariants (\refS{1.4}).
By \refC{secinv} for smooth maps this constraint can be given in terms of $u$ as $\int\limits_Mu^*\Theta\in\O_\pfi$
where $\Theta$ is an $\R^N$--valued $3$--form representing the basic class of $G$ (\refD{basic}) and 
\begin{equation}\label{Z-Ofi}
\O_\pfi:=\{\int\limits_Mw^*\Theta\mid w\pfi=\pfi\}<\Z^N.
\end{equation}
It is worth reminding that $\O_\pfi$ only depends on the $2$--homotopy type of $\pfi$.

We will need an explicit expression for the deRham representative $\Theta$. 
When $G$ is a simple group one can take the normalized Cartan $3$--form \cite{CE}:
$$
\Theta:=c_G\tr(g^{-1}dg\wedge g^{-1}dg \wedge g^{-1}dg),
$$
where $c_G$ is a numerical coefficient that ensures integrality. These coefficients are 
computed explicitly for all simple groups in \cite{AK1}. 
These authors also give a generalization of the above $\Theta$ to arbitrary compact groups.
For simply connected ones it reduces to the following. Let $\g=\g_1\oplus\dots\oplus\g_N$ be the decomposition of the Lie algebra of $G$ into simple components.
For any $\g$--valued form $\alpha$ let $\alpha^k:=\pr_{\g_k}(\alpha)$ denote its orthogonal projection to $\g_k$ with respect to some invariant metric on $\g$. Then let
$$
\Theta_k:=c_{G_k}\tr((g^{-1}dg)^k\wedge(g^{-1}dg)^k\wedge(g^{-1}dg)^k)
$$
and 
$$
\Theta:=(\Theta_1,\dots,\Theta_N)
$$
is the required representative. Therefore for smooth maps
\bee\label{Thk}
u^*\Theta_k:=c_{G_k}\tr((u^{-1}du)^k\wedge(u^{-1}du)^k\wedge(u^{-1}du)^k)=
c_{G_k}\tr(a^k\wedge a^k\wedge a^k),
\eee
where as usual $a=u^{-1}du$. Note that the expression on the right 
is defined almost everywhere as a form even if $u$ is just a $W^{1,2}$ map.

It is easy to see from the product rule and the definition of Sobolev norms that
\be
||\alpha^k||_{W^{l,p}}\leq ||\alpha||_{W^{l,p}}
\ee
for any form $\alpha$. Moreover, for any pair of forms $\alpha,\beta$
\be
(\alpha\wedge\beta)^k=\alpha^k\wedge\beta^k
\ee
since elements from different $\g_k$ always commute. Therefore if $a$ is admissible we have for each $k$:
\bee\label{kadmis}
\bal
{\rm 1)}\ &(a^\bot)^k \in L^2(\Lambda^1M\otimes\g)\\
{\rm 2)}\ &(a^\bot)^k\wedge(a^\bot)^k=(a^\bot\wedge a^\bot)^k\in L^2(\Lambda^2M\otimes\g)\\
{\rm 3)}\ &(a^\vert)^k\in W^{1,2}(\Lambda^1M\otimes\g).
\eal
\eee
By the way, each $a^k$ separately may not be admissible since in general
$(a^k)^\vert\neq(a^\vert)^k$, $(a^k)^\bot\neq(a^\bot)^k$. 

Even though $u^*\Theta$ is defined almost everywhere as a form in order 
to integrate it over $M$ we need it to be defined at least as a distribution. 
Since we only know that $a^k\in L^2$ the triple product $a^k\wedge a^k\wedge a^k$ 
is not even in $L^1$ and one can not use the expression \eqref{Thk} for integration directly. 
To take advantage of the conditions \eqref{kadmis} we decompose $a^k=(a^\vert+a^\bot)^k$,
plug it into $a^k\wedge a^k\wedge a^k$ and use the distributive law. 
The resulting sum will have terms like $(a^\bot)^k\wedge (a^\vert)^k\wedge (a^\bot)^k$ that are still
not in $L^1$. Fortunately, we only have to integrate {\it traces} of such
terms and the situation can be helped. 

Since $\tr(\xi_1\cdots\xi_n)$ is invariant under cyclic
permutations of $\xi_k$-s by definition of the wedge product \eqref{e0.34} we get for
any cyclic permutation $\sigma$ and $1$-forms $\alpha_k$:
\be
\tr(\alpha_{\sigma
(1)}\wedge\dots\wedge\alpha_{\sigma (n)})=(-1)^\sigma \mathop{\rm
tr}\nolimits(\alpha_1\wedge\dots\wedge\alpha_n)=(-1)^{n-1}\mathop{\rm
tr}\nolimits(\alpha_1\wedge\dots\wedge\alpha_n).
\ee 
As a corollary for any forms $\alpha$, $\beta$ the wedge cube $\tr((\alpha +\beta)^{\wedge 3})$ 
reduces to the binomial form 
\be 
\tr((\alpha +\beta)^{\wedge 3})
=\tr(\alpha^{\wedge 3})
+3\tr(\alpha^{\wedge 2}\wedge\beta)
+3(\alpha\wedge\beta^{\wedge 2})
+\tr(\beta^{\wedge 3}).
\ee
Applying it to $\alpha =(a^\vert)^k=:a^{\vert k}$, $\beta=(a^\bot)^k=:a^{\bot k}$ we get 
\bee\label{traaa}
\bal 
\tr&(a^k\wedge a^k\wedge a^k)=\\
&\tr(a^{\vert k}\wedge a^{\vert k}\wedge a^{\vert k})
+3\tr(a^{\vert k}\wedge a^{\vert k}\wedge a^{\bot k})
+3\tr(a^{\vert k}\wedge a^{\bot k}\wedge a^{\bot k})
+\tr(a^{\bot k}\wedge a^{\bot k}\wedge a^{\bot k}). 
\eal
\eee

From \eqref{kadmis} and the Sobolev multiplication theorems we derive
\bee\label{trspace}
\bal
{\rm 1)}\ &a^{\vert k}\wedge a^{\vert k}\wedge a^{\vert k}\in L^2\\
{\rm 2)}\ &a^{\vert k}\wedge a^{\vert k}\wedge a^{\bot k}\in L^{6/5}\\
{\rm 3)}\ &a^{\vert k}\wedge a^{\bot k}\wedge a^{\bot k}\in L^{3/2}\\
{\rm 4)}\ &a^{\bot k}\wedge a^{\bot k}\wedge a^{\bot k}\in L^1.
\eal
\eee
Overall $\tr(a^k\wedge a^k\wedge a^k)\in L^1$ and hence $u^*\Theta_k$ can be defined 
for admissible $u$ as an $L^1$ form by replacing $\tr(a^k\wedge a^k\wedge a^k)$ 
in \eqref{Thk} by the righthand side of \eqref{traaa}. If $u$ just has finite energy 
we only know $a^{\vert k}\in L^2$ and the first two terms are not in $L^1$. 
Thus, {\it in general the secondary invariants are not even defined for all finite energy maps}. 
There is a case when they actually are. If $G$ is a simple group and the subgroup $H$ is Abelian we have $[\h,\h]=0$ and hence $a^{\vert}\wedge a^{\vert}=0$ so the 'bad' terms vanish. In particular $X=SU_2/U_1$ is such a case or more generally, flag manifolds $X=SU_{n+1}/\T^n$, where $T^n$ is a maximal torus.

Even though secondary invariants are defined for all admissible maps they do not behave well. More exactly, it is unclear if one can approximate an admissible $u$ by smooth maps in such a way that $u^*\Theta$ is approximated in $L^1$ or even in $\mathcal{D}'$ by the corresponding smooth forms. For the latter one would need\footnote[1]{The other three terms converge trivially.} 
$$
a_n^{\bot k}\wedge a_n^{\bot k}\wedge a_n^{\bot k}\ovs{\mathcal{D}'}
\rightharpoonup a^{\bot k}\wedge a^{\bot k}\wedge a^{\bot k}.
$$  
By the Wedge product theorem this happens if $d(a_n^\bot\wedge a_n^\bot)$ is bounded in $L^{6/5+\ve}$.
But in general by \eqref{e0.37}(xi) and \refL{dcurv}
\begin{multline}\label{daa}
d(a^\bot\wedge a^\bot)=[da^\bot,a^\bot]\\
=-[d\Phi\wedge a^{\|},a^\bot]-[d\Phi\wedge a^\bot,a^\bot]-[[a^{\|},a^\bot],a^\bot]
-[(I-\Phi)(a^\bot\wedge a^\bot),a^\bot].
\end{multline}
The first term is in $L^{3/2}$ and so is the third one due to the cancellation formula \eqref{e0.37}(vi)
\be
[[a^{\|},a^\bot],a^\bot]=[a^{\|},a^\bot\wedge a^\bot]\in L^{3/2}.
\ee
However a priori we only have 
$$
[d\Phi\wedge a^\bot,a^\bot]\in L^1\quad \text{and}\quad [(I-\Phi)(a^\bot\wedge a^\bot),a^\bot]\in L^1,
$$ 
while $1<6/5$. Without smooth approximation we do not know if $\int\limits_Mu^*\Theta$ are still integral and the secondary constraint $\int\limits_Mu^*\Theta\in\O_\pfi<\Z^N$ makes sense. To deal with this problem we have to confine ourselves to symmetric spaces and strongly admissible maps.

Recall that $X=G/H$ is a {\it symmetric space} if there is a homomorphic involution $G\to G$ that fixes $H$ pointwise
\cite{Ar,Br2,Hl}. What is important to us is that in addition to the usual relations
\bee\label{shh}
[\h,\h]\subset\h\quad ,\quad [\h,\h^\bot]\subset\h^\bot
\eee
in a symmetric space one also has
\bee\label{snh}
[\h^\bot,\h^\bot]\subset\h
\eee
and therefore
\bee\label{snhfi}
[\h_\pfi^\bot,\h_\pfi^\bot]\subset\h_\pfi. 
\eee
Since $\Phi=\pr_{\h_\pfi}$ and $I-\Phi=\pr_{\h^\bot_\pfi}$ we have immediately
\bee\label{e0.109}
(I-\Phi)(a^\bot\wedge a^\bot)=0
\eee
and one of the singular terms vanishes altogether.
Differentiating \eqref{e0.109} gives a second relation
\bee\label{e0.110}
(I-\Phi)d(a^\bot\wedge a^\bot)=d\Phi\wedge(a^\bot\wedge a^\bot).
\eee
Formulas of \refL{dcurv} can now be improved.
\begin{lemma}\label{L:sdcurv}
Let $X=G/H$ be a symmetric space and $a\in L^2(\Lambda^1M\otimes\g)$
a distributionally flat gauge potential. Then
\bee\label{sdcurv}
\bal
&{\rm (i)}\ F(a^{\|})=d\pfi\wedge a^\bot-a^\bot\wedge a^\bot-\pfi^*\omega^\bot\wedge\pfi^*\omega^\bot\\
&{\rm (ii)}\ da^\bot=-d\Phi\wedge a^{\|}-d\Phi\wedge a^\bot-[a^{\|},a^\bot]\\
&{\rm (iii)}\ d(a^\bot\wedge a^\bot)=-[d\Phi\wedge a^{\|},a^\bot]+d\Phi\wedge(a^\bot\wedge a^\bot).
\eal
\eee
\end{lemma}
\begin{proof}
{\rm (i)}, {\rm (ii)} follow directly from \refL{dcurv} and
\eqref{e0.109}. 

\noindent {\rm (iii)} We need to simplify \eqref{daa} for symmetric spaces. 
Since $d\Phi\wedge a^\bot=-\Phi(da^\bot)$ takes values in $\h_\pfi$ and
$[a^{\|},a^\bot]$ in $[\h_\pfi,\h_\pfi^\bot]\subset\h_\pfi$ we have that
$$
[d\Phi\wedge a^\bot,a^\bot]+[[a^{\|},a^\bot],a^\bot]
$$
is $\h_\pfi^\bot$--valued. On the other hand 
$$
d\Phi\wedge a^{\|}=(I-\Phi)da^{\|}
$$ 
is $\h_\pfi^\bot$--valued and by \eqref{snh} $[d\Phi\wedge a^{\|},a^\bot]$ takes values in $\h_\pfi$. Thus
\begin{align*}
\Phi d(a^\bot\wedge a^\bot) &=-[d\Phi\wedge a^{\|},a^\bot]\\ 
(I-\Phi)d(a^\bot\wedge a^\bot) &=-[d\Phi\wedge a^\bot,a^\bot]-[[a^{\|},a^\bot],a^\bot]. 
\end{align*} 
Adding them together and using \eqref{e0.109} gives {\rm (iii)}.
\end{proof}

\refL{sdcurv}(iii) implies $d(a^\bot\wedge a^\bot)\in L^{3/2}$ for an admissible $a$ and the difficulty we had
with the convergence of $u^*\Theta$ is eliminated. Let us formalize this observation.
\begin{definition}[Convergence in $\mathcal{E}(M,G)$]
A sequence $u_n\in\mathcal{E}(M,G)$ weakly converges to $u$ in
$\mathcal{E}(M,G)$ if for $a_n:=u_n^{-1}du_n$ and $a:=u^{-1}du$ one has
\bee\label{e0.111}
\bal
&{\rm 1)}\ u_n\ovs{W^{1,2}}\rightharpoonup u\quad (\text{and hence} \quad a_n^\bot\ovs{L^2}\rightharpoonup a^\bot)\\
&{\rm 2)}\ a_n^\bot\wedge a_n^\bot\ovs{L^2}\rightharpoonup
a^\bot\wedge a^\bot\\
&{\rm 3)}\ a_n^{\|}\ovs{W^{1,2}}\rightharpoonup a^{\|}.
\eal
\eee
We denote this convergence by $u_n\ovs{\mathcal{E}}\rightharpoonup u$. The strong convergence $\xra{\mathcal{E}}$ is obtained by replacing the weak convergences above by the strong ones in the same Banach spaces.
\end{definition}
Keep in mind that although the notation does not reflect it the definition of the space $\mathcal{E}(M,G)$ does depend on the homogeneous space under consideration since this space determines the isotropic decomposition $a^{\|}+a^\bot$ of a potential $a$.
Now the above discussion yields:
\begin{lemma}\label{L:sconv}
If $X$ is a symmetric space then $u_n\ovs{\mathcal{E}}\rightharpoonup u$ implies
$u_n^*\Theta\ovs{\mathcal{D}'}\rightharpoonup u^*\Theta$ 
and therefore the secondary invariants of $u_n$ converge to those of $u$:
$$
\int\limits_Mu_n^*\Theta\to \int\limits_Mu^*\Theta.
$$ 
\end{lemma}
Our next observation is that the weak convergence also behaves well with respect to the gauge-fixing.
For two sequences of maps $u_n\ovs{\mathcal{E}}\rightharpoonup u$,
$v_n\ovs{\mathcal{E}}\rightharpoonup v$ does not necessarily imply
$u_nv_n\ovs{\mathcal{E}}\rightharpoonup uv$. As a matter of fact, 
$u_nv_n$ may not even belong to $\mathcal{E}(M,G)$. Recall that in the proof of \refT{3.1} we had to multiply $u_n$-s from a minimizing sequence by 'gauge transformations'
$$
w_n\in W^{2,2}(H_\pfi)=\{w\in W^{2,2}(M,G)|\ w\pfi=\pfi\}
$$ 
to control the norms of $a_n^\vert$.
\begin{lemma}\label{L:cow}
Let $u_n\ovs{\mathcal{E}}\rightharpoonup u$ and either
$w_n\ovs{C^\infty}\lra w$ or $w_n\in W^{2,2}(H_\pfi)$ and $w_n
\ovs{W^{2,2}}\lra w$ then $u_nw_n\ovs{\mathcal{E}}\rightharpoonup
uw$
\end{lemma}
\begin{proof}
$C^\infty$ case follows trivially from the definition. For the
second case note that {\rm 2)} in \eqref{e0.111} can be replaced by
\bee\label{e0.112}
D_\pfi a_n\wedge D_\pfi
a_n\ovs{L^2}\rightharpoonup D_\pfi a\wedge D_\pfi a
\eee
with $D_\pfi a:=a^\bot+\pfi^*\omega^\bot$ (see \refD{FadSkyrp}). The gain is that
for $a_n^{w_n}=(u_nw_n)^{-1}$ $d(u_nw_n)$ and $w_n\in W^{2,2}(H_\pfi)$
\bee\label{e0.113}
D_\pfi(a_n^{w_n})=\Ad_*(w_n^{-1})(D_\pfi a_n)\quad \text{a.e.}
\eee
Since $W^{2,2}(M,G)\subset C^0(M,G)$ by the Sobolev embedding theorems we have
$$
w_n\ovs{C^0}\lra w,\quad \Ad_*(w_n^{-1})\ovs{C^0}\lra \Ad_*(w^{-1})
$$
and therefore
\begin{multline*}
D_\pfi(a_n^{w_n})\wedge D_\pfi(a_n^{w_n})=\Ad_*(w_n^{-1})(D_\pfi a_n\wedge D_\pfi a_n)\\
\ovs{L^2}\rightharpoonup \Ad_*(w^{-1})(D_\pfi a\wedge D_\pfi a)=D_\pfi(a^w)\wedge D_\pfi(a^w).
\end{multline*}
The conditions {\rm 1)}, {\rm 3)} in \eqref{e0.111} can be checked similarly using in \eqref{e0.113}
and the fact that $\Ad_*(w^{-1})$ commutes with $\pr_{\h_\pfi}$, $\pr_{\h_\pfi^\bot}$ when
$w\pfi=\pfi$.
\end{proof}

There is one more thing that one would like to have for the secondary invariants. For smooth maps $M\xra{u,v}G$
\refL{1.2} and \eqref{deRham} imply 
\bee\label{additv}
\int\limits_M(uv)^*\Theta=\int\limits_Mu^*\Theta+\int\limits_Mv^*\Theta.
\eee
Of course one can not expect \eqref{additv} to hold when both $u,v$ are just admissible (the lefthand side may not be defined in this case) but even assuming that $v$ is smooth it is unclear if \eqref{additv} holds for all admissible $u$. 
Thus to have the secondary invariants behave 'reasonably' we need to work with maps that are
'closer' to smooth ones than arbitrary admissible maps.

\begin{definition}[Strongly admissible maps]\label{D:sadmis}
Denote by $\mathcal{E}'(M,G)$ the sequentially weak closure of
$C^\infty(M,G)$ in $\mathcal{E}(M,G)$. We call elements of
$\mathcal{E}'(M,G)$ strongly admissible. For maps into $X$ set $\psi\in\mathcal{E}'(M,X)$ if $\psi=u\pfi$ for
$u\in\mathcal{E}'(M,G)$, $\pfi\in C^\infty(M,X)$.
\end{definition}
Similarly constructed spaces have been used in \cite{Es1,GMS1} 
for similar minimization problems. It may well be that 
$$
\mathcal{E}(M,X)=\mathcal{E}'(M,X)
$$
but the question is still open even for $X=SU_2$ (see \cite{Es2}). 
If $\mathcal{E}'\neq\mathcal{E}$ one may ask whether the Lavrentiev phenomenon takes place, i.e.
\be
\inf_{\psi\in C^\infty}E(\psi)=\inf_{\psi\in\mathcal{E}'}E(\psi)<\inf_{\psi\in\mathcal{E}}E(\psi)?
\ee
This phenomenon is known to take place for the Dirichlet
energy \cite{GMS2}. Just from the definition we can only claim that
$W^{2,2}(M,G)\subset\mathcal{E}'(M,G)$ (in fact it is contained even
in the strong closure of $C^\infty$ in $\mathcal{E}$).

\begin{definition}[Homotopy sector]\label{D:hsector}
An element $\psi\in\mathcal{E}'(M,X)$ is in the homotopy sector of
$\pfi$ and we write $\psi\in\mathcal{E}_\pfi'$ if
\bee\label{e0.125}
\bal
&{\rm 1)}\ \psi=u\pfi\quad \text{with}\quad u\in\mathcal{E}'(M,G)\\
&{\rm 2)}\ \int\limits_Mu^*\Theta=0\mod\O_\pfi
\eal
\eee
\end{definition}
\noindent If $\psi\in C^1(M,X)$ then
$\psi\in\mathcal{E}_\pfi'$ if and only if $\psi$ is homotopic to
$\pfi$ by \refT{1.2}. 

The next Lemma shows that strongly admissible maps are 'topologically reasonable'.
\begin{lemma}\label{L:sadmis}
Let $X=G/H$ be a symmetric space, $M\ovs{\pfi}\lra G$ a smooth
reference map. Then

\noindent {\rm (i)\bf(integrality)} For a strongly admissible map $u\in\mathcal{E}'(M,G)$ the secondary invariants are integral:
$$
\int\limits_Mu^*\Theta\in\Z^N.
$$
\noindent {\rm (ii)\bf(stabilizer)} If $w\in W^{2,2}(M,G)$ stabilizes $\pfi$, i.e. $w\in W^{2,2}(H_\pfi)$ then
$$
\int\limits_Mw^*\Theta=0\mod\O_\pfi
$$
\noindent {\rm (iii)\bf(additivity)} If $u\in\mathcal{E}'(M,G)$ and either $w\in
C^\infty(M,G)$ or $w\in W^{2,2}(H_\pfi)$ then $uw\in\mathcal{E}'(M,G)$ and
\bee\label{e0.119}
\int\limits_M(uw)^*\Theta=\int\limits_M u^*\Theta+\int\limits_M w^*\Theta
\eee
\noindent {\rm (iv)\bf(change of reference)} If $\widetilde{\pfi}$ is smooth and homotopic to $\pfi$ then 
$$
\mathcal{E}_\pfi'=\mathcal{E}_{\widetilde{\pfi}}'
$$
\noindent {\rm (v)\bf(smooth representative)}
Every homotopy sector of strongly admissible maps contains a smooth representative, i.e.
\be
\mathcal{E}'(M,X)=\bigcup\limits_{\pfi\in C^\infty(M,X)}\mathcal{E}_\pfi '
\ee
\end{lemma}
\begin{proof}
{\rm (i)} Let $u_n$ be smooth and
$u_n\ovs{\mathcal{E}}\rightharpoonup u$. Then $\int\limits_M
u_n^*\Theta\in\Z^N$ and by \refL{sconv} $\int\limits_Mu^*\Theta\in\Z^N$.
\bigskip

\noindent {\rm (ii)} If $F$ is any manifold then $W^{2,2}(M,F)\subset
C^0(M,F)$ (recall that $\mbox{dim}M=3$) and therefore
$C^\infty(M,F)$ is dense in $W^{2,2}(M,F)$ \cite{Bt}. Since
the approximation property is local it extends to bundles and
$C^\infty(H_\pfi)$ is dense in $W^{2,2}(H_\pfi)$. Since
$\int\limits_Mw^*\Theta\in  \O_\pfi$ for $w\in C^\infty(H_\pfi)$
we get {\rm (ii)} by passing to limit.
\bigskip

\noindent {\rm (iii)} As we know \eqref{e0.119} holds for smooth $u$, $w$ (see \eqref{additv}). If
$u_n\ovs{\mathcal{E}}\rightharpoonup u$,
$w_n\ovs{W^{2,2}}\lra w$ then by \refL{cow}
$u_nw_n\ovs{\mathcal{E}}\rightharpoonup uw$ and by \refL{sconv} this
implies convergence of $\int\limits_M(u_nw_n)^*\Theta$. Hence \eqref{e0.119} holds in
the limit.
\bigskip

\noindent {\rm (iv)} Since $\widetilde{\pfi},\pfi$ are both smooth and homotopic 
it follows from \refC{1.2} that there is a smooth $v$ such that $\widetilde{\pfi}:=v\pfi$ 
and $v$ is nullhomotopic, in particular $\int\limits_Mv^*\Theta=0$.
Also $\O_{\widetilde{\pfi}}=\O_\pfi$ by \refC{1.5}. Let $\psi=u\pfi\in\mathcal{E}_\pfi'$ be arbitrary. By definition of $\mathcal{E}_\pfi'$ we have $\int\limits_Mu^*\Theta=0\mod\O_\pfi$. 
Set $\widetilde{u}:=uv^{-1}$ then $\psi=\widetilde{u}\widetilde{\pfi}$ and by (iii):
\be
\int\limits_M\widetilde{u}^*\Theta
=\int\limits_Mu^*\Theta+\int\limits_M(v^{-1})^*\Theta
=\int\limits_Mu^*\Theta-\int\limits_Mv^*\Theta=0\mod\O_\pfi=\O_{\widetilde{\pfi}}.
\ee
Thus $\psi\in\mathcal{E}_{\widetilde{\pfi}}'$ and $\mathcal{E}_\pfi'\subset\mathcal{E}_{\widetilde{\pfi}}'$. The other inclusion follows by switching $\pfi$ and $\widetilde{\pfi}$.
\bigskip

\noindent {\rm (v)} By definition of $\mathcal{E}'(M,X)$ for any map $\psi\in
C^\infty(M,X)$ there is $\widetilde{\pfi}\in C^\infty(M,X)$ and $\widetilde{u}\in\mathcal{E}'(M,G)$ with
$\psi=\widetilde{u}\widetilde{\pfi}$. Then the vector
$$
\nu:=\int\limits_M\widetilde{u}^*\Theta
$$ 
is in $\Z^N$ by (i). By the Eilenberg classification theorem \cite{St} there is a $v\in C^\infty(M,G)$
such that 
$$
\int\limits_Mv^*\Theta=\nu.
$$ 
Set $u:=\widetilde{u}v^{-1}$, $\pfi:=v\widetilde{\pfi}$ then still $\psi=u\pfi$. 
By (iii) $u\in\mathcal{E}'(M,G)$ and
\be
\int\limits_Mu^*\Theta=\int\limits_M\widetilde{u}^*\Theta-\int\limits_Mv^*\Theta=0
\ee
so $\psi\in\mathcal{E}_\pfi'$, where $\pfi$ is smooth by construction.
\end{proof}

Thus we found a class that is closed under both the gauge-fixing and
weak limits. It may even be argued (see \cite{GMS1}) that this class is
more 'natural' than $\mathcal{E}(M,X)$ for minimization since we
really want to minimize energy over smooth maps.
An essential restriction of course is that it only works for symmetric
spaces but this appears to be the natural generality. Our main secondary 
minimization result is next.
\begin{theorem}\label{T:11}
Let $X$ be a symmetric space. Then every homotopy sector of strongly
admissible maps contains a minimizer of Faddeev-Skyrme energy.
\end{theorem}
\begin{proof}
We proceed as in the proof of \refT{3.1} by choosing a
minimizing sequence $\psi_n=u_n\pfi$, $u_n\in\mathcal{E}'(M,G)$ and
$\int\limits_Mu_n^*\Theta\in\O_\pfi$. Gauge-fixing replaces
$u_n$ by $u_nw_n$ with $w_n\in W^{2,2}(H_\pfi)$ and by \refL{sadmis}(ii),(iii)
\be
\int\limits_M(u_nw_n)^*\Theta=\int\limits_Mu_n^*\Theta+\int\limits_Mw_n^*\Theta=0\mod\O_\pfi,
\ee
i.e. we may assume having $u_nw_n$ from the start
and drop $w_n$ from the notation. Now setting $a_n=u_n^{-1}du_n$ we have
$a_n^{\|}\ovs{W^{1,2}}\rightharpoonup a^{\|}$ since $u_n$ is
gauge-fixed. As in the proof of primary minimization we establish on
a subsequence
\be
\bal
&u_n\ovs{W^{1,2}}\rightharpoonup u\\
&a_n^\bot\ovs{L^2}\rightharpoonup a^\bot\\
&a_n^\bot\wedge a_n^\bot\ovs{L^2}\rightharpoonup a^\bot\wedge a^\bot,
\eal
\ee
where $a:=u^{-1}du$. But this means that
$u_n\ovs{\mathcal{E}}\rightharpoonup u$ and by \refL{sconv}
\be
u_n^*\Theta\ovs{\mathcal{D}'}\rightharpoonup u^*\Theta,
\ee
i.e. $\int\limits_Mu^*\Theta\in  \O_\pfi$. Since $u$ is a limit in
$\mathcal{E}$ of maps from $\mathcal{E}'$ it is in $\mathcal{E}'$
itself and hence $\psi=u\pfi\in\mathcal{E}_\pfi '$. As in the proof of \refT{3.1}
$$
E(\psi)\leq\liminf_{n\to\infty}E(\psi_n)
$$ 
and since $\psi_n$ was a minimizing sequence $\psi$ is a minimizer in $\mathcal{E}_\pfi'$.
\end{proof}

\backmatter

\chapter{Conclusions}

In this section we describe some directions for future work suggested by the study of Faddeev-Skyrme models.
Due to rich geometric and analytic structure Faddeev-Skyrme models manifest multiple connections with the geometric knot theory, the theory of harmonic maps, non-linear elastisity and other classical fields. Different interpretations of the energy functional lead to a number of non-trivial geometric, topological and analytic questions.

One of the central problems in the geometric knot theory is minimizing magnetic energy among all divergence-free fields (closed $2$--forms) with a given helicity \cite{CDG}. The Faddeev-Skyrme functional on homogeneous spaces can be considered as a non-Abelian generalization of this problem with vector fields replaced by pullbacks of the curvature forms and the secondary invariants playing the role of helicity. This suggests a study of minimizers involving subtle properties of a map, e.g., related to the knot type of its solitonic center as in \cite{FH}. This should help answer questions like: at what energy levels should one expect the appearence of a particular knot as the center of a minimizer? Are there several minimizers in the same homotopy class? Currently the fine geometry of the Faddeev-Skyrme minimizers remains purely conjectural even in the case of $S^2$ \cite{FN1,LY2}.

From analytical point of view the Faddeev-Skyrme functional is very similar to functionals in the theory of harmonic maps and non-linear elasticity \cite{EL,GMS4}. Indeed if in the expression for energy 
$$
E(u)=\int\frac12 |d u\,u^{-1}|^2\,
+\,\frac14 |d u\,u^{-1}\wedge d u\,u^{-1}|^2\;dx\,.
$$
the metric on $G$ is bi-invariant then $|du\,u^{-1}|= |du|$ and the first term describes Dirichlet energy of $u$. The same holds for homogeneous spaces as $|u^*\omega^\perp|=|du|$. The second term is reminiscent of expressions for elastic energy in non-linear models (in fact when $G=SU_2\simeq S^3$ it coincides with one of them).

It is well known that for harmonic maps with a target space $X$ the phenomenon of 'bubbling' occurs, i.e. spherical components split off at the limit when $\pi_2(X)\neq0$. Similar effects are known in the elasticity theory as 'cavitation'. When $\pi_2(X)=0$ (no bubbling) regularity theory for harmonic maps implies that solutions are H\"older continuous. Results of my thesis imply that when $X$ is a symmetric space bubbling does not happen for the Faddeev-Skyrme energy even if $\pi_2(X)\neq0$. In the presense of additional non-linear terms however even the absense of 'bubbling' or 'cavitation' is no guarantee that minimizers are H\"older continuous \cite{GMS4}. They may be mildly singular and behave 'like smooth maps' for the purposes of integration by parts. It is curious to find out what happens in the cases of simply connected Lie groups and non-simply connected symmetric spaces as the targets. 

Questions about bubbling underscore the absense of a regularity theory for Faddeev-Skyrme minimizers similar to the one for harmonic maps \cite{EL}. An important step in this direction would be proving Conjectures \ref{Cj:1}, \ref{Cj:2} that provide an explicit description of admissible maps (as finite energy maps) and ensure density of smooth maps among them in the topology dictated by the energy functional. Establishing these links is necessary for applying classical ideas of regularity theory to maps described via gauge potentials. For $S^2$ as the target equivalents of these conjectures are proved in \cite{AK3}.

It does not seem likely that no bubbling occurs for an arbitrary simply connected homogeneous $X$. However, the gauge methods seem to be well suited for proving that it is avoided when the target space is a flag manifold $X=G/\T$ ($\T$ is a maximal torus of a Lie group $G$). The flag manifold targets appear in the Faddeev-Niemi conjecture \cite{FN2} which states that the $SU_{n+1}/\T^n$ Faddeev-Skyrme model describes the low-energy limit of the $SU_{n+1}$ Yang-Mills theory. This motivates studying the topology of the configuration spaces of the $SU_{n+1}/\T^n$ Faddeev-Skyrme models and comparing it to the topology of the Yang-Mills configuration space. For the case of the $2$--sphere the fundamental group and the real cohomology ring of the configuration space were computed in \cite{AS} and it is instructive to generalize the computation to the case of flag manifolds.

One can also try to replace closed $3$-manifolds as domains of the maps in Faddeev-Skyrme models. Whereas the results of this thesis generalize to bounded domains in $\R^3$ rather straightforwardly, it is not the case with non-compact manifolds, unbounded domains in $\R^3$ or even $\R^3$ itself. As suggested by \cite{KV,LY2} an important step in analyzing Faddeev-Skyrme models on $\R^3$ is to obtain an asymptotic growth estimate for energy of minimizers as a function of their topological numbers (the degree, Hopf invariant, etc.). We know that the growth is linear for Lie groups and fractional with power $3/4$ for $SU_2/U_1$. It is interesting that for bounded domains there is a linear lower bound on energy even if the Dirichlet term $|du|^2$ is dropped \cite{CDG}. One would want to find analogous growth estimates for other homogeneous spaces $G/H$ and investigate the dependence of the power of the growth on a way $H$ sits inside of $G$ for both bounded and unbounded domains. 

The concentration-compactness method used in \cite{LY2} so far does not give complete solution to the existence of Faddeev-Skyrme minimizers on $\R^3$ or its unbounded domains. The minimization problem on $\R^3$ has a specific difficulty of maps 'jumping' from one homotopy class to another at the limit due to effects at infinity. On the other hand, the Uhlenbeck compactness theorem has been recently generalized to some non-compact manifolds in \cite{Wr}. Hopefully the gauge methods of this work combined with these new results will lead to a complete solution for $\R^3$.

\par\vfill\pagebreak

%%%%%%%%%%%%%%%%%%%%%%%%%%%%%%%%%%%%%%%%%%%%%

%BIBLIOGRAPHY

%%%%%%%%%%%%%%%%%%%%%%%%%%%%%%%%%%%%%%%%%%%%%

{
\renewcommand{\baselinestretch}{1}
%\renewcommand{\bibname} {}

%\cleardoublepage
\addcontentsline{toc}{chapter}{Bibliography}
%\bibliography{frooble}

}

\end{document}